\documentclass[aps,pra,twocolumn,superscriptaddress,longbibliography,10pt]{revtex4-2}

\usepackage[english]{babel}
\usepackage[utf8]{inputenc}

\usepackage{amsthm}
\usepackage{amsmath,bm,bbm}
\usepackage{amssymb}
\usepackage{amsfonts}
\usepackage{graphicx}
\graphicspath{{manuscript_figures/}}
\usepackage{txfonts}
\usepackage{xcolor}
\usepackage{float}
\usepackage{braket}

\usepackage{makecell}
\usepackage{multirow}
\usepackage[squaren]{SIunits}

\usepackage[colorlinks=true,linkcolor=blue,citecolor=blue,urlcolor=blue]{hyperref}

\usepackage{bbold}

\newcommand{\avg}[1]{\langle#1\rangle}

\newcommand{\mhe}{m_\text{He3}}
\newcommand{\vmean}{\overline{v}}	
\newcommand{\vrms}{v_\text{rms}}

\newcommand{\reTtt}{\Re\TT{2}{2}}
\newcommand{\imTtt}{\Im\TT{2}{2}}

\newcommand{\TT}[2]{T^{#1}_{#2}}

\newcommand{\Ncell}{{N_\text{cell}}}
\newcommand{\ncell}{{n_\text{cell}}}
\newcommand{\nph}{{n_\text{ph}}}

\newcommand{\Abeam}{{A_\text{beam}}}
\newcommand{\Acell}{{A_\text{cell}}}

\newcommand{\Npass}{{N_\text{pass}}}
\newcommand{\Gsq}{\Gamma_{\text{sq}}}

\newcommand{\he}{$^3$\text{He}\;}

\begin{document}

\title{Effective Faraday interaction between light and Helium-3 nuclear spins in a multi-pass cell}

\author{Kaiwen Yi}
\affiliation{School of Electronics, Peking University, Beijing 100871, China}
\author{Yida Sha}
\affiliation{School of Electronics, Peking University, Beijing 100871, China}
\author{Zejia Lin}
\affiliation{School of Electronics, Peking University, Beijing 100871, China}
\author{Matteo Fadel}
\email{fadelm@phys.ethz.ch}
\affiliation{Department of Physics, ETH Z\"{u}rich, 8093 Z\"{u}rich, Switzerland}
\author{Xiang Peng}
\email{xiangpeng@pku.edu.cn}
\affiliation{School of Electronics, Peking University, Beijing 100871, China}

\begin{abstract}
Helium-3 nuclear spins form an exceptionally stable quantum system with extremely long coherence time, offering exciting opportunities for quantum technologies. 
In particular, nuclear spin-squeezed states promise enhanced precision for sensing tasks and tests of new physics.
A central challenge for all these applications is the realization of a controllable light–nuclear spin interface. 
Here we experimentally demonstrate such an interface by exploiting metastability-exchange collisions in a low-pressure helium-3 gas cell at room temperature. 
A radio-frequency discharge produces a small population of metastable atoms that both enables efficient optical pumping and mediates an effective Faraday interaction between the collective nuclear spin and an optical probe. 
We quantitatively characterize the strength of this interaction as a function of the nuclear polarization, applied magnetic field, and probe-beam parameters.
Moreover, we show that using a multi-pass cell enhances this interaction by effectively increasing the optical depth. Extrapolating to a tenfold increase of the probe power used in the present experiment, we project a measurement-induced squeezing rate of $\unit{0.52}{s^{-1}}$. Our results provide a practical pathway for optical access to helium-3 nuclear spins and open prospects for generating long-lived, macroscopic nuclear spin-squeezed states for quantum metrology.
\end{abstract}

\maketitle

The nuclear spin of ground-state helium-3 atoms forms a remarkably well-isolated two-level quantum system, protected by the closed electronic shell and separated from the nearest electronic excitation by roughly \unit{20}{eV}. 
This isolation underlies the exceptionally long coherence times, extending to several days~\cite{2017-TRGentile-RMP}, that have been demonstrated in macroscopic ensembles, making \he nuclear spins extraordinarily stable and sensitive probes for precision measurements. 
Consequently, spin-polarized \he gas has become a key resource across a wide range of technologies and fundamental-physics applications, including atomic magnetometry~\cite{2010-CGemmel-EPJD,2017-Heil-book}, medical magnetic-resonance imaging~\cite{2015-MJcouch-MIB,1996-Ebert-Lancet,2013-GCollier-JAP}, inertial navigation via nuclear-spin gyroscopes~\cite{2011-JKitching-IEEE}, and neutron-spin filtering and analysis in scattering experiments~\cite{2020-Okudaira-NIMPRSA,2014-Chen-JPCS,2014-CYJiang-RSI,2011-CBFu-PhysicaB}.
Moreover, the combination of long intrinsic lifetime, low magnetic sensitivity to environmental perturbations, and room-temperature operation has established macroscopic \he nuclear polarization as a promising medium for generating and controlling collective quantum correlations~\cite{2005-ADantan-PRL,2007-GReinaudi-JMO}. 
These properties open a pathway toward entangled spin states with enhanced robustness and scalability, providing substantial potential for applications in quantum metrology and information storage.

\begin{figure}[t!]
    \centering
    \includegraphics[width=0.95\linewidth]{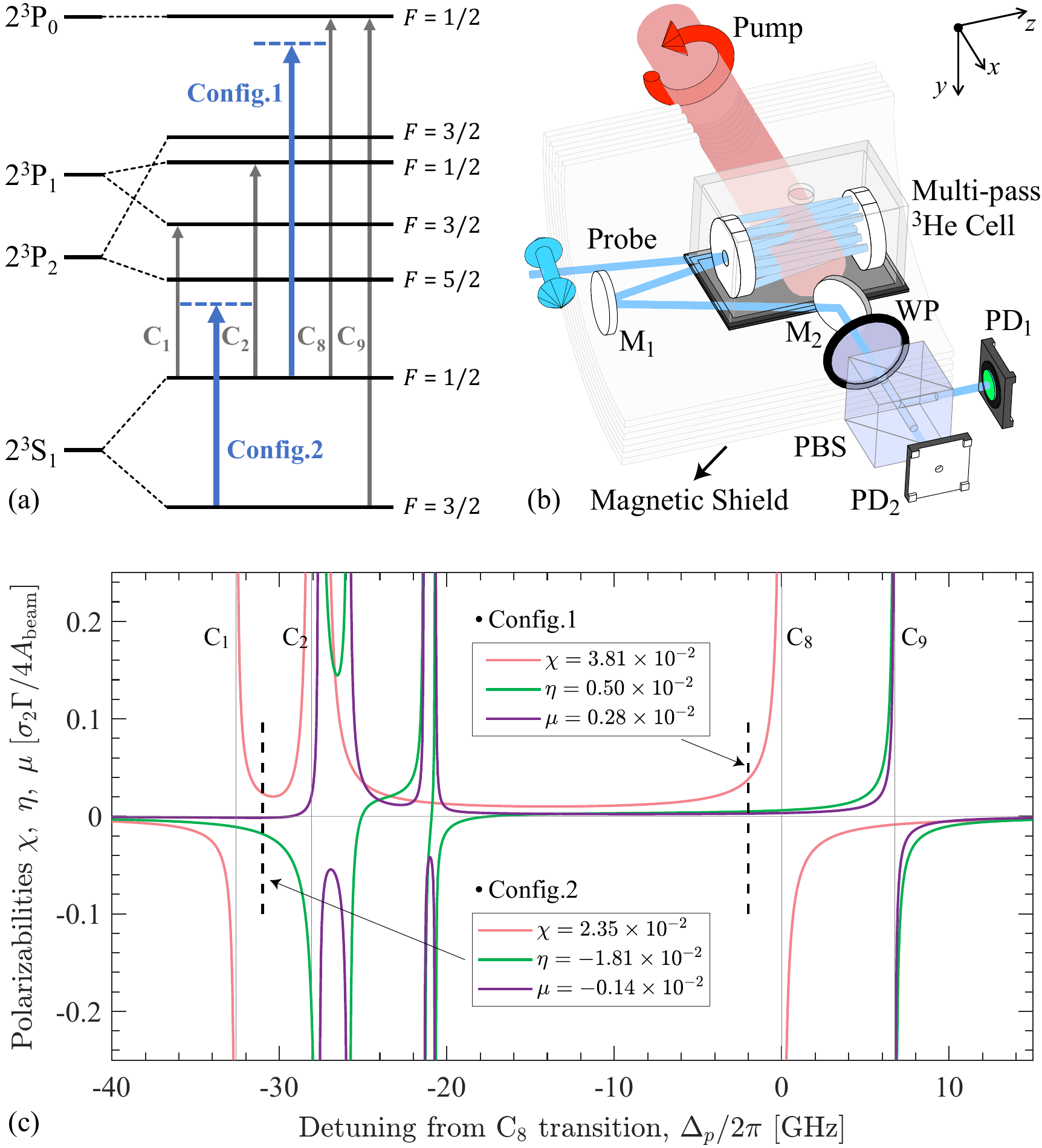}
    \caption{
    \textbf{\he transitions and experimental setup.}
    (a) Fine structure and hyperfine structure of states $2^3 \mathrm{S}_1$ and $2^3 \mathrm{P}$ in ${ }^3 \mathrm{He}$ (not to scale). The two probe-frequency operating points employed in this work are marked as ``Config.1" and ``Config.2". (b) Schematic of the experimental setup. $\mathrm{PBS}$: polarization beam splitter; M: mirrors; $\mathrm{PD}$: photodiodes; WP: half-wave plate for $S_y$ measurement. (c) Coupling constants $\chi$, $\eta$, $\mu$ for the metastable $F=1/2$ and $F=3/2$ levels, plotted in units of $\sigma_2 \Gamma /(4A_{\text{beam}})$ as a function of the light-frequency detuning $\Delta_p$ referenced to the $C_8$ transition.
    The values of coupling constants in the two probe detuning configurations are highlighted.}
    \label{fig:1}
\end{figure}

Despite being crucial for the extraordinarily long coherence times of \he nuclear spins, their strong decoupling from the environment also makes preparation and control technically challenging. 
Nuclear polarization is typically generated through two well-established indirect optical-pumping techniques~\cite{2017-TRGentile-RMP}: spin-exchange optical pumping (SEOP)~\cite{2011-TGWalker-JPCS} and metastability-exchange optical pumping (MEOP)~\cite{1963-FDColegrove-PR,2011-MBatz-JPCS}. 
In both approaches, optically driven electronic polarization is transferred to the nuclear spin by spin-exchange collisions, either between different atomic species in SEOP or between ground-state and metastable \he atoms in MEOP.
For nuclear-spin detection, standard readout methods rely on nuclear magnetic resonance, where the collective magnetization is monitored via pickup coils that detect the free-induction decay (FID) signal~\cite{2019-MFarooq-thesis}. 
In addition, metastability-exchange collisions (MECs) enable optical probing of the ground-state spin dynamics by mapping the nuclear-spin polarization onto the optically accessible metastable state, allowing all-optical detection of nuclear magnetic resonance~\cite{2024-Yuting-SCPMA}.

In recent years, the optical quantum control of noble-gas nuclear-spin ensembles has attracted considerable interest due to its potential for generating nuclear spin-squeezed states with exceptionally long coherence times~\cite{2007-GReinaudi-JMO,2021-ASerafin-PRL,2021-ASerafin-CRPhys,2020-OKatz-PRL}. 
Such states offer new opportunities for quantum-enhanced sensing \cite{2018-RMP-Pezze} and broader quantum-technology applications. 
Central to many optical control protocols is the Faraday interaction between the collective atomic spin and the Stokes vector of light, which provides a powerful and versatile spin–light quantum interface~\cite{1998-AKuzmich-EPL,TakahashiPRA99,2010-KHammerer-RMP}. 
This mechanism is well established in alkali-metal ensembles, where it has been used to generate strong spin squeezing via quantum-non-demolition measurements~\cite{TakanoPRL09,SewellPRL12,2015-GVasilakis-NP,Kong2020,2020-HBao-Nature}. 
Extending such interactions to \he nuclear spins, however, requires an intermediary that mediates an effective coupling between the optical field and the nuclear spin.

One established pathway employs noble-gas–alkali hybrid systems, where spin-exchange collisions between the two species mediate the interaction~\cite{2020-OKatz-PRL,2022-OKatz-PRX,2022-RShaham-NP,2021-OKatz-SA}. 
These systems typically operate at elevated temperatures (200$^\circ$C) to maintain sufficient alkali-metal vapor density and at helium-3 pressures of several atmospheres. 
Moreover, the nuclear polarization in such mixtures relies on SEOP, whose efficiency is generally lower than that of MEOP~\cite{2017-TRGentile-RMP,2024-Yuting-SCPMA}. 
These constraints have motivated the exploration of alternative schemes based on metastable \he, where MECs coherently map the optically addressable electronic polarization onto the nuclear-spin degree of freedom, thereby establishing an effective Faraday interaction directly with the nuclear spins~\cite{2021-ASerafin-CRPhys,2024-MFadel-NJP,2021-ASerafin-PRL}. 
Such approaches can operate at room temperature and millibar pressures, and the interaction strength can be switched simply by turning the sustaining discharge on or off. To date, however, this concept remains purely theoretical, with no experimental demonstration.

In this work, we investigate the effective interaction between light and the ground-state \he nuclear spin mediated by metastable \he atoms. 
Our experiments are carried out in either a single-pass or a multi-pass gas cell containing pure \he at room temperature and low pressure. A radio-frequency discharge maintains a small fraction of atoms in the metastable state, enabling optical polarization of the ground-state nuclear spin via standard MEOP. We demonstrate that, under these conditions, metastability-exchange collisions can also mediate an effective Faraday interaction between an off-resonant probe light and the nuclear spin polarization.
We systematically characterize the strength of this interaction as a function of the nuclear polarization, the applied magnetic field, the probe-beam parameters, and the effective optical depth controlled by multiple passes of the probe through the cell. These measurements represent a crucial first step toward optical quantum control of ground-state \he nuclear spins, including the generation of nuclear-spin–squeezed states, and open a pathway to a range of applications.

\vspace{2mm}
\textbf{Theoretical model.---} 
We consider an ensemble of \he atoms, composed by $\Ncell$ atoms in the ground state and a small fraction $\ncell\approx 10^{-6}\Ncell$ in the metastable state.
The collective nuclear spin state of the ground state atom is described by the operator $\mathbf{I}$, while the collective spin states in the $2^3S_1$ $F=1/2$ and $F=3/2$ manifolds are described by the operators $\mathbf{K}$ and $\mathbf{J}$, respectively \cite{2021-ASerafin-PRL,2021-ASerafin-CRPhys}. The relevant energy level diagram is shown in Fig.~\ref{fig:1}(a).

Illuminating the ensemble with a light field propagating in the $z$ direction and addressing the $2^3 S - 2^3 P$ transition with a large detuning results in the interaction Hamiltonian with the metastable atoms $H_{\text{LA}}=H_{1/2}^V+H_{3/2}^V + H_{3/2}^T$ \cite{2024-MFadel-NJP}. 
Here, 
\begin{equation}
H_{1/2}^V=\hbar \chi K_z S_z \qquad \text{and} \qquad H_{3/2}^V = \hbar \eta J_z S_z \;
\label{eqHv1/2and3/2}
\end{equation}
describe the vector Faraday interaction with the $F=1/2$ and $F=3/2$ manifolds with strengths $\chi$ and $\eta$, respectively, and $S_z=(a_x^{\dagger} a_y-a_y^{\dagger} a_x) / 2 i$ is the $z$-component of the light Stokes operator as a function of the $x$- and $y$-polarized modes.
The term $H_{3/2}^T$ describes the tensor interaction with the $F=3/2$ manifold and is proportional to the constant $\mu$. We leave its expression to Sec.~\ref{supp:secHLA} of \cite{SM}, as in the chosen parameter regime $\mu\approx 0$, which allows us to neglect this term.

The coupling constants $\chi$, $\eta$ and $\mu$ depend on the specific atomic structure and on the light beam waist and frequency~\cite{2024-MFadel-NJP}, see Fig.~\ref{fig:1}(c) and Sec.~\ref{supp:secHLA} of \cite{SM}.
In our study, we are interested in investigating two complementary configurations.
The first, marked as ``Config.1'' in Figs.~\ref{fig:1}(a) and \ref{fig:1}(c), is at a light frequency detuning of $\Delta_p/(2\pi)=\unit{-2}{GHz}$ with respect to the $C_8$ transition, where $\chi \gg \eta,\mu$ allows us to consider only $H_{1/2}^V$. 
The second, marked as ``Config.2'', is for $\Delta_p/(2\pi)=\unit{-31}{GHz}$,
around a local minimum of absorption~\cite{2024-MFadel-NJP} and where one could work with a large nuclear spin polarization for which the $F = 1/2$ spin manifold is empty.
Together with the fact that $\mu\ll \chi,\eta$, this configuration allows us to consider only $H_{3/2}^V$.

Besides interacting with the light, metastable atoms are interacting with the ground state atoms through MECs. 
These are rapid processes in which the metastable excitation is exchanged between a metastable and a ground-state atom, thereby transferring the associated electronic–nuclear spin degrees of freedom between the two atoms \cite{Partridge, 2017-TRGentile-RMP}.
Remarkably, it was shown in Refs.~\cite{2021-ASerafin-PRL,2021-ASerafin-CRPhys} that MECs can mediate a coherent interaction between the light and the nuclear spin, that takes the form of the effective Faraday Hamiltonian $H_\text{eff} \propto I_z S_z$. 
For $x$-polarized light and nuclear spins, $\avg{S_x}=\nph/2$ and $\avg{I_x}=M \Ncell/2$, where $n_\text{ph}$ is the photon flux, and $M\in[0,1]$ is the nuclear polarization, it is possible to consider the Holstein-Primakoff approximation $P_S=S_z/\sqrt{\avg{S_x}}$, $X_S=S_y/\sqrt{\avg{S_x}}$, $P_I=I_z/\sqrt{\avg{I_x}}$, $X_I=I_y/\sqrt{\avg{I_x}}$, to write
\begin{equation}\label{eq:effHPSPI12}
    H_{\text{eff}} = \hbar \Omega^{(i)}  P_S P_I .
\end{equation}
Here, we have introduced the effective coupling strength
\begin{equation}\label{eq:OmegaTh}
\Omega^{(i)} = v^{(i)} \dfrac{\ncell}{\Ncell} \sqrt{\nph \Ncell} f^{(i)}(M) ,
\end{equation}
where $v^{(i)}\in\{\chi,\eta \}$ depending on the spin manifold $i$ relevant in the chosen configuration, and $f^{(i)}(M)$ a polarization-dependent scaling function \cite{2024-MFadel-NJP}.
Eq.~(\ref{eq:effHPSPI12}) is derived assuming a weak Faraday coupling between light and metastable atoms, a metastable MEC time much shorter than the metastable Larmor period, and a quasi-static nuclear polarization during probing~\cite{2021-ASerafin-CRPhys,2021-ASerafin-PRL}.
Under the approximation where the irrelevant spin manifold is adiabatically eliminated (see the simplified model described in Sec.~\ref{supp:secEffCoup} of \cite{SM}),
\begin{subequations}\label{eqs:fMscalings}
\begin{align}
    f^{(1/2)}(M) &= \dfrac{1}{2} \left(\dfrac{1-M^2}{3+M^2}\right)\sqrt{M} , \\
    f^{(3/2)}(M) &= \left(\dfrac{5+M^2}{3+M^2}\right)\sqrt{M} .
\end{align}
\end{subequations}
However, the exact dependence does not have a known analytical expression and requires a numerical simulation taking into account the full level structure to be computed \cite{2024-MFadel-NJP}, see the full model described in Sec.~\ref{supp:secEqsatomic} of \cite{SM}.

At this point, let us mention that Eq.~\eqref{eq:OmegaTh} has been derived under the assumption that all metastable atoms in the cell are illuminated homogeneously by the probe beam.
In typical experimental situations, however, this is not the case, for example because the transversal area of the beam, $\Abeam$, is smaller than the one of the cell, $\Acell$, or due to the beam's intensity profile.
Nevertheless, if the atomic motion is fast enough all atoms experience the same average interaction with the light, a phenomenon known in the alkali-atom community as motional averaging \cite{2010-KHammerer-RMP}.
In the current experiment, although metastable atoms do not diffuse fast enough compared to their average lifetime, they just serve as an interface with the ground state atoms that do experience motional averaging (see Sec.~\ref{supp:secDiffuse} of \cite{SM}).
For this reason, the coupling strength $\Omega^{(i)}$ needs to be rescaled by $\Abeam/\Acell < 1$.
In order to compensate for this reduction in the interaction strength, a common approach is to use optical cavities or, as we do here, multi-pass cells. In summary, the coupling in Eq.~\eqref{eq:OmegaTh} gets changed into 
\begin{equation}\label{eq:OmegaThCorrection}
\Omega^{(i)} = N_\text{pass}  \left(\dfrac{A_{\text{beam}}}{A_{\text{cell}}} \right) \, v^{(i)} \dfrac{\ncell}{\Ncell} \sqrt{\nph \Ncell} f^{(i)}(M),
\end{equation}
where $N_\text{pass}$ is the number of light beam passes through the atomic ensemble.
Interestingly, since $\Omega^{(i)}\propto\nu^{(i)}\propto 1/\Abeam$, the coupling does not depend on the area of the probe beam but on the one of the cell.

The goal of our work is to measure the value of $\Omega^{(i)}$ for different experimental configurations and compare it with the theortical prediction Eq.~\eqref{eq:OmegaThCorrection}, demonstrating the effective Faraday interaction between light and ground state \he nuclear spins. 
To this end, we start by polarizing the nuclear spins using MEOP and then apply at $t=0$ a small rotation from the $x$ to the $z$ direction. In the presence of a weak static magnetic field applied along $x$, the slightly-tilted nuclear spins undergo Larmor precession on the $yz$-plane at frequency $\omega_I$. This is described by $P_I(t) = P_I(0) \cos \left(\omega_I t \right)$, where $P_I(0)= \sin\theta \sqrt{M \Ncell/2}$ depends on the tilt angle $\theta$. According to Eq.~\eqref{eq:effHPSPI12}, this evolution is mapped into the light field by the Faraday interaction, thus resulting in the signal (see Sec.~\ref{supp:secEffCoup} of \cite{SM})
\begin{equation}\label{eq:lightsig}
X_S(t) = \Omega^{(i)} P_I(0)  \cos \left(\omega_I t \right).
\end{equation}
As we are going to show below, this expression allows us to extract the coupling strength $\Omega^{(i)}$ from a measurement of $X_S$.

\begin{figure*}[t]
    \centering
    \includegraphics[width=1\linewidth]{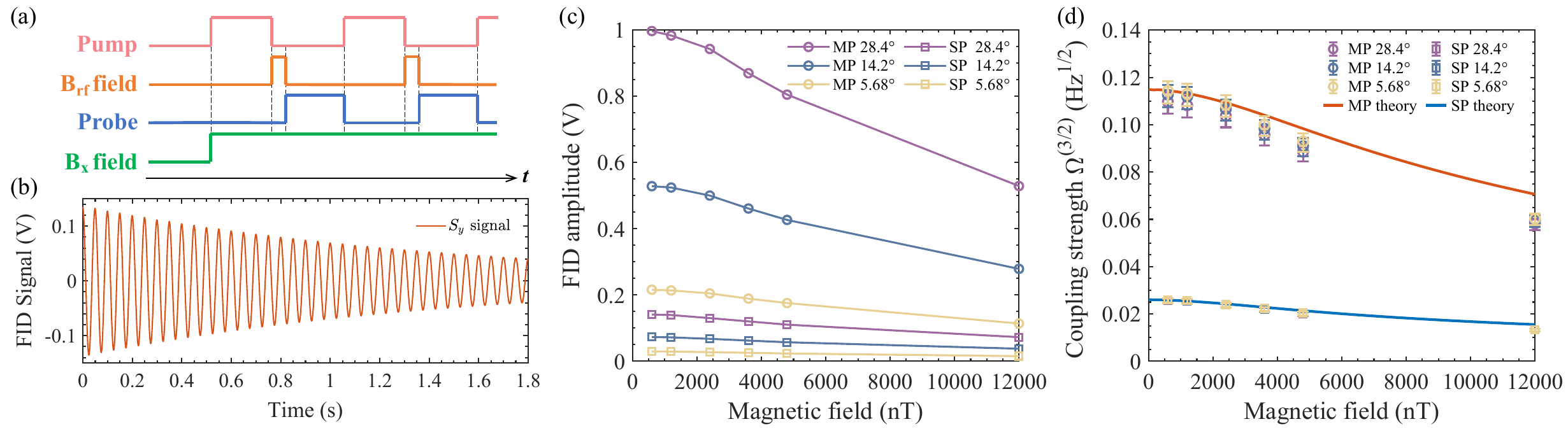}
    \caption{\textbf{Effective coupling measurement sequence and magnetic field dependence.} (a) Timing diagram for the pump beam, rf Rabi pulse $\text{B}_{\text{rf}}$, probe readout, and bias magnetic field $\text{B}_{\text{x}}$. High (low) level denotes on (off) state. (b) Typical $S_y$ FID signal measured for the single-pass \he cell at a $B_x=\unit{600}{nT}$, $M=0.38$, and $\theta=28.4^\circ$ under Config.2. Dependence of the FID signal amplitude (c) and of the corresponding effective coupling strength $\Omega^{(3/2)}$ (d) on the bias magnetic field, for different tilt angles $\theta$ under Config.2. Square (circle) markers represent measurement from the single-(multi-)pass cell with $M=0.38$ ($M=0.67$), while the curves show the simulation results based on the full model. The simulation parameters are the same as the experimental parameters. MP: multi-pass; SP: single-pass. Error bars mostly come from uncertainty in the value of $M$.}
    \label{fig:2}
\end{figure*}

\vspace{2mm}
\textbf{Experimental setup.---}
The experimental setup is schematically shown in Fig.~\ref{fig:1}(b), where only the multi-pass \he cell is drawn for simplicity. The configuration used for the single-pass cell is analogous, and more detailed descriptions of the two implementations are provided in Sec.~\ref{supp:secCells} and Sec.~\ref{supp:secExpsetup} of \cite{SM}. The glass cell (either single-pass or multi-pass) filled with pure \he gas at a pressure of 100 Pa is placed inside a seven-layer magnetic shield. Two electrodes on the cell, connected to an rf source with an output power of 0.2 W (not shown), allows for igniting a discharge inside the cell which excites a fraction of ground-state atoms into the metastable $2{ }^3 S_1$ state. Inside the shield, a guiding magnetic field is applied along the $x$-direction, while an rf magnetic field used for Rabi rotation of the nuclear spins is applied along the $z$-direction.

A circularly polarized pump beam, propagating along $x$-direction and resonant with the $C_8$ transition, is used to polarize the ground-state nuclear spins through MEOP. An $x$-polarized probe beam, tuned either to Config.1 or Config.2, passes through the cell and is analyzed by a balanced polarimeter composed of a wave plate (WP), a polarization beam splitter (PBS), and two photodiodes (PDs). For the multi-pass cell, the probe beam passes $N_\text{pass}=22$ times through the cell before exiting and being detected by the balanced polarimeter. Here, WP denotes a half-wave plate for $S_y$ measurement (see Sec.~\ref{supp:secStokesMea} of \cite{SM} for details).

\vspace{2mm}
\textbf{Measurement sequence and results.---} 
To experimentally measure the effective coupling strength $\Omega^{(i)}$ we proceed in the following way, see Fig.~\ref{fig:2}(a). 
First, after having polarized the nuclear spin along the $x$-axis through conventional MEOP, we slightly tilt it by an angle $\theta$ by applying a resonant rf Rabi pulse. 
Keeping the modulation frequency $\omega_m$ tuned to the nuclear Larmor frequency $\omega_I$, the pulse amplitude $B_{rf}$ and duration $t$ are jointly adjusted to produce a desired angle $\theta=\gamma_I B_{rf} t$ at different holding magnetic fields (see Sec.~\ref{calrffield} of \cite{SM} for details), where $\gamma_I \approx \unit{2\pi \times 3.24}{kHz/G}$ is the gyromagnetic ratio of the nuclear spin.
$\theta$ is chosen to be small (e.g.~$<\unit{30}{\degree}$) to ensure the validity of the Holstein-Primakoff approximation. 
This tilt results in a precession of the nuclear spin around the holding magnetic field $B_x$ at Larmor frequency $\omega_I =\gamma_I B_x$. 
To measure this precession, we send a probe pulse for time duration 2 s and continuously measure the evolution of $S_y$.
A typical signal we record is shown in Fig.~\ref{fig:2}(b), showing the typical free-induction-decay (FID) signal. We fit the data with the damped sinusoidal model $A e^{-\gamma t}\sin(\omega t+\phi)$ and extract the FID signal amplitude $A_\text{FID}$ associated to $S_y$. The values obtained for different tilt angles $\theta\in\{5.68^\circ,14.2^\circ,28.4^\circ\}$ and magnetic field $B_x$ are shown in Fig.~\ref{fig:2}(c), for both the single- and multi-pass cells.

A precise calibration of our photodiodes allows us to convert the FID amplitude $A_\text{FID}$ to $X_S(t)$ and, using Eq.~\eqref{eq:lightsig}, to extract the coupling strength $\Omega^{(i)}$ (see Sec.~\ref{supp:secOmegaMea} of \cite{SM}).
The expression reads
\begin{equation}
\Omega^{(i)}=\frac{A_\text{FID} }{\sqrt{\hbar \omega_{p} R_\text{pd} G} \cdot \sqrt{V_1+V_2} \cdot \sqrt{M N_{\text{cell}}} \cdot \sin \theta},
\label{expOmega}
\end{equation}
where $R_\text{pd}$ is the responsivity of the photodiode, $G=\unit{10^4}{V/A}$ is the transimpedance gain of the amplifier, $\omega_{p}$ is the probe light frequency and $V_i$ is the output voltage of the $i$th photodiode.
The results we obtain in Config.2 are shown in Fig.~\ref{fig:2}(d).
Within the explored magnetic-field range, the effective coupling strength calibrated using the multi-pass cell is significantly larger than that obtained with the single-pass cell. In addition, we can compare the experimentally calibrated values at 600 nT with the theoretical values predicted by Eq.~\eqref{eq:OmegaThCorrection}. To this end, besides computing the number of ground state atoms $\Ncell$, we measure the density of metastable atoms through a standard absorption measurement of a $x$-polarized probe beam resonant with the $C_{8}$ transition \cite{batzthesis}.
This gives $\ncell=7.2 \times 10^{11} (5.5 \times 10^{11})$ atoms for the single-pass (multi-pass) cell (see Sec.~\ref{supp:secMetaNumMea} of \cite{SM}).
From the probe beam power we obtain $n_{\mathrm{ph}}=P_L/\hbar\omega_p=2.9\times 10^{15} ~\text{s}^{-1}~(2.8 \times 10^{15}~\text{s}^{-1})$ for the measurement of single-pass (multi-pass) cell, and from its area we estimate $\chi=3.5 \times 10^{-11}$ ($\eta=1.7 \times 10^{-11}$) under Config.1 (Config.2) for the single-pass cell, while $\chi=3.0 \times 10^{-11}$ ($\eta=1.4 \times 10^{-11}$) is under Config.1 (Config.2) for the multi-pass cell. Inserting the above parameters into Eq.~\eqref{eq:OmegaThCorrection}, we find good agreement between the experimental and theoretical values of $\Omega^{(i)}$ for the single-pass cell, whereas the experimental value for the multi-pass cell is approximately a factor of two smaller than the theoretical prediction. The reason is likely due to uncertainties in the calibration of the metastable density for the multi-pass cell.

As shown in Fig.~\ref{fig:2}(c), the measured FID amplitude $A_\text{FID}$ decreases with increasing magnetic field, which in turn leads to a reduction in the calibrated coupling strength $\Omega^{(i)}$. 
This behavior originates from the gyromagnetic-ratio mismatch between the metastable and ground-state manifolds~\cite{2025-PRR-Lu}, whose importance grows with magnetic field. Although the frequent MECs lock the two spin ensembles to a common Larmor precession frequency, this mismatch suppresses the metastable transverse spin components $K_z$ and $J_z$ (see the simulation results in Sec.~\ref{supp:secMagdepend} of \cite{SM}), thereby reducing the observed FID amplitude. Further, as shown in Fig.~\ref{fig:2}(d), the magnetic-field dependence of the experimental $\Omega^{(i)}$ exhibits excellent agreement with the full model simulation (see Sec.~\ref{supp:secMagdepend} of \cite{SM} for details) in the low-field regime, while at higher magnetic fields the
experimental values decrease more rapidly than the theoretical prediction, which may be attributed to an additional reduction of $K_z$ and $J_z$ caused by the magnetic-field gradients.

\begin{figure*}[t]
    \centering
    \includegraphics[width=1\linewidth]{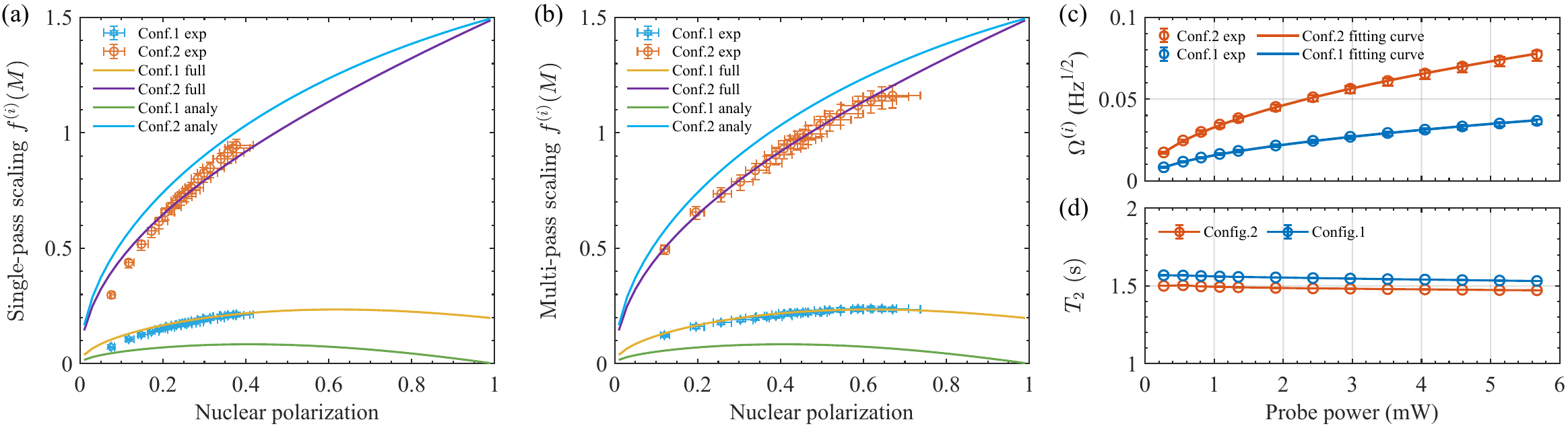}
    \caption{\textbf{Coupling dependence on the nuclear polarization and on the probe power.} (a,b) Comparison of the scaling function $f^{(i)}(M)$ obtained in the single- and multi-pass experiments (dots) with the analytical/full models (lines, see Eqs.~\eqref{eqs:fMscalings} and Sec.~\ref{supp:secEqsatomic} of \cite{SM}, including a factor 1/2 to correct for a typo in Ref.~\cite{2024-MFadel-NJP}). Error bars mostly come from uncertainty in the value of $M$. (c) Dependence of the effective coupling strength $\Omega^{(i)}$ on the probe power in the single-pass cell. Solid lines are fits using the function $\Omega^{(i)} = k\sqrt{P_L}$. (d) Dependence of the $^3$He ground-state transverse relaxation time $T_2$ on the probe power in the single-pass cell.}
    \label{fig:3}
\end{figure*}

Next, we measure the effective coupling strength $\Omega^{(i)}$ at different nuclear polarizations by varying the pump-beam power. The dependence of $\Omega^{(i)}$ on the scaling functions $f^{(i)}(M)$ can then be extracted by inverting Eq.~\eqref{eq:OmegaThCorrection}. 
After the inversion, the experimental values of $f^{(i)}(M)$ are compared to their theoretical predictions under adiabatic elimination Eq.~\eqref{eqs:fMscalings} and to their numerical computation using a full model taking into account all spin manifolds \cite{2024-MFadel-NJP}.
The results are shown in Figs.~\ref{fig:3}(a,b), for the single- and multi-pass cell, respectively, and show excellent agreement with the full model simulation. In addition, it can be seen that $f^{(i)}(M)$ is much larger for Config.2 than for Config.1. As a result, although the coupling constant $\eta$ in Config.2 is only about 0.48 times the coupling constant $\chi$ in Config.1, Eq.~\eqref{eq:OmegaThCorrection} shows that the effective coupling $\Omega^{(i)}$ is nevertheless larger in Config.2, with the difference becoming more pronounced at high nuclear polarization. This makes Config.2 more favorable for achieving \he nuclear-spin squeezing.

To further enhance the effective coupling strength, we investigate the effect of increasing the Faraday probe power. Taking the single-pass cell as an example, we vary the probe power up to $P_L=5.67~\mathrm{mW}$ and obtain the results shown in Figs.~\ref{fig:3}(c,d). For both Config.1 and Config.2, the measured $\Omega^{(i)}$ follows $\Omega^{(i)} = k\sqrt{P_L}$, confirming the expected $\sqrt{n_{\mathrm{ph}}}$ scaling and indicating that probe-induced changes in $\ncell$ are negligible within the investigated range. 
We further extract the ground-state transverse relaxation time and observe no appreciable dependence on the probe power. Therefore, increasing the probe power can enhance the effective light–nuclear-spin interaction without immediately compromising the ground-state coherence, making probe-power optimization important toward improving the spin-squeezing rate in future work.

Finally, we demonstrate that the effective Faraday interaction can be switched on/off 
by controlling the rf-discharge power, as detailed in Sec.~\ref{supp:secSwitchability} of \cite{SM}. 
This switchability allows the light–spin coupling to be activated only when needed for state preparation and readout, while being suppressed during free evolution or storage. Because the metastable atoms that mediate the optical coupling also introduce additional decoherence channels, turning off the rf discharge after state preparation reduces 
MEC-induced relaxation and thereby extends the useful coherence lifetime of nuclear-spin-squeezed states~\cite{2021-ASerafin-PRL}. Such controllable coupling is particularly advantageous for realizing quantum-metrology protocols with long interrogation times, as well as for nuclear-spin-based quantum memories.

\vspace{2mm}
\textbf{Towards nuclear spin squeezing.---} 
The effective Faraday interaction between light and \he nuclear spins can allow for the measurement-based preparation of spin-squeezed states \cite{2021-ASerafin-PRL}.
This occurs at a rate $\Gamma_\text{sq} \approx 4(\Omega^{(i)})^2$ and results in a squeezed spin variance that can be reduced up to $\text{Var}[P_I] \approx \frac{1}{2}\sqrt{\gamma_0’/\Gamma_\text{sq}}$, with the limit set by wall collisions through the rate $\gamma_0’= \gamma_0 (\gamma_f^{(i)}/\gamma_m^{(i)})$.
For our multi-pass cell, we have effective MEC rates $\gamma_f^{(3/2)} \approx \unit{14.7}{s^{-1}}$ and $\gamma_m^{(3/2)} \approx \unit{1.1\times 10^{6}}{s^{-1}}$, for the ground and metastable state, respectively, and a metastable relaxation rate due to wall collisions of $\gamma_0 \approx \unit{8.6 \times 10^3}{s^{-1}}$ (see Sec.~\ref{supp:secEffCoup} of \cite{SM}).
This gives $\gamma_0’\approx \unit{0.11}{s^{-1}}$.
Considering our measured $\Omega^{(3/2)}$, we can currently achieve $\Gamma_\text{sq} \approx \unit{5.2 \times 10^{-2}}{s^{-1}}$. 
Since the probe power used in this experiment was arbitrarily chosen to only $\unit{0.55}{mW}$, we can increase it tenfold to a level comparable to Ref.~\cite{2021-ASerafin-PRL} and thus enhance the squeezing rate to possibly $\Gamma_\text{sq} \approx \unit{0.52}{s^{-1}}$.
This power increase is not expected to appreciably compromise the ground-state coherence, since in Config.2 the corresponding off-resonant scattering rate is estimated to be $\gamma_\mathrm{sc}^{(m)}\approx \unit{1}{s^{-1}}$ for the metastable atoms, which is projected onto the ground-state nuclear spin yields $\gamma_\mathrm{opt}^{(I)}\approx \gamma_\mathrm{sc}^{(m)} (\gamma_f^{(3/2)}/\gamma_m^{(3/2)}) \approx \unit{1.3\times10^{-5}}{s^{-1}}$ of additional relaxation.
It implies that we are promised to enter a regime satisfying the requirement for the effective squeezing dynamics, $\gamma_f^{(3/2)} \gg \Gamma_\text{sq}$, and the requirement of squeezing faster than the nuclear spin decoherence rate $\Gamma_\text{sq} > \gamma_0’$. 
However, if we take into account the additional nuclear-spin decoherence induced by MEC under a nonzero magnetic field~\cite{2025-MST-Wu}, rather than only the decoherence projected from the metastable wall relaxation as mentioned in Ref.~\cite{2021-ASerafin-PRL}, a higher squeezing rate might be required. 
For this reason, in order to realize nuclear spin squeezing additional effort needs to be put in increasing further $\Gamma_\text{sq}$, for example by increasing further the Faraday probe beam power, optimizing the gas pressure and the rf discharge power, and increasing the number of passes in the cell.
Crucially, compared to optical cavities, multi-pass cells are typically more robust to mechanical vibrations and avoid complex requirements such as active cavity-length stabilization~\cite{2011-PRA-Li}, making them well suited for practical applications.

\vspace{2mm}
\textbf{Conclusions and outlook.---} 
We have shown the experimental realization of an effective Faraday interaction between light and ground-state \he nuclear spins, mediated by metastability-exchange collisions in a low-pressure gas cell at room temperature. 
By relying on a small, discharge-driven population of metastable atoms, our approach enables both efficient MEOP and controllable optical probing of the collective nuclear spin state. 
Through systematic measurements of the Faraday rotation as a function of nuclear polarization, magnetic field, probe-beam parameters, and optical depth, we have quantitatively established the dependence and tunability of this interaction. 
These results confirm recent theoretical proposals~\cite{2021-ASerafin-PRL,2024-MFadel-NJP} and provide experimental evidence that MECs can mediate an effective, switchable, Faraday coupling between noble-gas nuclear spins and light fields.

The demonstrated coupling offers a practical pathway towards optical quantum control of \he nuclear-spin ensembles. 
In particular, the ability to coherently imprint and read out nuclear spin information optically, without compromising its intrinsic long coherence time, opens the door to generating macroscopic nuclear-spin-squeezed states through quantum-nondemolition measurement protocols. 
Such capabilities would enable a new class of quantum-enhanced sensors based on \he atoms, including precision magnetometers, inertial sensors, and probes for fundamental physics. 
Future work will focus on increasing the interaction strength through optimized experimental parameters, with the goal of achieving nuclear spin squeezing.
Together, these advances position metastable-mediated light–spin interactions in \he as a powerful platform for long-lived and high-precision quantum technologies.

\vspace{2mm}
\textbf{Acknowledgments.---}
This work was supported by the National Natural Science Foundation of China (Grant No. 62375002). M.F. was supported by the Swiss National Science Foundation Ambizione Grant No. 208886, and by The Branco Weiss Fellowship -- Society in Science, administered by the ETH Z\"{u}rich. We thank Prof. Dong Sheng for helpful assistance in the design of the multi-pass cell. We also acknowledge Peking Nanofab for access to facilities used to clean the raw materials of the multi-pass cell.


\let\oldaddcontentsline\addcontentsline
\renewcommand{\addcontentsline}[3]{}
\bibliography{manuscript}
\let\addcontentsline\oldaddcontentsline

\clearpage
\newpage

\renewcommand{\thetable}{S\arabic{table}}  
\renewcommand{\thepage}{\arabic{page}}  
\renewcommand{\thefigure}{S\arabic{figure}}
\renewcommand{\theHfigure}{S\arabic{figure}}
\renewcommand{\theequation}{S\arabic{equation}}
\setcounter{page}{1}
\setcounter{figure}{0}
\setcounter{table}{0}
\setcounter{section}{0}
\setcounter{equation}{0}

\widetext

{\centering\textbf{Supplementary Materials for:} \\} 
{\centering\textbf{Effective Faraday interaction between light and Helium-3 nuclear spins in a multi-pass cell}\\} 
\normalsize
\vspace{.3cm}
{\centering Kaiwen Yi$^1$, Yida Sha$^1$, Zejia Lin$^1$, Matteo Fadel$^{2,}$$^\ast$, and Xiang Peng$^{1,}$$^\dagger$\\
\vspace{2mm}
\textit{$^1$ School of Electronics, Peking University, Beijing 100871, China\\}
\textit{$^2$ Department of Physics, ETH Z\"{u}rich, 8093 Z\"{u}rich, Switzerland\\}
\vspace{2mm}
\textit{$^\ast$ fadelm@phys.ethz.ch\\}
\textit{$^\dagger$ xiangpeng@pku.edu.cn}\\
}

\suppressfloats

\tableofcontents

\clearpage
\newpage

\section{Table of main symbols used in the manuscript}

\renewcommand{\arraystretch}{1.8} 
\setlength{\tabcolsep}{12pt}

\begin{table*}[h!]
    \centering
    \label{tab:table_main}
    \begin{tabular}{|c|c|c|}
        \hline
        \textbf{symbol} & \textbf{definition} & \textbf{reference} \\
        \hline
        $\Ncell$, $\ncell$  & Number of ground-state and metastable-state \he atoms in the cell & Eq.~\eqref{eq:OmegaTh} and Eq.~\eqref{eq:OmegaThCorrection} \\
        $\mathbf{I}$  & Collective nuclear spin operator & Eq.~\eqref{eq:effHPSPI12} \\  
        $\mathbf{K}$  & Collective spin operator for the 2$^3S_1$ $F=1/2$ manifold & Eq.~\eqref{eqHv1/2and3/2} \\        
        $\mathbf{J}$  & Collective spin operator for the 2$^3S_1$ $F=3/2$ manifold & Eq.~\eqref{eqHv1/2and3/2} \\ 
        $\mathbf{S}$  & Light Stokes operator &  Eq.~\eqref{eqHv1/2and3/2} \\ 
        $H_{\text{LA}}$  & \makecell{Full Hamiltonian describing \he atom-light interaction \\ that is decomposed as $H_{\text{LA}}=H_{1/2}^V+H_{3/2}^V + H_{3/2}^T$ with terms \\ for the $F=1/2,3/2$ vector ($V$) and tensor ($T$) interactions}  & Eq.~\eqref{eqHv1/2and3/2} and \cite{2024-MFadel-NJP} \\         
        $\nu^{(i)}\in\{\chi,\eta \}$, and $\mu$ & Vector and tensor atom-light coupling constants & Eqs.~(\ref{eqHv1/2and3/2},~\ref{eq:OmegaTh},~\ref{eq:OmegaThCorrection}) and \cite{2024-MFadel-NJP} \\
        $\omega_p$, $\Delta_p$  & Probe light frequency and its detuning relative to \he $C_8$ transition &  $\Delta_p=\omega_p - \omega_{C_8}$ \\
        $P_L$, $\nph$  & Probe beam power and photon flux & $\nph=P_L/\hbar \omega_p$ \\
        $M$ & Nuclear spin polarization $M\in[0,1]$ & Eq.~\eqref{eq:OmegaTh}, Eq.~\eqref{eq:OmegaThCorrection} and Eq.~\eqref{expOmega} \\
        $X_S,P_S,X_I,P_I$ & \makecell{Quadrature observables associated to the Stokes ($S$) and nuclear ($I$) \\ operators in the Holstein-Primakoff approximation} & \makecell{$X_S=S_y/\sqrt{\avg{S_x}}$, $P_S=S_z/\sqrt{\avg{S_x}}$,\\ $X_I=I_y/\sqrt{\avg{I_x}}$, $P_I=I_z/\sqrt{\avg{I_x}}$} \\
        $H_{\text{eff}}$  & Effective nuclear spin-light Faraday Hamiltonian & Eq.~\eqref{eq:effHPSPI12} \\
        $\Omega^{(i)}$ & Effective nuclear spin-light coupling strength &  Eqs.~(\ref{eq:effHPSPI12},~\ref{eq:OmegaTh},~\ref{eq:OmegaThCorrection},~\ref{eq:lightsig},~\ref{expOmega}) \\
        $f^{(i)}(M)$ & Polarization-dependent scaling function & Eq.~\eqref{eq:OmegaTh}, Eq.~\eqref{eqs:fMscalings}, and Eq.~\eqref{eq:OmegaThCorrection} \\ 
        $\Abeam$, $\Acell$  & Transverse areas of probe beam and gas cell & Eq.~\eqref{eq:OmegaThCorrection} \\
        $N_\text{pass}$  & Number of probe-beam passes through the atomic ensemble & Eq.~\eqref{eq:OmegaThCorrection} \\
        $\omega_I$  & Nuclear Larmor precession frequency & $\omega_I=\gamma_I {B}_x$, Eq.~\eqref{eq:lightsig} \\
        $\omega_m$  & Modulation frequency of the rf magnetic field & Set to $\omega_m=\omega_I$ in this work \\
        $\theta$  & Tilt angle of the nuclear spin at time $t=0$ & $\theta=\gamma_I B_{r f} t$, Eq.~\eqref{expOmega} \\
        $\gamma_I$  & Gyromagnetic ratio of the nuclear spin & $\gamma_I \approx \unit{2\pi \times 3.24}{kHz/G}$ \\
        $B_{rf}$, $B_{x}$ & Magnitude of the rf magnetic field and the guiding field & Experimental parameters \\
        $A_{\mathrm{FID}}$  & Amplitude of the FID signal & Eq.~\eqref{expOmega} \\
        $R_\text{pd}$ & Responsivity of the photodiode & Eq.~\eqref{expOmega} \\
        $G$ & Transimpedance gain of the amplifier & Eq.~\eqref{expOmega} \\
        $V_i$ & Output voltage of the photodiode & Eq.~\eqref{expOmega} \\
        $\Gamma_{\text{sq}}$ & Measurement-based squeezing rate from Faraday interaction & \cite{2021-ASerafin-PRL} \\
        $\gamma_0,$ $\gamma_0^{\prime}$ & \makecell{Metastable relaxation rate due to wall collisions \\ and resulting effective ground-state relaxation rate} & $\gamma_0’= \gamma_0 (\gamma_f^{(i)}/\gamma_m^{(i)})$ from \cite{2021-ASerafin-PRL}\\
        $\gamma_f^{(i)}, \gamma_m^{(i)}$ & Effective MEC rates for ground-state and metastable-state \he atoms & \cite{2024-MFadel-NJP} \\
        \hline
    \end{tabular}
    \caption{\textbf{Definition of the symbols used in the main text.}}
\end{table*}

\begin{table*}[h!]
    \centering
    \label{tab:table_supp}
    \begin{tabular}{|c|c|c|}
        \hline
        \textbf{symbol} & \textbf{definition} & \textbf{reference} \\
        \hline
        $R$, $d$, $\tau_D$ & \makecell{Characteristic container dimension, spatial dimensions, \\ and characteristic diffusion time} & $\tau_D=R^2 / 2 d D$ \\
        $n(\vec{r}, t)$ & Atomic density distribution in the cell & $n(\vec{r}, t)=n(\vec{r}) e^{-t / \tau_D}$ \\
        $\tau_D^{\text {cyl }}$, $\tau_D^{\text {cuboid }}$ & Characteristic diffusion time for a cylinder and cuboid & Eq.~\eqref{cylandcub} \\        
        $L$ & Length of the cell & Throughout the text \\ 
        $r_{\mathrm{MEC}}^*$ & Diffusion distance of a metastable atom between two ME collisions & $r_{\mathrm{MEC}}^*=\sqrt{6 D^*(p) \tau_{\mathrm{MEC}}}$ \\
        $h^{(i)}$ & \makecell{Microscopic Hamiltonian describing the Faraday interaction \\ between the $i$th spin ${j}^{(i)}_z$ at position $\vec{r}^{(i)}$ and the probe light}& Eq.~\eqref{microham} \\
        $\alpha$ & Single-atom-light coupling constant & Eq.~\eqref{alphaeq} and Eq.~\eqref{eq:Fex} \\ 
        $\Gamma$, $\Delta$, $\sigma_2$ & \makecell{Excited-state spontaneous decay rate, probe light detuning \\ and atomic resonant scattering cross section} & Eq.~\eqref{alphaeq} \\
        $\mathcal{J}_z(\vec{r})$, $J_z$ & Collective spin density operator, collective spin operator & Eq.~\eqref{Collspindensity}, Eq.~\eqref{Collspinoperator} \\   
        $\rho$, $k_z,$ $j_z$ & \makecell{Atomic density, angular-momentum density operators of \\ the $F=1/2$ and $F=3/2$ manifolds} & Eq.~\eqref{supp:eqsHV} \\     
        $T_0^2, \Re T_2^2, \Im T_2^2$ & Collective tensor operators of the $F=3/2$ manifold & Eq.~\eqref{supp:eqsHT} \\
        $\Delta_i$ & Frequency offsets of transitions $C_1-C_9$ with respect to transition $C_8$ & $\Delta_i=\omega_{F F^{\prime}}-\omega_{C_8}$ \\      
        $T_{\mathrm{MEC}}$ & Characteristic time interval between two MECs for a ground-state atom & Eq.~\eqref{MECrateConf1} and Eq.~\eqref{MECrateConf2} \\ 
        $\text{OD}$ & Optical depth of a general atomic ensemble & Sec.~\ref{multienhance} \\
        $I(t)$ & Photocurrent produced by a continuous homodyne measurement of $X_S^{\text {out }}$ & Eq.~\eqref{photocurrent} and Eq.~\eqref{photocurrent2} \\ 
        $\eta_{\operatorname{det}}, d W(t) $ & Detection efficiency and Wiener noise increment & Eq.~\eqref{photocurrent2} \\
        $\rho_c$, $\mathcal{D}$, $\mathcal{H}$ & Atomic density operator and Lindblad superoperators & Eq.~\eqref{stochastic} \\
        $\Gamma_m$ & Measurement rate for the QND observable $P_J$ & Eq.~\eqref{measrate} \\
        $f_m$, $f_L$ & \makecell{Modulation frequency of the rf magnetic field \\ and nuclear Larmor precession frequency}  & Sec.~\ref{calrffield} \\
        $V_{cont}$ & Sinusoidal voltage output from the signal generator & Sec.~\ref{calrffield} \\
        $I(z)$, $k_a(z)$ & Local light intensity and absorption rate & Eq.~\eqref{lambbeer} \\
        $I_T$, $I_0$, $T_s$ & \makecell{Transmitted and incident light intensities, \\ Transmission coefficient} & Eq.~\eqref{Ts} \\
        $n_{\mathrm{m}}(z)$, $\omega_{i j}$, $a_i^{\mathrm{ST}}(M)$ & \makecell{Local metastable density, angular frequency for $\mathrm{A}_i \rightarrow \mathrm{B}_j$ line component, \\ population of the state $\mathrm{A}_i$ in spin-temperature distribution} & Eq.~\eqref{localabsrate} \\
        $\bar{\gamma}_{i j}$, $\bar{\Gamma}_{i j}$ & \makecell{Maxwell-averaged optical transition rate for $\mathrm{A}_i \rightarrow \mathrm{B}_j$ line component \\ and its normalized coefficient} & Eq.~\eqref{normaltranrate} \\
        $\alpha_{\mathrm{FS}}$, $f$, $D$ & \makecell{Fine structure constant, oscillator strength and Doppler width \\ for the $2^3 \mathrm{S} \rightarrow 2^3 \mathrm{P}$ transition} & Eq.~\eqref{normaltranrate} \\
        $\omega$, $\delta_{\mathrm{L}}^{i j}$, $T_{ij}$ & \makecell{Light frequency and its detuning with respect to the $\mathrm{A}_i \rightarrow \mathrm{B}_j$ transition, \\ and the transition matrix element} & Eq.~\eqref{normaltranrate} \\
        $n_{\mathrm{m}}^{\mathrm{S}}$ & Average metastable number density along the probe beam path & Eq.~\eqref{nmMulti} \\
        $A_k(M)$ & Absorption of light with $k=\pi, \sigma$ polarization & Eq.~\eqref{lightabseq} \\
        $V_k(M)$, $V_{k, \text { off }}$ & Readout voltages for $k=\pi, \sigma$ polarization with the discharge on and off &  Eq.~\eqref{absdef} \\
        $R_L$ & Ratio between the light absorptions $A_\pi(M)$ and $A_\sigma(M)$ & Eq.~\eqref{lightabsratio} \\
        \hline
    \end{tabular}
    \caption{\textbf{Definition of the symbols used in the supplementary materials}. Symbols that are identical to those defined in the main text, as well as those listed in Table~\ref{tab}, are not repeated here.}
\end{table*}

\clearpage
\newpage

\section{Details on the single-pass and multi-pass cells}\label{supp:secCells}

In the experiments presented in this work, we used either a single-pass or a multi-pass cell, filled with \he gas at a pressure of \unit{100}{Pa}. 
A detailed description of the shape, dimensions, and other relevant parameters of the cells used is provided in the following.

\begin{figure}[h!]
    \centering
    \includegraphics[width=0.7\linewidth]{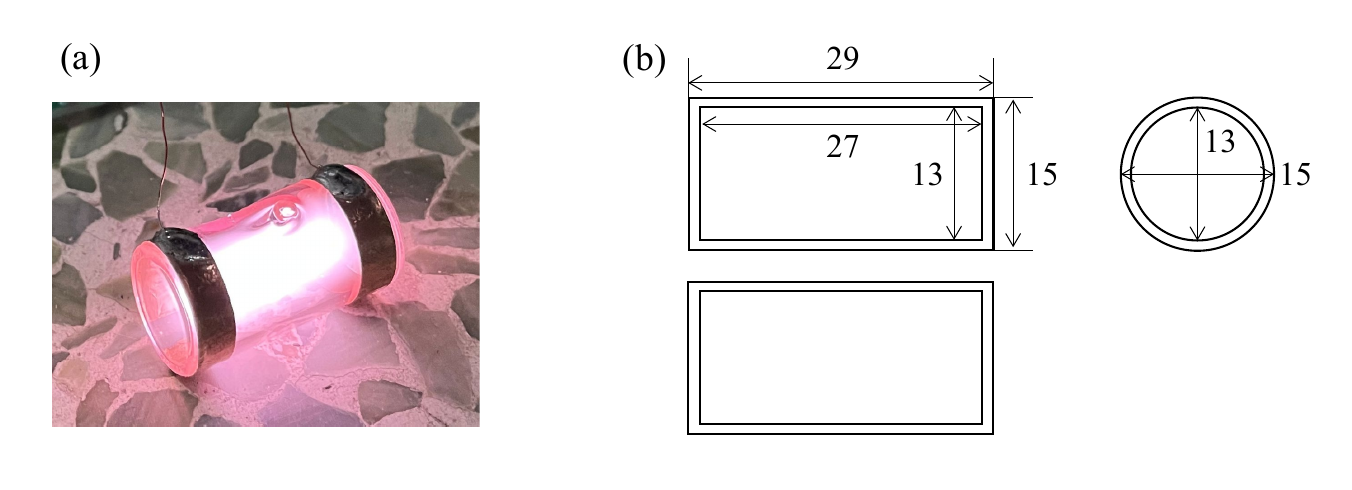}
    \caption{\textbf{Single-pass cell.} (a) Photograph of the single-pass cell. (b) Dimensions of the single-pass cell (in millimeters).}
    \label{singlecell}
\end{figure}

The single-pass cell used in our experiments is shown in Fig.~\ref{singlecell}(a). 
This is a cylindrical cell with dimensions specified in Fig.~\ref{singlecell}(b). 
The cell has an outer diameter of 15 mm and an inner diameter of 13 mm, with an overall length of 29 mm and an internal length of 27 mm.

\begin{figure}[h!]
    \centering
    \includegraphics[width=0.7\linewidth]{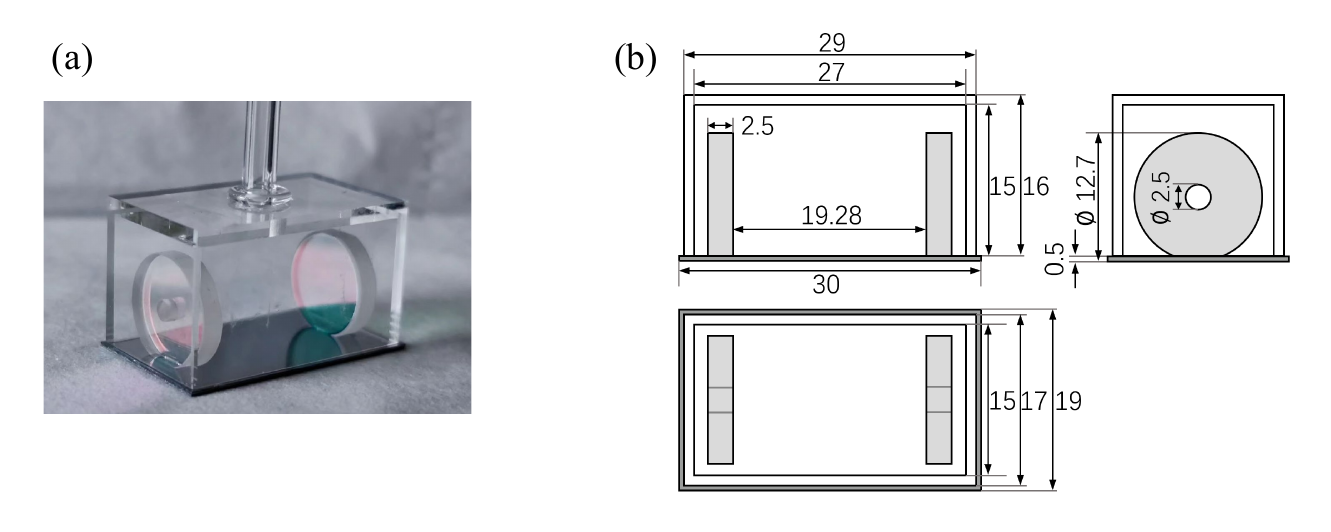}
    \caption{\textbf{Multi-pass cell.} (a) Photograph of the multi-pass cell. The base of the cell is made of a silicon wafer, and two mirrors are installed inside, one of which has an aperture for light transmission. (b) Dimensions of the multi-pass cell (in millimeters).}
    \label{multicell}
\end{figure}

The multi-pass cell used in our experiments is shown in Fig.~\ref{multicell}(a). The cell consists of three main components: a silicon base, a Herriott cavity formed by two cylindrical mirrors, and a glass enclosure. The mirrors and the glass cover are securely mounted onto the silicon base. It should be noted that, in contrast to the layer of aluminum oxide ($\mathrm{Al}_2 \mathrm{O}_3$) used in mirror coatings of typical multi-pass alkali-metal cells, we utilize here the layers of tantalum pentoxide ($\mathrm{Ta}_2 \mathrm{O}_5$) and silicon dioxide ($\mathrm{SiO}_2$) to increase the mirror reflectivity, while reducing contamination due to impurities generated by collisions between gas particles and the coating materials. One of the mirrors has a central hole through which the laser beam enters the cell. The beam then passes $\Npass=22$ times between the mirrors and exits through the same hole. Both the incident and exit angles are $5^{\circ}$. In addition, it is worth noting that in all of our experiments, the beam diameter entering the multi-pass cell is fixed at approximately 1.5 mm.

The three-view diagram of the multi-pass cell is shown in Fig.~\ref{multicell}(b). The overall dimensions of the cell are 30 mm × 19 mm × 16.5 mm, with the silicon base having a thickness of 0.5 mm. The dimensions of the glass enclosure are 29 mm × 17 mm × 16 mm, and the inner dimensions of the cell are 27 mm × 15 mm × 15 mm. Each of the two mirrors has a diameter of 12.7 mm and a thickness of 2.5 mm, with a spacing of 19.28 mm between their reflective surfaces. The diameter of the hole is 2.5 mm, and the vertical distance from the hole center to the silicon base is 5.85 mm. The total optical path length accumulated over the 22 passes between the two mirrors is approximately 424 mm.

The multi-pass \he cell is fabricated using anodic bonding technology. The overall fabrication procedure consists of three main steps: (1) preparation and cleaning of the bonding materials, (2) formation of the cell preform via a two-step anodic bonding process, and (3) gas filling and sealing of the cell through a glass stem. In the first step, the glass cover and silicon wafer are thoroughly cleaned to remove residual organic contaminants and particulate impurities. The cleaning procedure includes immersion in a piranha solution, followed by ultrasonic treatment in acetone, isopropanol, and deionized water, and finally drying with nitrogen. In the second step, anodic bonding is performed using high-precision machined fixtures to complete the bonding of the cavity mirrors to the silicon wafer and subsequently the bonding of the glass cover to the silicon wafer. In the third step, the cell is filled and sealed using an ultrahigh-purity helium gas filling technique developed in our laboratory, which enhances gas purity and extends the operational lifetime of the cell.

\section{Details on the experimental setup}\label{supp:secExpsetup}

\begin{figure}[H]
    \centering
    \includegraphics[width=0.8\linewidth]{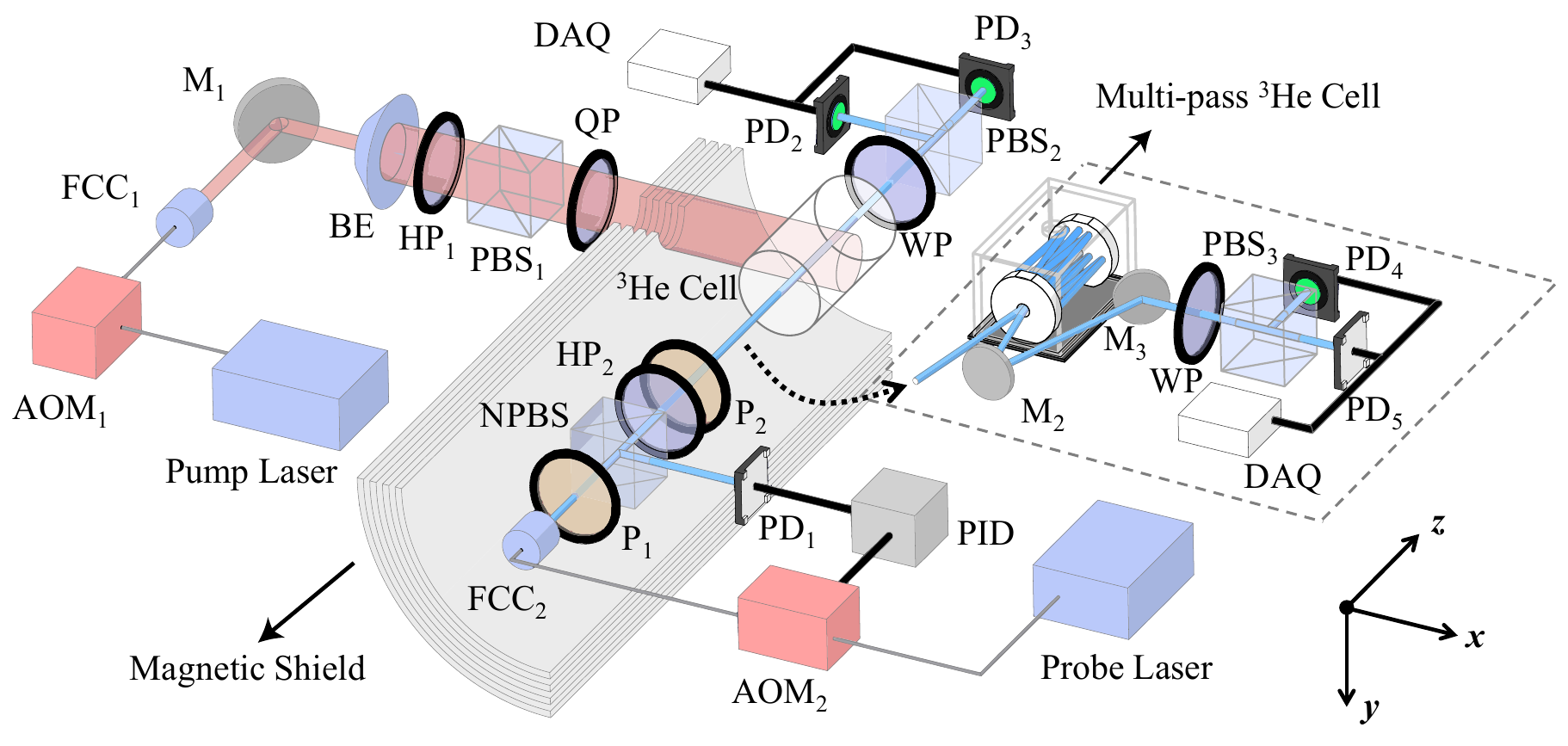}
    \caption{\textbf{Detailed optical layout of the experimental setups used for the single- and multi-pass \he cell}. The sketched elements are as follows. $\mathrm{AOM}$: acousto-optic modulator; $\mathrm{FCC}$: fiber-coupled collimator; $\mathrm{M}$: mirrors; BE: beam expander; $\mathrm{HP}$: half-wave plate; $\mathrm{PBS}$: polarization beam splitter; $\mathrm{NPBS}$: non-polarizing beam splitter; QP: quarter-wave plate; $\mathrm{PD}$: photodiodes; PID: proportional-integral-derivative controller; $\mathrm{P}$: polarizer; DAQ: data acquisition; WP: half-wave plate for $S_y$ measurement.}
    \label{Detailed setup}
\end{figure}

In this section, we provide a detailed description of the experimental setups used for the single-pass and multi-pass \he cells. The corresponding optical layout is shown in Fig.~\ref{Detailed setup}. This figure is a functional layout of the setup: it combines in a single drawing the optical elements relevant to the single-pass and multi-pass configurations, while some differences in their actual spatial arrangement are clarified below.

We first describe the part of the setup that is common to both configurations. Inside the magnetic shield, a guiding magnetic field in the $x$-direction is generated by a two-axis Helmholtz coil driven by a stable dc current source (Krohn-Hite Model 523), while the rf magnetic field along the $z$-direction used for Rabi rotations of the nuclear spins is generated by a coil driven by a voltage-controlled current source (Stanford Research Systems CS580). The pump beam used for MEOP of the ensemble is emitted from a 1083 nm fiber laser (NKT Koheras BASIK Y10 with an OEM amplifier) and tuned on resonance with the \he $C_8$ transition. Rapid switching on/off of the pump light intensity is achieved by an acousto-optic modulator ($\mathrm{AOM}_1$). Before illuminating the cell, the pump beam is expanded to a 20 mm ($1 / e^2$) waist, sent through a half-wave plate ($\mathrm{HP}_1$) and a polarization beam splitter ($\mathrm{PBS}_1$) for adjustment of the light intensity, and finally converted into circular polarization using a quarter-wave plate (QP). The probe beam, whose frequency is set to Config.1 or Config.2, originates from a second fiber laser. After passing through a linear polarizer $\mathrm{P}_1$, this beam is split into two using $\mathrm{NPBS}$. One of these beams is sent to a photodiode for laser power stabilization through a proportional-integral-derivative controller (PID) acting on $\mathrm{AOM}_2$. 
The other part of the probe beam is the one used for the actual Faraday measurements. 
The half-wave plate $\mathrm{HP}_2$ and linear polarizer $\mathrm{P}_2$ are used to adjust the probe intensity and align its polarization with the $x$-axis. 
The probe beam power is set to approximately $P_L=\unit{0.55}{mW}$.

The two implementations differ in the way the probe beam is delivered to and collected from the cell. For the single-pass cell, the probe beam propagates in free space, enters the magnetic shield through an aperture, traverses the cell, exits the shield, and is then analyzed outside the shield by a balanced polarimeter composed of WP, $\mathrm{PBS}_2$, $\mathrm{PD}_2$, and $\mathrm{PD}_3$. The corresponding differential signal is recorded by the data acquisition system (DAQ). For the multi-pass cell, part of the probe-light delivery optics is mounted inside the magnetic shield on a dedicated 3D-printed support structure. The probe beam first enters the multi-pass \he cell through the aperture of one cavity mirror and undergoes $N_\text{pass}=22$ passes inside the cell. After exiting from the cell, the transmitted beam is redirected by mirrors $\mathrm{M}_2$ and $\mathrm{M}_3$ so that it is guided out through the aperture on the right side of the magnetic shield. Outside the shield, the beam is analyzed by a second balanced polarimeter composed of WP, $\mathrm{PBS}_3$, $\mathrm{PD}_4$ and $\mathrm{PD}_5$, with the differential signal also recorded by DAQ.

Finally, we comment on the coordinate system shown in Fig.~\ref{Detailed setup}. The coordinate axes displayed in the figure are directly applicable to the single-pass configuration. For the multi-pass configuration, it should be noted that the probe beam is redirected by mirrors $\mathrm{M}_2$ and $\mathrm{M}_3$ after it exits the multi-pass cell. Accordingly, for the multi-pass configuration, the probe polarization and WP orientations are defined with respect to the propagation direction before the probe enters the cell, not with respect to the redirected beam path after the cell.

\clearpage
\newpage

\section{Detailed derivation of the atom-light coupling}\label{supp:secatomlight}

\subsection{Atom-light interaction Hamiltonian}\label{supp:secHLA}

In the weak-saturation regime, the interaction between light and metastable $2^3 \mathrm{S}_1$ atoms in either hyperfine state $F=1 / 2$ or $F=3 / 2$ can be described by an effective Hamiltonian, which is obtained by adiabatically eliminating both the optical coherences and the excited-state populations~\cite{2007-Cviklinski-PRA,1999-Kuzmich-PRA}. Following the derivation presented in Appendix A of Ref.~\cite{2024-MFadel-NJP}, but keeping the spatial dependence of the operators, we find that the Hamiltonian describing the interaction between a propagating light beam and \he metastable atoms is
\begin{equation}
    H_\text{LA} = H_{1/2}^V + H_{3/2}^V + H_{3/2}^T,
    \label{HLA}
\end{equation}
where $H_{1/2}^V$ (resp. $H_{3/2}^V$) describes the vector coupling to the collective spin $\mathbf{K}$ ($\mathbf{J}$) of the $F=1/2$ ($F=3/2$) manifold defined as
\begin{equation}\label{supp:eqsHV}
H_{1/2}^V=\hbar \chi \int_0^L  k_z S_z\; \rho A_\text{cell} d z,\qquad\qquad \quad H_{3/2}^V = \hbar \eta \int_0^L j_z S_z\; \rho A_\text{cell} d z,
\end{equation}
where $\rho$ is the atomic density, $L$ is the length of the cell, $A_\text{cell}$ is the cross-sectional area of the cell, $k_z$ and $j_z$ are the angular-momentum density operators, satisfying $K_z(t)=\int_0^L k_z(z, t) \rho A_\text{cell} d z$ and $J_z(t)=\int_0^L j_z(z, t) \rho A_\text{cell} d z$, respectively. $H_{3/2}^T$ describes the tensor coupling to the collective alignment tensor of the $F=3/2$ manifold defined as
\begin{align}\label{supp:eqsHT}
    H_{3/2}^T = \hbar\mu \int_0^L \left[ - 2 t^{2}_{0} S_0 + \sqrt{12} \left( \Re t_2^2 S_x + \Im t_2^2 S_y \right) \right] \rho A_\text{cell} d z,
\end{align} 
with $\TT{2}{0}$, $\reTtt$, $\imTtt$ collective tensor operators. These are defined as $\TT{2}{0}(t)=\int_0^L t^{2}_{0}(z, t) \rho A_\text{cell} d z$, $\reTtt(t)=\int_0^L \Re t_2^2(z, t) \rho A_\text{cell} d z$, and $\imTtt(t)=\int_0^L \Im t_2^2(z, t) \rho A_\text{cell} d z$, where $t^2_0 = (3 j_z^2 - \mathbf{j}^2 )/6$, $\Re t_2^2=(t_2^2+t_2^{2\dagger})/\sqrt{2}$, $\Im t_2^2=(t_2^2-t_2^{2\dagger})/i\sqrt{2}$, and $t^2_{2} = j_{+}^2/2\sqrt{6}$.

The dimensionless coupling constants $\chi$, $\eta$ and $\mu$ representing the strength of the different contributions in the effective atom-light Hamiltonian $H_{\text{LA}}$ have the form \cite{2024-MFadel-NJP}
\begin{eqnarray}
\chi &=& \dfrac{\sigma_2}{4A_{\text{beam}}} \Gamma \left(  \dfrac{2}{9 (\Delta_p - \Delta_{1})} - \dfrac{8}{9 (\Delta_p - \Delta_{2})} + \dfrac{10}{9 (\Delta_p - \Delta_{4})} - \dfrac{4}{9 \Delta_p}   \right), \label{eq:chi} \\
\eta &=& \dfrac{\sigma_2}{4A_{\text{beam}}} \Gamma \left(  \dfrac{3}{5 (\Delta_p - \Delta_{3})} - \dfrac{2}{9 (\Delta_p - \Delta_{5})} - \dfrac{1}{9 (\Delta_p - \Delta_{6})} - \dfrac{2}{45 (\Delta_p - \Delta_{7})} - \dfrac{2}{9 (\Delta_p - \Delta_{9})} \right),  \label{eq:eta} \\
\mu &=& \dfrac{\sigma_2}{10 A_{\text{beam}}} \Gamma \left(  - \dfrac{1}{4 (\Delta_p - \Delta_{3})} + \dfrac{5}{9 (\Delta_p - \Delta_{5})} - \dfrac{5}{36 (\Delta_p - \Delta_{6})} + \dfrac{1}{9 (\Delta_p - \Delta_{7})} - \dfrac{5}{18 (\Delta_p - \Delta_{9})} \right)  \label{eq:mu} .
\end{eqnarray} 
In these equations, $\sigma_2=3\lambda_\text{TLS}^2/2\pi$ is the resonant scattering cross section of a two-level transition with wavelength $\lambda_\text{TLS}$, $A_{\text{beam}}$ is the cross sectional area of the light mode,  $\Gamma\approx\unit{10^7}{s^{-1}}$ is the excited-state spontaneous decay rate, and taking the $C_8$ transition as a reference, we have defined $\Delta_p=\omega_p-\omega_{C_8}$ and $\Delta_{i} = \omega_{F F'}-\omega_{C_8}$.
In Fig.~\ref{fig:1}(c) of the main text, we plot the three coupling constants divided by the constant $\sigma_2\Gamma/(4A_{\text{beam}})$, as a function of probe detuning.

\subsection{Equations of motion for the Stokes operators}\label{supp:secEqsStokes}

For a free-propagating beam, the evolution of the Stokes operators under slowly varying envelope and paraxial approximations ($L/c$ is much shorter than the characteristic timescale of the atomic dynamics, so the envelope changes negligibly during propagation) is described by \cite{2010-KHammerer-RMP}
\begin{equation}
    \left( c\dfrac{\partial}{\partial z} + \dfrac{\partial}{\partial t} \right) S_i = \dfrac{i}{\hbar} [H_\text{LA}, S_i].
\end{equation}
In the limit $c\rightarrow\infty$, which corresponds to treating the light field as interacting instantaneously with the atoms (i.e. neglecting finite-speed propagation during the interaction), the time derivative can be neglected. The right-hand side of this expression is computed using the commutation relations 
\begin{equation}
\begin{aligned}
& {\left[{S}_x(z, t), {S}_z(z^{\prime}, t)\right]=-i {S}_y(z, t) \; \delta(z-z^{\prime})}\; c, \\
& {\left[{S}_y(z, t), {S}_z(z^{\prime}, t)\right]=i {S}_x(z, t) \; \delta(z-z^{\prime})}\; c, \\
& {\left[{S}_z(z, t), {S}_z(z^{\prime}, t)\right]=0}.
\end{aligned}
\label{commu2}
\end{equation}

To give a concrete example for the derivation, let us first consider Config.1, where the Hamiltonian $H_\text{LA}$ reduces to
\begin{equation}\label{eqsupp:HLA}
H_\text{LA}=\hbar \chi \int_0^L k_z(z,t) S_z(z,t) \rho A_\text{cell} d z.
\end{equation}
The Heisenberg equation is then
\begin{equation}
c\frac{\partial}{\partial z} {S}_i(z, t)=-i\left[{S}_i(z, t), \chi\int_0^L {S}_z(z^{\prime}, t){k}_z(z^{\prime}, t) \rho A_\text{cell} d z^{\prime}\right],
\end{equation}
which, using the commutation relations Eqs.~\eqref{commu2}, gives
\begin{equation}
\begin{aligned}
& \frac{\partial}{\partial z} S_x(z, t)=-\chi k_z(z, t) S_y(z, t) \rho A_\text{cell}, \\
& \frac{\partial}{\partial z} S_y(z, t)=\chi k_z(z, t) S_x(z, t) \rho A_\text{cell}.
\end{aligned}
\end{equation}
Under the assumption of spatially uniform spin density $k_z(z,t)$, motivated by the fact that in the regime we are considering MECs lock the metastable spin state to the spatially uniform ground-state spin polarization (see Sec.~\ref{supp:secDiffuse}), we can perform the replacement ${k}_z(z, t) ={K}_z(t) / \rho L A_\text{cell}$, where ${K}_z(t)$ is the collective spin operator and $L$ is the length of the gas cell. 
This gives us
\begin{equation}
\begin{aligned}
& \frac{\partial}{\partial z} S_x(z, t)=-\frac{\chi}{L} K_z(t) S_y(z, t), \\
& \frac{\partial}{\partial z} S_y(z, t)=\frac{\chi}{L}K_z(t) S_x(z, t).
\end{aligned}
\label{Stokeseqconf1}
\end{equation}
Now, we can easily extend Eq.~\eqref{Stokeseqconf1} to the general case, that is, considering the contribution of the full Hamiltonian (Eq.~\eqref{HLA}), thereby obtaining the full evolution equations for the Stokes operators~\cite{2024-MFadel-NJP}
\begin{align}
    \dfrac{\partial S_x}{\partial z} \bigg\vert_{\text{L}} &= - \frac{1}{L} \left(\chi K_z S_y + \eta J_z S_y - \sqrt{12} \mu \imTtt S_z \right), \\
    \dfrac{\partial S_y}{\partial z} \bigg\vert_{\text{L}} &= \frac{1}{L} \left(\chi K_z S_x + \eta J_z S_x - \sqrt{12} \mu \reTtt S_z \right),\label{eq:LMSyEOM}\\
    \dfrac{\partial S_z}{\partial z} \bigg\vert_{\text{L}} &= \frac{1}{L} \left[\sqrt{12} \mu \left( \reTtt S_y - \imTtt S_x \right)\right]\label{eq:LMSzEOM}.
\end{align}

\subsection{Equations of motion for the atomic operators}\label{supp:secEqsatomic}

Having obtained the evolution equations for the Stokes operators, we next consider the evolution equations for the atomic operators. These include the collective ground-state nuclear-spin operator $\mathbf{I}$, the collective spin operator $\mathbf{K}$ of the metastable $F=1/2$ manifold, the collective spin operator $\mathbf{J}$ of the metastable $F=3/2$ manifold, and the collective tensor operators $T_q^{(k)}$ of the metastable $F=3/2$ manifold. Except for the ground-state nuclear-spin operators, which do not interact directly with the light field, the evolution of all atomic operators is governed by the combined effects of three interactions: the light-metastable-atom interaction, the magnetic-field interaction, and metastability-exchange collisions (MECs). Denoting a generic atomic operator by $O$, its semiclassical evolution equation can be written as~\cite{2024-MFadel-NJP}
\begin{equation}\label{atomiceqs}
\frac{\mathrm{d}\langle O\rangle}{\mathrm{d} t}=\left.\frac{\mathrm{d}\langle O\rangle}{\mathrm{d} t}\right|_{\mathrm{L}}+\left.\frac{\mathrm{d}\langle O\rangle}{\mathrm{d} t}\right|_{\mathrm{B}}+\left.\frac{\mathrm{d}\langle O\rangle}{\mathrm{d} t}\right|_{\mathrm{MEC}},
\end{equation}
where the semiclassical approximation consists of replacing all operators by their expectation values. In Eq.~\eqref{atomiceqs}, the first term on the right-hand side describes the contribution from the light-metastable-atom interaction, the second term accounts for the interaction between the atoms and the magnetic field, and the third term represents the effect of MECs. Ref.~\cite{2024-MFadel-NJP} derives in detail the resulting set of 21 coupled evolution equations, including three ground-state nuclear-spin operators, three metastable $F=1/2$ spin operators, three metastable $F=3/2$ spin operators, and twelve metastable $F=3/2$ tensor operators (five rank-2 tensors and seven rank-3 tensors). In this manuscript, we refer to this set of equations as the \textbf{``full model''}. For completeness and to facilitate understanding of the present work, we briefly review these equations below.

According to the interaction Hamiltonian between light and metastable atoms given in Eq.~\eqref{HLA}, together with the Heisenberg equation of motion $\mathrm{d} O / \mathrm{d} t=i[H, O] / \hbar$, the evolution equations of the collective metastable atomic operators under the action of the light field are given by~\cite{2024-MFadel-NJP}:
\begingroup
\allowdisplaybreaks[4]
\begin{align}
\left.\frac{\mathrm{d}K_x}{\mathrm{d}t}\right|_{\mathrm{L}}
&=-\chi K_yS_z,\qquad 
\left.\frac{\mathrm{d}K_y}{\mathrm{d}t}\right|_{\mathrm{L}}
=\chi K_xS_z,\qquad 
\left.\frac{\mathrm{d}K_z}{\mathrm{d}t}\right|_{\mathrm{L}}
=0 ,\\
\left.\frac{\mathrm{d}J_x}{\mathrm{d}t}\right|_{\mathrm{L}}
&=-\eta J_yS_z+
\sqrt{12}\mu\left[\Im T^{2}_{1}(S_x-S_0)-\Re T^{2}_{1}S_y\right], \\
\left.\frac{\mathrm{d}J_y}{\mathrm{d}t}\right|_{\mathrm{L}}
&=\eta J_xS_z+
\sqrt{12}\mu\left[\Re T^{2}_{1}(S_x+S_0)+\Im T^{2}_{1}S_y\right], \\
\left.\frac{\mathrm{d}J_z}{\mathrm{d}t}\right|_{\mathrm{L}}
&=2\sqrt{12}\mu\left(\Im T^{2}_{2}S_x-\Re T^{2}_{2}S_y\right), \\
\left.\frac{\mathrm{d}}{\mathrm{d}t}\Re T^{2}_{2}\right|_{\mathrm{L}}
&=-2\eta S_z\Im T^{2}_{2}
+\sqrt{12}\mu\left[\frac{1}{\sqrt{5}}S_y\left(2T^{1}_{0}-T^{3}_{0}\right)
+\frac{1}{\sqrt{3}}S_0\Im T^{3}_{2}\right], \\
\left.\frac{\mathrm{d}}{\mathrm{d}t}\Im T^{2}_{2}\right|_{\mathrm{L}}
&=2\eta S_z\Re T^{2}_{2}
-\sqrt{12}\mu\left[\frac{1}{\sqrt{5}}S_x\left(2T^{1}_{0}-T^{3}_{0}\right)
+\frac{1}{\sqrt{3}}S_0\Re T^{3}_{2}\right], \\
\left.\frac{\mathrm{d}}{\mathrm{d}t}\Re T^{2}_{1}\right|_{\mathrm{L}}
&=-\eta S_z\Im T^{2}_{1}
+\sqrt{6}\mu\Bigg[
S_x\left(-\sqrt{\frac{3}{5}}\Im T^{3}_{1}-\frac{\sqrt{2}}{5}J_y+\Im T^{3}_{3}\right) \notag\\
&
+S_0\left(\frac{2}{\sqrt{15}}\Im T^{3}_{1}-\frac{\sqrt{2}}{5}J_y\right)
+S_y\left(\sqrt{\frac{3}{5}}\Re T^{3}_{1}+\frac{\sqrt{2}}{5}J_x-\Re T^{3}_{3}\right)\Bigg]
, \\
\left.\frac{\mathrm{d}}{\mathrm{d}t}\Im T^{2}_{1}\right|_{\mathrm{L}}
&=\eta S_z\Re T^{2}_{1}
+\sqrt{6}\mu\Bigg[
S_x\left(\sqrt{\frac{3}{5}}\Re T^{3}_{1}+\frac{\sqrt{2}}{5}J_x+\Re T^{3}_{3}\right) \notag\\
&
+S_0\left(\frac{2}{\sqrt{15}}\Re T^{3}_{1}-\frac{\sqrt{2}}{5}J_x\right)
+S_y\left(\sqrt{\frac{3}{5}}\Im T^{3}_{1}+\frac{\sqrt{2}}{5}J_y+\Im T^{3}_{3}\right)
\Bigg], \\
\left.\frac{\mathrm{d}}{\mathrm{d}t}T^{2}_{0}\right|_{\mathrm{L}}
&=\sqrt{12}\mu\left(S_x\Im T^{3}_{2}-S_y\Re T^{3}_{2}\right), \\
\left.\frac{\mathrm{d}}{\mathrm{d}t}\Re T^{3}_{3}\right|_{\mathrm{L}}
&=-3\eta S_z\Im T^{3}_{3}
+\sqrt{6}\mu\left(S_x\Im T^{2}_{1}+S_y\Re T^{2}_{1}\right), \\
\left.\frac{\mathrm{d}}{\mathrm{d}t}\Im T^{3}_{3}\right|_{\mathrm{L}}
&=3\eta S_z\Re T^{3}_{3}
-\sqrt{6}\mu\left(S_x\Re T^{2}_{1}-S_y\Im T^{2}_{1}\right), \\
\left.\frac{\mathrm{d}}{\mathrm{d}t}\Re T^{3}_{2}\right|_{\mathrm{L}}
&=-2\eta S_z\Im T^{3}_{2}
+2\mu\left(\sqrt{3}S_yT^{2}_{0}+S_0\Im T^{2}_{2}\right), \\
\left.\frac{\mathrm{d}}{\mathrm{d}t}\Im T^{3}_{2}\right|_{\mathrm{L}}
&=2\eta S_z\Re T^{3}_{2}
-2\mu\left(\sqrt{3}S_xT^{2}_{0}+S_0\Re T^{2}_{2}\right), \\
\left.\frac{\mathrm{d}}{\mathrm{d}t}\Re T^{3}_{1}\right|_{\mathrm{L}}
&=-\eta S_z\Im T^{3}_{1}
+\sqrt{\frac{2}{5}}\mu\left[(2S_0+3S_x)\Im T^{2}_{1}-3S_y\Re T^{2}_{1}\right], \\
\left.\frac{\mathrm{d}}{\mathrm{d}t}\Im T^{3}_{1}\right|_{\mathrm{L}}
&=\eta S_z\Re T^{3}_{1}
-\sqrt{\frac{2}{5}}\mu\left[(2S_0-3S_x)\Re T^{2}_{1}-3S_y\Im T^{2}_{1}\right], \\
\left.\frac{\mathrm{d}}{\mathrm{d}t}T^{3}_{0}\right|_{\mathrm{L}}
&=-\frac{2}{5}\sqrt{15}\mu\left(S_x\Im T^{2}_{2}-S_y\Re T^{2}_{2}\right).
\end{align}
\endgroup

In the presence of an external static magnetic field $\mathbf{B}$, the ground-state and metastable atomic operators evolve according to the magnetic-interaction Hamiltonian $H_B=-\hbar(\gamma_{1 / 2} \mathbf{K} \cdot \mathbf{B}+\gamma_{3 / 2} \mathbf{J} \cdot \mathbf{B}+\gamma_{\text {nuc }} \mathbf{I} \cdot \mathbf{B})$, where $\gamma_{1 / 2}=4\gamma_{\mathrm{ms}}/3$ is the gyromagnetic ratio of the metastable $F=1/2$ manifold, $\gamma_{3 / 2}=2\gamma_{\mathrm{ms}}/3$ is the gyromagnetic ratio of the metastable $F=3/2$ manifold, $\gamma_{\mathrm{ms}}=-2 \pi \times 28.02 \mathrm{~Hz} / \mathrm{nT}$ is the electronic gyromagnetic ratio, and $\gamma_{\mathrm{nuc}}=-2 \pi \times 32.43 \mathrm{~mHz} / \mathrm{nT}$ is the nuclear gyromagnetic ratio. Accordingly, the evolution equations of the ground-state and metastable atomic operators in the magnetic field can be derived as~\cite{2024-MFadel-NJP}
\begingroup
\allowdisplaybreaks[4]
\begin{align}
\left.\frac{\mathrm{d}}{\mathrm{d}t}\mathbf{I}\right|_{\mathrm{B}}
&=\gamma_{\mathrm{nuc}}\mathbf{I}\times\mathbf{B},\qquad
\left.\frac{\mathrm{d}}{\mathrm{d}t}\mathbf{K}\right|_{\mathrm{B}}
=\gamma_{1/2}\mathbf{K}\times\mathbf{B},\qquad
\left.\frac{\mathrm{d}}{\mathrm{d}t}\mathbf{J}\right|_{\mathrm{B}}
=\gamma_{3/2}\mathbf{J}\times\mathbf{B}, \\
\left.\frac{\mathrm{d}}{\mathrm{d}t}\Re T^{2}_{2}\right|_{\mathrm{B}}
&=\gamma_{3/2}\left(B_x\Im T^{2}_{1}+B_y\Re T^{2}_{1}+2B_z\Im T^{2}_{2}\right), \\
\left.\frac{\mathrm{d}}{\mathrm{d}t}\Im T^{2}_{2}\right|_{\mathrm{B}}
&=\gamma_{3/2}\left(-B_x\Re T^{2}_{1}+B_y\Im T^{2}_{1}-2B_z\Re T^{2}_{2}\right), \\
\left.\frac{\mathrm{d}}{\mathrm{d}t}\Re T^{2}_{1}\right|_{\mathrm{B}}
&=\gamma_{3/2}\left[B_x\Im T^{2}_{2}+B_y\left(\sqrt{3}T^{2}_{0}-\Re T^{2}_{2}\right)+B_z\Im T^{2}_{1}\right], \\
\left.\frac{\mathrm{d}}{\mathrm{d}t}\Im T^{2}_{1}\right|_{\mathrm{B}}
&=\gamma_{3/2}\left[-B_x\left(\sqrt{3}T^{2}_{0}+\Re T^{2}_{2}\right)-B_y\Im T^{2}_{2}-B_z\Re T^{2}_{1}\right], \\
\left.\frac{\mathrm{d}}{\mathrm{d}t}T^{2}_{0}\right|_{\mathrm{B}}
&=\gamma_{3/2}\sqrt{3}\left(B_x\Im T^{2}_{1}-B_y\Re T^{2}_{1}\right), \\
\left.\frac{\mathrm{d}}{\mathrm{d}t}\Re T^{3}_{3}\right|_{\mathrm{B}}
&=\gamma_{3/2}\left(B_x\sqrt{\frac{3}{2}}\Im T^{3}_{2}+B_y\sqrt{\frac{3}{2}}\Re T^{3}_{2}+3B_z\Im T^{3}_{3}\right), \\
\left.\frac{\mathrm{d}}{\mathrm{d}t}\Im T^{3}_{3}\right|_{\mathrm{B}}
&=\gamma_{3/2}\left(-B_x\sqrt{\frac{3}{2}}\Re T^{3}_{2}+B_y\sqrt{\frac{3}{2}}\Im T^{3}_{2}-3B_z\Re T^{3}_{3}\right), \\
\left.\frac{\mathrm{d}}{\mathrm{d}t}\Re T^{3}_{2}\right|_{\mathrm{B}}
&=\gamma_{3/2}\Bigg[
B_x\left(\sqrt{\frac{10}{4}}\Im T^{3}_{1}+\sqrt{\frac{3}{2}}\Im T^{3}_{3}\right)
+B_y\left(\sqrt{\frac{10}{4}}\Re T^{3}_{1}-\sqrt{\frac{3}{2}}\Re T^{3}_{3}\right)+2B_z\Im T^{3}_{2}
\Bigg], \\
\left.\frac{\mathrm{d}}{\mathrm{d}t}\Im T^{3}_{2}\right|_{\mathrm{B}}
&=\gamma_{3/2}\Bigg[
B_x\left(-\sqrt{\frac{10}{4}}\Re T^{3}_{1}-\sqrt{\frac{3}{2}}\Re T^{3}_{3}\right)
+B_y\left(\sqrt{\frac{10}{4}}\Im T^{3}_{1}-\sqrt{\frac{3}{2}}\Im T^{3}_{3}\right)-2B_z\Re T^{3}_{2}
\Bigg], \\
\left.\frac{\mathrm{d}}{\mathrm{d}t}\Re T^{3}_{1}\right|_{\mathrm{B}}
&=\gamma_{3/2}\left[B_x\sqrt{\frac{5}{2}}\Im T^{3}_{2}+B_y\left(\sqrt{6}T^{3}_{0}-\sqrt{\frac{5}{2}}\Re T^{3}_{2}\right)+B_z\Im T^{3}_{1}\right], \\
\left.\frac{\mathrm{d}}{\mathrm{d}t}\Im T^{3}_{1}\right|_{\mathrm{B}}
&=\gamma_{3/2}\left[-B_x\left(\sqrt{6}T^{3}_{0}-\sqrt{\frac{5}{2}}\Re T^{3}_{2}\right)-B_y\sqrt{\frac{5}{2}}\Im T^{3}_{2}-B_z\Re T^{3}_{1}\right], \\
\left.\frac{\mathrm{d}}{\mathrm{d}t}T^{3}_{0}\right|_{\mathrm{B}}
&=\gamma_{3/2}\sqrt{6}\left(B_x\Im T^{3}_{1}-B_y\Re T^{3}_{1}\right).
\end{align}
\endgroup

In addition to the two interactions discussed above, both the ground-state and metastable atoms are subject to MECs. The evolution equations of the ground-state and metastable atomic operators due to MECs are given by~\cite{2024-MFadel-NJP}
\begingroup
\allowdisplaybreaks[4]
\begin{align}
\left.\frac{\mathrm{d}}{\mathrm{d}t}\mathbf{I}\right|_{\mathrm{MEC}}
&=-\frac{1}{T_{\mathrm{MEC}}}\mathbf{I}
+\frac{1}{3T_{\mathrm{MEC}}}\frac{N_{\text{cell}}}{n_{\text{cell}}}
\left(\mathbf{J}-\mathbf{K}\right), \\
\left.\frac{\mathrm{d}}{\mathrm{d}t}\mathbf{K}\right|_{\mathrm{MEC}}
&=-\frac{7}{9\tau_{\mathrm{MEC}}}\mathbf{K}
+\frac{1}{9\tau_{\mathrm{MEC}}}\mathbf{J}
-\frac{1}{9\tau_{\mathrm{MEC}}}\frac{n_{\text{cell}}}{N_{\text{cell}}}\mathbf{I}
-\frac{4}{3\tau_{\mathrm{MEC}}}\frac{1}{N_{\text{cell}}}\mathbf{Q}\cdot\mathbf{I}, \\
\left.\frac{\mathrm{d}}{\mathrm{d}t}\mathbf{J}\right|_{\mathrm{MEC}}
&=-\frac{4}{9\tau_{\mathrm{MEC}}}\mathbf{J}
+\frac{10}{9\tau_{\mathrm{MEC}}}\mathbf{K}
+\frac{10}{9\tau_{\mathrm{MEC}}}\frac{n_{\text{cell}}}{N_{\text{cell}}}\mathbf{I}
+\frac{4}{3\tau_{\mathrm{MEC}}}\frac{1}{N_{\text{cell}}}\mathbf{Q}\cdot\mathbf{I}, \\
\left.\frac{\mathrm{d}}{\mathrm{d}t}\Re T^{2}_{2}\right|_{\mathrm{MEC}}
&=-\frac{2}{3\tau_{\mathrm{MEC}}}\Re T^{2}_{2}
+\frac{1}{\sqrt{3}\tau_{\mathrm{MEC}}}\frac{1}{N_{\text{cell}}}
\left(I_x\Sigma_x-I_y\Sigma_y\right), \\
\left.\frac{\mathrm{d}}{\mathrm{d}t}\Im T^{2}_{2}\right|_{\mathrm{MEC}}
&=-\frac{2}{3\tau_{\mathrm{MEC}}}\Im T^{2}_{2}
+\frac{1}{\sqrt{3}\tau_{\mathrm{MEC}}}\frac{1}{N_{\text{cell}}}
\left(I_x\Sigma_y+I_y\Sigma_x\right), \\
\left.\frac{\mathrm{d}}{\mathrm{d}t}\Re T^{2}_{1}\right|_{\mathrm{MEC}}
&=-\frac{2}{3\tau_{\mathrm{MEC}}}\Re T^{2}_{1}
-\frac{1}{\sqrt{3}\tau_{\mathrm{MEC}}}\frac{1}{N_{\text{cell}}}
\left(I_x\Sigma_z+I_z\Sigma_x\right), \\
\left.\frac{\mathrm{d}}{\mathrm{d}t}\Im T^{2}_{1}\right|_{\mathrm{MEC}}
&=-\frac{2}{3\tau_{\mathrm{MEC}}}\Im T^{2}_{1}
-\frac{1}{\sqrt{3}\tau_{\mathrm{MEC}}}\frac{1}{N_{\text{cell}}}
\left(I_y\Sigma_z+I_z\Sigma_y\right), \\
\left.\frac{\mathrm{d}}{\mathrm{d}t}T^{2}_{0}\right|_{\mathrm{MEC}}
&=-\frac{2}{3\tau_{\mathrm{MEC}}}T^{2}_{0}
+\frac{1}{3\tau_{\mathrm{MEC}}}\frac{1}{N_{\text{cell}}}
\left(3I_z\Sigma_z-\mathbf{I}\cdot\boldsymbol{\Sigma}\right), \\
\left.\frac{\mathrm{d}}{\mathrm{d}t}\Re T^{3}_{3}\right|_{\mathrm{MEC}}
&=-\frac{1}{\tau_{\mathrm{MEC}}}\Re T^{3}_{3}
-\frac{\sqrt{6}}{3\tau_{\mathrm{MEC}}}\frac{1}{N_{\text{cell}}}
\left(I_x\Re T^{2}_{2}-I_y\Im T^{2}_{2}\right), \\
\left.\frac{\mathrm{d}}{\mathrm{d}t}\Im T^{3}_{3}\right|_{\mathrm{MEC}}
&=-\frac{1}{\tau_{\mathrm{MEC}}}\Im T^{3}_{3}
-\frac{\sqrt{6}}{3\tau_{\mathrm{MEC}}}\frac{1}{N_{\text{cell}}}
\left(I_x\Im T^{2}_{2}-I_y\Re T^{2}_{2}\right), \\
\left.\frac{\mathrm{d}}{\mathrm{d}t}\Re T^{3}_{2}\right|_{\mathrm{MEC}}
&=-\frac{1}{\tau_{\mathrm{MEC}}}\Re T^{3}_{2}
-\frac{2}{3\tau_{\mathrm{MEC}}}\frac{1}{N_{\text{cell}}}
\left(I_x\Re T^{2}_{1}-I_y\Im T^{2}_{1}-I_z\Re T^{2}_{2}\right), \\
\left.\frac{\mathrm{d}}{\mathrm{d}t}\Im T^{3}_{2}\right|_{\mathrm{MEC}}
&=-\frac{1}{\tau_{\mathrm{MEC}}}\Im T^{3}_{2}
-\frac{2}{3\tau_{\mathrm{MEC}}}\frac{1}{N_{\text{cell}}}
\left(I_x\Im T^{2}_{1}+I_y\Re T^{2}_{1}-I_z\Im T^{2}_{2}\right), \\
\left.\frac{\mathrm{d}}{\mathrm{d}t}\Re T^{3}_{1}\right|_{\mathrm{MEC}}
&=-\frac{1}{\tau_{\mathrm{MEC}}}\Re T^{3}_{1}
+\frac{2}{3\sqrt{10}\tau_{\mathrm{MEC}}}\frac{1}{N_{\text{cell}}}
\left[I_x\left(\Re T^{2}_{2}-2\sqrt{3}T^{2}_{0}\right)+I_y\Im T^{2}_{2}+4I_z\Re T^{2}_{1}\right], \\
\left.\frac{\mathrm{d}}{\mathrm{d}t}\Im T^{3}_{1}\right|_{\mathrm{MEC}}
&=-\frac{1}{\tau_{\mathrm{MEC}}}\Im T^{3}_{1}
+\frac{2}{3\sqrt{10}\tau_{\mathrm{MEC}}}\frac{1}{N_{\text{cell}}}
\left[I_x\Im T^{2}_{2}-I_y\left(\Re T^{2}_{2}-2\sqrt{3}T^{2}_{0}\right)+4I_z\Im T^{2}_{1}\right], \\
\left.\frac{\mathrm{d}}{\mathrm{d}t}T^{3}_{0}\right|_{\mathrm{MEC}}
&=-\frac{1}{\tau_{\mathrm{MEC}}}T^{3}_{0}
+\frac{2}{\sqrt{5}\tau_{\mathrm{MEC}}}\frac{1}{N_{\text{cell}}}
\left[\frac{2\sqrt{3}}{6}\left(I_x\Re T^{2}_{1}+I_y\Im T^{2}_{1}\right)+I_zT^{2}_{0}\right],
\end{align}
\endgroup
where $\mathbf{Q}$ denotes the collective rank-2 tensor operator expressed in the Cartesian basis, and $\mathbf{\Sigma}=\frac{2}{3}(\mathbf{J}+2\mathbf{K})$ is the metastable electronic-spin operator. In summary, combining the light-interaction, magnetic-interaction, and MEC contributions gives the full semiclassical model summarized by Eq.~\eqref{atomiceqs}.

\subsection{Effective coupling between light and \he nuclear spins}\label{supp:secEffCoup}

Consider a nuclear polarization $M\in[0,1]$, a holding magnetic field applied along the $x$ direction, $\mathbf{B}=B_0\mathbf{e}_x$, and fixed probe-light intensity and fixed input polarization. Assuming that both the Stokes spin and the nuclear spin are polarized along the holding-field direction, the corresponding steady-state values can be written as
\begin{equation}
\langle S_x\rangle_s=\frac{n_{\mathrm{ph}}}{2}, \quad\langle S_y\rangle_s=\langle S_z\rangle_s=0, \quad\langle I_x\rangle_s=M \frac{N_{\mathrm{cell}}}{2}, \quad\langle I_y\rangle_s=\langle I_z\rangle_s=0.
\end{equation}
When MEOP reaches steady state, the metastable populations are well described by a spin-temperature distribution~\cite{2017-TRGentile-RMP}. 
The steady-state values of the metastable atomic operators can be expressed as~\cite{2024-MFadel-NJP}
\begin{subequations}
\begin{align}
\langle K_x\rangle_s&=\frac{M}{2}\left(\frac{1-M^2}{3+M^2}\right) n_{\text {cell }}, \quad\langle J_x\rangle_s=M\left(\frac{5+M^2}{3+M^2}\right) n_{\text {cell }},\\
\langle T_0^2\rangle_s&=-\left(\frac{M^2}{3+M^2}\right) n_{\text {cell }}, \quad \langle\Re T_2^2\rangle_s=\sqrt{3}\left(\frac{M^2}{3+M^2}\right) n_{\text {cell }},\\
\langle\Re T_1^3\rangle_s&=\sqrt{\frac{3}{10}}\left(\frac{M^3}{3+M^2}\right) n_{\text {cell }}, \quad \langle\Re T_3^3\rangle_s=-\frac{1}{\sqrt{2}}\left(\frac{M^3}{3+M^2}\right) n_{\text {cell }},
\end{align}
\end{subequations}
while the steady-state values of all other metastable atomic operators not listed above vanish. Since the present work focuses on the regime where the nuclear spin is tilted only slightly away from the longitudinal axis, Eq.~\eqref{atomiceqs} can be linearized. Specifically, we substitute $\langle O\rangle=\langle O\rangle_s+\delta O$, where $\delta O$ denotes a small fluctuation of $\langle O\rangle$ around its steady-state value $\langle O\rangle_s$. Retaining only terms up to first order in the fluctuations yields a set of linear differential equations for the fluctuation variables~\cite{2024-MFadel-NJP}.

For the two probe-detuning configurations shown in Fig.~\ref{fig:1} of the main text, the set of linear differential equations can be further reduced to a \textbf{``simplified model''} involving only three coupled spin degrees of freedom, corresponding to the Stokes spin, the metastable spin, and the ground-state nuclear spin. In Config.1 and Config.2, the light-metastable-atom interaction Hamiltonian $H_{\mathrm{LA}}$ is dominated by $H_{1 / 2}^V$ and $H_{3 / 2}^V$, respectively. One may therefore neglect the non-dominant terms in $H_{\mathrm{LA}}$. Furthermore, among the metastable degrees of freedom, those atomic operators whose dynamics are governed solely by the magnetic-field interaction and MECs can be adiabatically eliminated~\cite{2024-MFadel-NJP}. We now describe this procedure in detail.

We first consider Config.1. Starting from the linearized equations, the coupling between the light field and the metastable $F=3/2$ manifold can be neglected by setting $\eta=\mu=0$. We then adiabatically eliminate the degrees of freedom $\delta J_\alpha$ and $\delta Q_{\alpha x}$ by solving the algebraic equations $\mathrm{d}(\delta J_\alpha)/\mathrm{d}t=0$ and $\mathrm{d}(\delta Q_{\alpha x})/\mathrm{d}t=0$ (where $\alpha=y,z$, and $\delta Q_{\alpha x}$ denotes a collective rank-2 tensor operator in the Cartesian basis~\cite{2024-MFadel-NJP}). The resulting solutions are substituted back into the evolution equations of the remaining variables. Since our primary interest is the atom-light coupling, we further take the limit $B_0\rightarrow0$, which yields~\cite{2021-ASerafin-CRPhys,2024-MFadel-NJP}
\begin{subequations}
\begin{align}
\frac{\partial}{\partial z} \delta S_y&=\frac{1}{L}\langle S_x\rangle_s \chi \delta K_z, \label{deltaSypz}\\
\frac{\partial}{\partial z} \delta S_z&=0, \\
\frac{\mathrm{d}}{\mathrm{d} t} \delta I_y&=-\gamma^{(1/2)}_f \delta I_y+\gamma^{(1/2)}_m \delta K_y, \label{deltaIydt}\\
\frac{\mathrm{d}}{\mathrm{d} t} \delta I_z&=-\gamma^{(1/2)}_f \delta I_z+\gamma^{(1/2)}_m \delta K_z, \\
\frac{\mathrm{d}}{\mathrm{d} t} \delta K_y&=-\gamma^{(1/2)}_m \delta K_y+\gamma^{(1/2)}_f \delta I_y+\chi\langle K_x\rangle_s \delta S_z,\\
\frac{\mathrm{d}}{\mathrm{d} t} \delta K_z&=-\gamma^{(1/2)}_m \delta K_z+\gamma^{(1/2)}_f \delta I_z,
\end{align}
\end{subequations}
where the effective MEC rates of the ground-state and metastable atoms are given by~\cite{2021-ASerafin-CRPhys,2024-MFadel-NJP}
\begin{equation}\label{MECrateConf1}
\gamma_f^{(1 / 2)}=\frac{1}{T_{\mathrm{MEC}}} \frac{\left(4+M^2\right)\left(1-M^2\right)}{\left(8-M^2\right)\left(3+M^2\right)}, \quad \gamma_m^{(1 / 2)}=\frac{1}{\tau_{\mathrm{MEC}}} \frac{\left(4+M^2\right)}{\left(8-M^2\right)}.
\end{equation}
Following the approach of Ref.~\cite{2024-MFadel-NJP}, the effective coupling strength between the light field and the ground-state nuclear spin can be obtained by further adiabatically eliminating the metastable atomic operators $\delta K_y$ and $\delta K_z$. Specifically, we set $\mathrm{d}(\delta K_y)/\mathrm{d}t=0$ and $\mathrm{d}(\delta K_z)/\mathrm{d}t=0$, which yields
\begin{subequations}
\begin{align}
\gamma_m^{(1 / 2)} \delta K_y-\gamma_f^{(1 / 2)} \delta I_y&=\chi\langle K_x\rangle_s \delta S_z, \label{deltadengshi} \\
\delta K_z=\frac{\gamma_f^{(1 / 2)}}{\gamma_m^{(1 / 2)}} \delta I_z&=\frac{\langle K_x\rangle_s}{\langle I_x\rangle_s} \delta I_z. \label{deltaKz}
\end{align}
\end{subequations}
Substituting Eq.~\eqref{deltaKz} into Eq.~\eqref{deltaSypz}, and Eq.~\eqref{deltadengshi} into Eq.~\eqref{deltaIydt}, one obtains
\begin{subequations}\label{deltaSydeltaIy}
\begin{align}
\frac{\partial}{\partial z} \delta S_y&=\frac{\chi}{L} \frac{\langle K_x\rangle_s}{\langle I_x\rangle_s}\langle S_x\rangle_s \delta I_z,\\
\frac{\mathrm{d}}{\mathrm{d} t} \delta I_y&=\chi\langle K_x\rangle_s \delta S_z.
\end{align}
\end{subequations}
Using the Holstein-Primakoff approximation, we introduce the canonical operators
\begin{subequations}\label{zhengzesuanfu}
\begin{align}
\frac{\delta S_y}{\sqrt{\langle S_x\rangle_s}} &\simeq X_S, \quad \frac{\delta I_y}{\sqrt{\langle I_x\rangle_s}} \simeq X_I, \\
\frac{\delta S_z}{\sqrt{\langle S_x\rangle_s}} &\simeq P_S, \quad \frac{\delta I_z}{\sqrt{\langle I_x\rangle_s}} \simeq P_I,
\end{align}
\end{subequations}
which satisfy the commutation relations $[X_S, P_S]=[X_I, P_I]=i$. Substituting Eq.~\eqref{zhengzesuanfu} into Eq.~\eqref{deltaSydeltaIy}, we obtain
\begin{subequations}\label{XsXieqs}
\begin{align}
\frac{\partial}{\partial z} X_S&=\frac{\chi}{L} \frac{\left\langle K_x\right\rangle_s}{\left\langle I_x\right\rangle_s} \sqrt{\left\langle S_x\right\rangle_s\left\langle I_x\right\rangle_s} P_I, \label{Xspartialz} \\
\frac{\mathrm{d}}{\mathrm{d} t} X_I&=\chi \frac{\left\langle K_x\right\rangle_s}{\left\langle I_x\right\rangle_s} \sqrt{\left\langle S_x\right\rangle_s\left\langle I_x\right\rangle_s} P_S.
\end{align}
\end{subequations}
From the Heisenberg equations satisfied by the light-field and atomic operators, one verifies that the form of Eq.~\eqref{XsXieqs} allows one to construct an effective interaction Hamiltonian between the light field and the ground-state nuclear spin, $H_{\text {eff}}=\hbar \Omega^{(1 / 2)} P_S P_I$, where the effective coupling strength is given by~\cite{2024-MFadel-NJP}
\begin{equation}\label{Conf1Omegaeq}
\begin{aligned}
\Omega^{(1 / 2)}&=\chi \frac{\left\langle K_x\right\rangle_s}{\left\langle I_x\right\rangle_s} \sqrt{\left\langle S_x\right\rangle_s\left\langle I_x\right\rangle_s} \\
&=\chi \frac{n_{\text {cell }}}{N_{\text {cell }}} \sqrt{n_{\mathrm{ph}} N_{\text {cell }}} f^{(1 / 2)}(M).
\end{aligned}
\end{equation}
In the second line of this expression, we have defined the polarization-dependent scaling function
\begin{equation}
f^{(1 / 2)}(M)=\frac{1}{2}\left(\frac{1-M^2}{3+M^2}\right) \sqrt{M}.
\end{equation}
Furthermore, integrating both sides of Eq.~\eqref{Xspartialz} yields the input-output relation for the light field
\begin{equation}\label{XsoutConf1}
\begin{aligned}
X_S^{\text {out}}(t) & =X_S^{\mathrm{in}}(t)+\chi \frac{\left\langle K_x\right\rangle_s}{\left\langle I_x\right\rangle_s} \sqrt{\left\langle S_x\right\rangle_s\left\langle I_x\right\rangle_s} P_I(t) \\
& =X_S^{\mathrm{in}}(t)+\Omega^{(1 / 2)} P_I(t).
\end{aligned}
\end{equation}

We next consider Config.2. Starting again from the linearized equations, the contribution of $H_{1 / 2}^V$ can be neglected at high nuclear polarization, and we therefore set $\chi=0$ to ignore the coupling between the light field and the metastable $F=1/2$ manifold. In addition, Fig.~\ref{fig:1} of the main text shows that the tensor coupling is small in Config.2, allowing us to further set $\mu=0$. We then adiabatically eliminate the degrees of freedom $\delta K_\alpha$ and $\delta Q_{\alpha x}$ by solving the algebraic equations $\mathrm{d}(\delta K_\alpha)/\mathrm{d}t=0$ and $\mathrm{d}(\delta Q_{\alpha x})/\mathrm{d}t=0$. Substituting the resulting solutions into the evolution equations of the remaining variables, and taking the limit $B_0\rightarrow0$, one obtains~\cite{2024-MFadel-NJP}
\begin{subequations}
\begin{align}
\frac{\partial}{\partial z} \delta S_y&=\frac{1}{L}\langle S_x\rangle_s \eta \delta J_z, \label{deltaSypz2} \\
\frac{\partial}{\partial z} \delta S_z&=0, \\
\frac{\mathrm{d}}{\mathrm{d} t} \delta I_y&=-\gamma^{(3/2)}_f \delta I_y+\gamma^{(3/2)}_m \delta J_y, \label{deltaIydt2}\\
\frac{\mathrm{d}}{\mathrm{d} t} \delta I_z&=-\gamma^{(3/2)}_f \delta I_z+\gamma^{(3/2)}_m \delta J_z, \\
\frac{\mathrm{d}}{\mathrm{d} t} \delta J_y&=-\gamma^{(3/2)}_m \delta J_y+\gamma^{(3/2)}_f \delta I_y+\eta\langle J_x\rangle_s \delta S_z,\\
\frac{\mathrm{d}}{\mathrm{d} t} \delta J_z&=-\gamma^{(3/2)}_m \delta J_z+\gamma^{(3/2)}_f \delta I_z,
\end{align}
\end{subequations}
where the effective MEC rates of the ground-state and metastable atoms are given by~\cite{2024-MFadel-NJP}
\begin{equation}\label{MECrateConf2}
\gamma_f^{(3 / 2)}=\frac{1}{T_{\mathrm{MEC}}} \frac{\left(4+M^2\right)\left(5+M^2\right)}{\left(7+M^2\right)\left(3+M^2\right)}, \quad \gamma_m^{(3 / 2)}=\frac{1}{\tau_{\mathrm{MEC}}} \frac{\left(4+M^2\right)}{2\left(7+M^2\right)}.
\end{equation}
To obtain the effective coupling strength between the light field and the ground-state nuclear spin, we follow the same procedure as for Config.1. By adiabatically eliminating the metastable atomic operators and subsequently introducing the Holstein-Primakoff approximation, one can obtain
\begin{subequations}
\begin{align}
\frac{\partial}{\partial z} X_S&=\frac{\eta}{L} \frac{\left\langle J_x\right\rangle_s}{\left\langle I_x\right\rangle_s} \sqrt{\left\langle S_x\right\rangle_s\left\langle I_x\right\rangle_s} P_I, \label{partialXspz} \\
\frac{\mathrm{d}}{\mathrm{d} t} X_I&=\eta \frac{\left\langle J_x\right\rangle_s}{\left\langle I_x\right\rangle_s} \sqrt{\left\langle S_x\right\rangle_s\left\langle I_x\right\rangle_s} P_S.
\end{align}
\end{subequations}
This allows one to construct an effective interaction Hamiltonian between the light field and the ground-state nuclear spin, $H_{\mathrm{eff}}=\hbar\Omega^{(3/2)} P_S P_I$, where the effective coupling strength is given by~\cite{2024-MFadel-NJP}
\begin{equation}\label{Conf2Omegaeq}
\begin{aligned}
\Omega^{(3 / 2)}&=\eta \frac{\left\langle J_x\right\rangle_s}{\left\langle I_x\right\rangle_s} \sqrt{\left\langle S_x\right\rangle_s\left\langle I_x\right\rangle_s} \\
&=\eta \frac{n_{\text {cell }}}{N_{\text {cell }}} \sqrt{n_{\mathrm{ph}} N_{\text {cell }}} f^{(3 / 2)}(M).
\end{aligned}
\end{equation}
The polarization-dependent scaling function now reads
\begin{equation}
f^{(3 / 2)}(M)=\left(\frac{5+M^2}{3+M^2}\right) \sqrt{M}.
\end{equation}
Integrating both sides of Eq.~\eqref{partialXspz} yields the input–output relation for the light field
\begin{equation}\label{XsoutConf2}
\begin{aligned}
X_S^{\text {out}}(t) & =X_S^{\mathrm{in}}(t)+\eta \frac{\left\langle J_x\right\rangle_s}{\left\langle I_x\right\rangle_s} \sqrt{\left\langle S_x\right\rangle_s\left\langle I_x\right\rangle_s} P_I(t) \\
& =X_S^{\mathrm{in}}(t)+\Omega^{(3 / 2)} P_I(t).
\end{aligned}
\end{equation}

\subsection{Measurement-based spin squeezing and multi-pass enhancement}
\label{multienhance}

To illustrate how measurement-based spin squeezing can be achieved, consider the Faraday Hamiltonian for a single light pass
\begin{equation}\label{eq:Fex}
    H = \hbar\, \alpha \, {S}_z {J}_z \;.
\end{equation}
Here, $\alpha$ is the single-atom-light coupling constant, which is a dimensionless number that depends on the atomic cross section, light detuning, transition linewidth (see Eq.~\eqref{alphaeq}), but independent of the number of atoms $n_\text{at}$ or, in other words, of the optical depth of the ensemble $\text{OD} \propto n_\text{at}$.
Note that we have chosen to use in Eq.~\eqref{eq:Fex} the collective spin operator ${J}_z$, to follow the standard derivation given for alkali-atom systems, but it should be emphasized that the same idea will follow when considering the effective Faraday Hamiltonian between light and \he nuclear spins.

Using the the HP approximation $P_S=S_z/\sqrt{\avg{S_x}}$, $P_J=J_z/\sqrt{\avg{J_x}}$, with $\avg{S_x}=\nph/2$, $\avg{J_x}=n_\text{at}/2$, we have
\begin{align}
    H &= \hbar\, \underbrace{\alpha \, \frac{1}{2} \sqrt{\nph n_\text{at}}}_{\Omega} P_S P_J \\
    &= \hbar \, \Omega \, P_S P_J \;.
\end{align}
Since here $\nph$ is a photon flux, $\Omega$ has units of sec$^{-1/2}=\sqrt{\text{Hz}}$. 
From now on, the discussion we present can be directly applied also to the effective Faraday Hamiltonian Eq.~\eqref{eq:effHPSPI12} of the main text, namely $H_{\text{eff}} = \hbar \Omega^{(i)}  P_S P_I$, by replacing $P_J \rightarrow P_I$ and $\Omega \rightarrow \Omega^{(i)}$.

From the above Eq.~\eqref{XsoutConf1} and Eq.~\eqref{XsoutConf2}, the input/output relation is
\begin{equation}
\begin{aligned}
X_S^{\text {out}}(t) & =X_S^{\text {in}}(t)+ \alpha \sqrt{\langle S_x\rangle_s \langle J_x\rangle_s} P_J(t) \\
& =X_S^{\text {in}}(t)+\Omega P_J(t).
\end{aligned}
\end{equation}
A continuous homodyne measurement of $X_S^{\rm out}$ produces the photocurrent \cite{Carmichael1999Statistical1,Carmichael2008Statistical2}
\begin{equation}
    I(t)\, dt = \sqrt{2\eta_{\rm det}}\,
    X_S^{\rm out}(t)\, dt + dW(t),
\label{photocurrent}
\end{equation}
where $0 \le \eta_{\rm det} \le 1$ is the detection efficiency and
$dW(t)$ is a Wiener noise increment satisfying
\begin{equation}
    \mathbb E[dW(t)] = 0, \qquad dW(t)^2 = dt .
\end{equation}
Substituting the input-output relation gives
\begin{equation}
    I(t)\, dt = \sqrt{2\eta_{\rm det}}
    \left[ X_S^{\rm in}(t) + \Omega P_J(t) \right] dt + dW(t),
\label{photocurrent2}
\end{equation}
so that the measurement record contains a signal proportional to $P_J(t)$ and white shot noise from $dW(t)$. Note that in the above equation it is possible to absorb the term proportional to $X_S^{\rm in}(t)$ inside the Wiener increment $dW(t)$, such that the latter represents the total shot noise of the homodyne current.

Conditioned on the photocurrent $I(t)$, the atomic density operator
$\rho_c(t)$ evolves according to the stochastic master equation \cite{Carmichael1999Statistical1,Carmichael2008Statistical2}
\begin{equation}
    d\rho_c
    = - i [H, \rho_c]\, dt
      + \mathcal{D}\!\left[\sqrt{\Gamma_m}\, P_J \right]\rho_c\, dt
      + \sqrt{\Gamma_m}\,\mathcal{H}[P_J]\rho_c\, dW(t).
\label{stochastic}
\end{equation}
Here $\mathcal{D}$ and $\mathcal{H}$ are the standard Lindblad
superoperators
\begin{align}
    \mathcal{D}[L]\rho 
    &= L \rho L^\dagger - \frac{1}{2} \left( L^\dagger L \rho + \rho L^\dagger L \right), \\[4pt]
    \mathcal{H}[L]\rho
    &= L\rho + \rho L^\dagger
       - \mathrm{Tr}(L\rho + \rho L^\dagger)\,\rho ,
\end{align}
for operator $L$ and state $\rho$, and we have introduced the measurement rate
\begin{equation}
    \Gamma_m = 2\,\eta_{\rm det}\,\Omega^2 .
\label{measrate}
\end{equation}

Define the conditional mean and variance of the atomic quadrature $P_J$ as
\begin{equation}
    x(t) = \langle P_J \rangle_c , 
    \qquad 
    V(t) = 
    \langle P_J^2 \rangle_c - \langle P_J \rangle_c^2 .
\end{equation}
Using $\rho_c$ from the stochastic master equation and the fact that $P_J$ is a QND observable (\,$[P_J , H] = 0$\,), the conditional mean obeys
\begin{equation}
    dx(t) = 2\sqrt{\Gamma_m}\, V(t)\, dW(t),
\end{equation}
while the conditional second moment satisfies
\begin{equation}
    d\langle P_J^2 \rangle_c
    = 4\sqrt{\Gamma_m}\,
      \langle P_J^2 - x(t) P_J \rangle_c \, dW(t).
\end{equation}
Combining these expressions yields the evolution equation for the conditional variance
\begin{equation}
    dV(t) = - 2 \Gamma_m\, V(t)^2\, dt ,
\end{equation}
showing that the QND measurement results in a reduction of the variance at the squeezing rate
\begin{equation}
    \Gsq = 2\Gamma_m = 4\,\eta_{\rm det}\, \Omega^2 .
\end{equation}
The stronger the atom-light coupling $\Omega$, the faster the
measurement extracts information about the QND observable $P_J$, and
the faster the conditional variance is reduced.

\vspace{5mm}
In the case of a multiple number of light passes through the ensemble, $\Npass$, we have
\begin{align}
    H &= \hbar\, \alpha \Npass \, {S}_z {J}_z \\
    &= \hbar\, \underbrace{\alpha \,\Npass\, \frac{1}{2} \sqrt{\nph n_\text{at}}}_{\Omega} P_S P_J
\end{align}
Repeating the above calculation gives us a squeezing rate (for $\eta_{\rm det}=1$)
\begin{align}
    \Gsq &= 4 \Omega^2   \\
    &= \alpha^2 \,\Npass^2\, \nph n_\text{at} \;.
\end{align}
This expression is consistent with Eq.~(S39) of Ref.~\cite{2015-GVasilakis-NP}, where there is a Finesse$^2$ ($\propto\Npass^2$ here) and a dependence on $\kappa_0^2$ with $\kappa_0\propto\sqrt{\nph n_\text{at}}\propto\sqrt{\text{OD}}$.

\clearpage
\newpage

\section{Diffusion considerations}\label{supp:secDiffuse}

\renewcommand{\arraystretch}{1.4} 
\setlength{\tabcolsep}{12pt} 

\begin{table}[t]
\begin{center}
\begin{tabular}{ |c|c|c| } 
 \hline
 \textbf{quantity} & \textbf{value} & \textbf{reference} \\ 
 \hline
 temperature & $T=\unit{300}{K}$ &  \\ 
 \he gas pressure in the cells & $p=\unit{100}{Pa}=\unit{1}{mbar}=\unit{0.75}{Torr}$ &  \\ 
 probe beam power & $P_L=\unit{0.55}{mW}$ & \\
 \hline
 single-pass cell volume & $V=\unit{3.6}{cm^3}$ & $V=\pi R^2 L$ \\
 single-pass cell GS \he number &  $\Ncell= 8.6\times 10^{16}$  & $p V = \Ncell k_B T$ \\
 single-pass cell MS \he number &  $\ncell= 7.2\times 10^{11}$ & from measurement, Fig.~\ref{nMeaSinC8} \\
 \hline
 multi-pass cell volume & $V=\unit{5.5}{cm^3}$ & $V= H^2 L$ minus mirrors volume \\
 multi-pass cell GS \he number & $\Ncell= 1.3\times 10^{17}$ & $p V = \Ncell k_B T$ \\
 multi-pass cell MS \he number &  $\ncell= 5.5\times 10^{11}$ & from measurement, Fig.~\ref{nMeaMultiC8} \\
 \hline
 \he mass & $\mhe = \unit{5.008 \times 10^{-27}}{kg}$ & \\
 \he mean thermal velocity & $\vmean=\unit{1450}{m/s}$  &  $\vmean=\sqrt{8 k_B T / \pi \mhe }$ \\
 \he rms thermal velocity & $\vrms=\unit{1580}{m/s}$ &  $\vrms=\sqrt{3 k_B T / \mhe}$ \\
 \he diffusion constant & $D_0=\unit{1430(60)}{cm^2/s}$ $@$ \unit{1}{Torr}, \unit{300}{K} & \cite{Barbe74}, \cite{Hayden2004} and \cite{Tastevin2005HeDiffusion} \\ 
 metastable \he diffusion constant & $D_0^\ast=\unit{550(30)}{cm^2/s}$ $@$ \unit{1}{Torr}, \unit{300}{K} & \cite{TastevinPhD} Pg.~109, \cite{batzthesis} Pg.~314 and \cite{Fitzsimmons68}\\ 
 \he mean free path & $\lambda=\unit{3.9 \times 10^{-4}}{m}$ & $\lambda=3D(p)/\vmean$ \\
 metastable \he mean free path & $\lambda=\unit{1.5 \times 10^{-4}}{m}$ & $\lambda=3D^\ast(p)/\vmean$ \\
 \he coherence time & $T_2 = \unit{1.5}{s}$ $@$ \unit{600}{nT} & from measurement, Fig.~\ref{fig:3} \\
 metastable \he lifetime & $\tau_M = 0.1 - \unit{1}{ms}$ & \cite{1960-PR-Colegrove} \\
 \he diffusion distance & $r=\unit{1.3}{m}$ & $r=\sqrt{6D(p)T_2}$ \\
 metastable \he diffusion distance & $r^\ast=\unit{2.1}{cm}$ & $r^\ast=\sqrt{6D^\ast(p)\tau_M}$ \\
 metastability-exchange rate coefficient & $k_{\text{MEC}}=\unit{154 \times 10^{-12}}{cm^3/s}$ $@$ \unit{300}{K} & Tab.~II in \cite{2017-TRGentile-RMP} \\
 metastable \he MEC time constant & $\tau_{\text{MEC}}=\unit{2.7 \times 10^{-7}}{s}$ & $\tau_{\text{MEC}} = 1/(k_{\text{MEC}} \Ncell/V)$ \\
 \hline
\end{tabular}
\end{center}
\caption{\textbf{Experimental parameters.} }\label{tab} 
\end{table}

\subsection{Analysis of \he motion in our cell}

In this subsection, we conduct a theoretical analysis of the motion of \he atoms in our cell. 
To this end, we start with introducing a few fundamental concepts from kinetic theory. 
The first is the \textbf{mean~free~path} $\lambda$, which represents the average distance that a gas atom (or molecule) travels between two successive collisions.
Besides depending on the particles' collisional cross section, it depends on pressure $p$ and temperature $T$ as $\lambda\propto T/p$.
The motion is said to be in the \textbf{ballistic regime} if the mean free path is much larger than the characteristic container dimension $R$, meaning $\lambda \gg R$, or in the \textbf{diffusive regime} if $\lambda \ll R$.

The diffusive regime is characterized by the \textbf{diffusion~coefficient}, which for a diffusion process in $d$ spatial dimensions is given by $D=\vmean \lambda / d$, where $\vmean$ is the mean thermal velocity of the particles. 
From this, we introduce the \textbf{diffusion~distance} $r \equiv \sqrt{\avg{r(t)^2}}=\sqrt{2d D t}$ as the the root-mean-square displacement of the atoms due to diffusion over a time $t$. 
A rough estimate of the time it takes for a particle to reach the container walls is found by solving for $r=R$, which gives the \textbf{characteristic~diffusion~time} $\tau_D = R^2/2d D$.
For a more precise estimate, however, one needs to solve the diffusion equation $\partial n(\vec{r},t)/\partial t = D \nabla^2 n(\vec{r},t)$ for the atomic density distribution $n(\vec{r},t)$ within the specific container under consideration.
Taking as ansatz $n(\vec{r},t)=n(\vec{r})e^{-t/\tau_D}$, one finds for example
\begin{eqnarray}
    \dfrac{1}{\tau_D^\text{cyl}} = D \left[ \left(\dfrac{\alpha_0}{R}\right)^2 + \left(\dfrac{\pi}{L} \right)^2 \right] \;,\qquad\qquad  \dfrac{1}{\tau_D^\text{cuboid}} = D \left[ \left(\dfrac{\pi}{R_x}\right)^2 + \left(\dfrac{\pi}{R_y}\right)^2 + \left(\dfrac{\pi}{L} \right)^2 \right] \;,
    \label{cylandcub}
\end{eqnarray}
for a cylinder and cuboid, respectively, where $\alpha_0 \approx 2.405$ is the first zero of the cylindrical Bessel function $J_0(x)$, $L$ is the cell length, and $R$ is either the cell radius (for the cylinder) or the length of the edges (for the cuboid).

At constant temperature, the diffusion constant is inversely proportional to the pressure, $D\propto 1/p$. We know that the diffusion constant of metastable atoms at $\unit{300}{K}$ and pressure of $\unit{1}{Torr}$, is $D^\ast_0=\unit{550}{cm^2/s}$, see Tab.~\ref{tab}. 
For some other gas pressure $p$ we thus have
\begin{equation}\label{Dstar1}
D^\ast(p)=D^\ast_0 \cdot \dfrac{\unit{1}{Torr}}{p}.
\end{equation}
Taking the pressure of our cell, and considering the typical metastable lifetime $\tau_M \approx \unit{1}{ms}$, we can estimate that the diffusion distance is $r^\ast=\sqrt{6D^\ast(p)\tau_M}\approx\unit{2.1}{cm}$.
This shows us that, during their brief lifetime, metastable atoms can diffuse only over a short distance comparable to the cell diameter, $r^\ast\propto R$.
As a result, a metastable atom typically encounters the probe beam only once during its lifetime.
Note that, if we consider the characteristic time interval between two ME collisions, $\tau_{\text{MEC}}$ in Tab.~\ref{tab}, we obtain that the diffusion distance is $r^\ast_{\text{MEC}}=\sqrt{6D^\ast(p)\tau_{\text{MEC}}}\approx\unit{345}{\mu m}$.

In contrast, the relevant timescale to consider for ground-state \he atoms is the nuclear spin coherence time $T_2$, which in our experiments is in the order of a second.
Within this time, we can estimate the diffusion distance to be $r=\sqrt{6D(p)T_2}\approx\unit{1.3}{m}$, much larger than the cell dimensions.
This allows the nuclear spins to explore the entire cell volume, meaning that the collective nuclear spin becomes spatially homogenized.

Metastability-exchange collisions provide an effective interface between light and \he nuclear spins. In each collision, a metastable atom acquires the nuclear spin state of a ground-state atom at the collision location. Because the ground-state atoms diffuse over long distances, their spin polarization is spatially averaged across the cell, and metastable atoms locally sample this averaged spin. However, the light–matter interaction remains spatially localized, as metastable atoms do not diffuse sufficiently fast to experience motional averaging over the beam profile. In this sense, the role of a metastable atom is to sample the collective nuclear spin state at the point where MEC occurs and then convey that information to the light.

\subsection{Motional averaging considerations}

To clarify the role of motional averaging in the interaction between light and atoms, let us start by presenting a very general scenario where a probe light beam propagating along the $z$ direction interacts with an atomic ensemble described as a collection of spin-$1/2$ (i.e. two-level) systems.
The microscopic Hamiltonian for the Faraday interaction between the $i$th spin ${j}^{(i)}_z$ at position $\vec{r}^{(i)}$ and the light beam is
\begin{equation}
    h^{(i)} = \hbar\, \alpha\, {j}^{(i)}_z {S}_z \;,
\label{microham}
\end{equation}
where ${S}_z$ is the Stokes operator of the probe light and $\alpha$ is the single-atom-light coupling constant.
Note that the Stokes operator has dimension [s$^{-1}$], such that $h^{(i)}$ has dimensions of energy, and that it is assumed to take a constant non-zero value only inside a cylindrical volume of length $L$ and area $\Abeam$.
The coupling constant scales as
\begin{equation}
    \alpha \propto \dfrac{\Gamma}{\Delta}\dfrac{\sigma_2}{\Abeam} \qquad\qquad \text{[dimensionless]} \;,
\label{alphaeq}
\end{equation}
with $\Gamma$ the spontaneous decay rate of the considered transition, $\Delta$ the probe light detuning, and $\sigma_2=3\lambda_\text{TLS}^2/2\pi$ the resonant scattering cross-section for an ideal electric dipole transition with wavelength $\lambda_\text{TLS}$.

Now, introduce the collective spin density operator
\begin{equation}
    {\mathcal{J}}_z(\vec{r}) = \sum_i {j}^{(i)}_z \delta(\vec{r}-\vec{r}^{(i)}) \qquad\qquad \text{[$1/m^3$]} \;,
\label{Collspindensity}
\end{equation}
and the collective spin operator summing over all spins contained in the cell
\begin{align}
    {J}_z &= \sum_{i\in\text{cell}} {j}^{(i)}_z\\
    &= \int_{V_{\rm cell}} d^3\vec{r}\, {\mathcal{J}}_z(\vec{r}) \;.
\label{Collspinoperator}
\end{align}
Note that, in the presence of motional averaging the collective spin density is spatially uniform, meaning ${\mathcal{J}}_z(\vec{r})=J_z / \Acell L$.

We can write the total interaction Hamiltonian as
\begin{alignat}{2}
    H &= \sum_i h^{(i)} &&\\
    &= \hbar\, \alpha \int_{\mathbb{R^3}} d^3\vec{r} \,  {\mathcal{J}}_z(\vec{r}) {S}_z && \\ 
    &= \hbar\, \alpha \int_{\text{beam}} d^3\vec{r} \, {\mathcal{J}}_z(\vec{r}) {S}_z &&\qquad\qquad \text{since $S_z\neq0$ only inside the beam volume,}\\ 
    &= \hbar\, \alpha \int_{\text{beam}} d^3\vec{r} \, \frac{J_z}{\Acell L} {S}_z &&\qquad\qquad \text{assuming ${\mathcal{J}}_z(\vec{r}) = \frac{J_z}{\Acell L}$ due to motional averaging,}\\ 
    &= \hbar\, \alpha \, \left(\dfrac{\Abeam}{\Acell}\right) \, {J}_z {S}_z &&\qquad\qquad \text{using $\int_{\text{beam}} d^3\vec{r} = L \Abeam$}\;. 
\end{alignat}
In the above derivation we have assumed fast motional averaging, which results in a spatially uniform collective spin density ${\mathcal{J}}_z(\vec{r})$. However, if this process does not take place, the substitution ${\mathcal{J}}_z(\vec{r}) = {J}_z/\Acell L$ is in general not valid. For example, if atoms are completely frozen in space, we would have that the light field only interacts with the collective spin defined by $\int_{\text{beam}} d^3\vec{r} \,{\mathcal{J}}_z(\vec{r}) = \sum_{i\in\text{beam}} {j}^{(i)}_z$.

\vspace{5mm}
After this illustrative discussion, we can show in more detail how the  above consideration applies to our experiment. 
For concreteness, let us first focus on Config.1.
Following the discussion presented in Sec.~\ref{supp:secEqsStokes}, in the scenario where the the light beam only interact with a fraction of metastable atoms Eq.~\eqref{eqsupp:HLA} needs to be replaced by
\begin{equation}\label{HLAconf1}
H_{\mathrm{LA}}=\hbar \chi \int_0^L k_z(z, t) S_z(z, t) \rho A_{\text {beam}} \mathrm{d} z.
\end{equation}
With the substitution ${k}_z(z,t)={K}_z(t)/\rho A_\text{cell} L$, which is motivated by the fact that in the regime under consideration MEC keep the metastable spin state locked to the uniform ground-state spin state as  discussed above Eq.~\eqref{Stokeseqconf1}, this Hamiltonian becomes
\begin{equation}
H_{\mathrm{LA}}=\hbar \chi \int_0^L \dfrac{\Abeam}{\Acell L} {K}_z(t) S_z(z, t) \mathrm{d} z.
\end{equation}
Using the commutation relations Eqs.~\eqref{commu2}, the equations of motion for the Stokes operators become
\begin{subequations}
\begin{align}
\frac{\partial}{\partial z} S_x(z, t)&=-\frac{\chi}{L}\frac{A_\text{beam}}{A_\text{cell}} K_z(t) S_y(z, t),\\
\frac{\partial}{\partial z} S_y(z, t)&=\frac{\chi}{L}\frac{A_\text{beam}}{A_\text{cell}} K_z(t) S_x(z, t) .
\end{align}
\end{subequations}
This shows that one can consider the derivation given in Sec.~\ref{supp:secEffCoup} with the rescaling $\chi \rightarrow \chi\frac{A_\text{beam}}{A_\text{cell}}$. The same result applies to Config. 2. Consequently, the expression for the effective coupling strength $\Omega^{(i)}$ becomes
\begin{equation}
\Omega^{(i)}=\nu^{(i)} \left( \frac{A_\text{beam}}{A_\text{cell}} \right) \frac{n_{\text {cell}}}{N_{\text {cell}}} \sqrt{n_{\mathrm{ph}} N_{\text {cell}}} f^{(i)}(M) \;.
\end{equation}

\clearpage
\newpage

\section{Measurement of Stokes operators and photodiode calibration}

\subsection{Measurement of Stokes operators}\label{supp:secStokesMea}

\begin{figure}[H]
    \centering
    \includegraphics[width=0.6\linewidth]{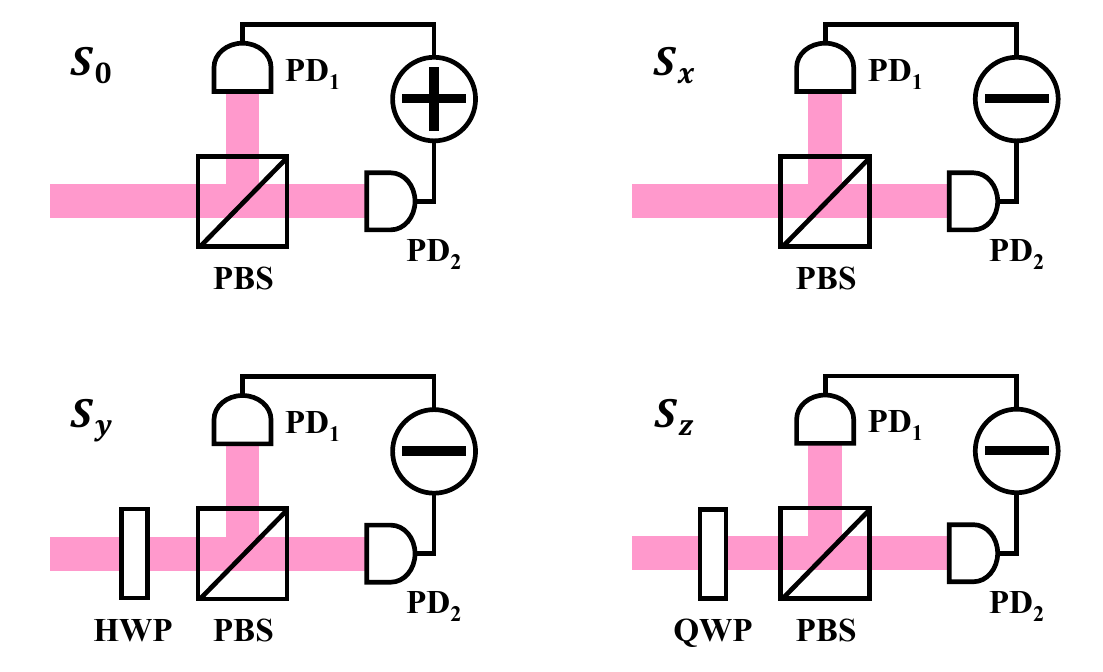}
    \caption{\textbf{Measurement method for different Stokes operators.} The transmission axis of the polarization beam splitter (PBS) is defined as the $x$-axis (corresponding to $0^\circ$), and the reflection axis is defined as the $y$-axis (corresponding to $\pi/2$). The optical axis of the half-wave plate (HWP) is set to $22.5^\circ$, and that of the quarter-wave plate (QWP) is set to $45^\circ$.}
    \label{StokesMeaSche}
\end{figure}

To illustrate how different Stokes operators are measured, let us start from their definition in terms of photon number (or flux) for different polarizations
\begin{equation}\label{supp:stokes}
\begin{aligned}
{S}_0 & =\frac{1}{2}\left({n}_{\mathrm{ph}}(x)+{n}_{\mathrm{ph}}(y)\right) \\
{S}_x & =\frac{1}{2}\left({n}_{\mathrm{ph}}(x)-{n}_{\mathrm{ph}}(y)\right) \\
{S}_y & =\frac{1}{2}\left({n}_{\mathrm{ph}}\left(+45^{\circ}\right)-{n}_{\mathrm{ph}}\left(-45^{\circ}\right)\right) \\
{S}_z & =\frac{1}{2}\left({n}_{\mathrm{ph}}\left(\sigma_{+}\right)-{n}_{\mathrm{ph}}\left(\sigma_{-}\right)\right).
\end{aligned}
\end{equation}
The measurement scheme of different Stokes operator~\cite{Thomasthesis} is shown in Fig.~\ref{StokesMeaSche}. We define the polarization axis of the light transmitted through the PBS as the $x$ axis (corresponding to $0^{\circ}$) and that of the reflected light as the $y$ axis (corresponding to $90^{\circ}$). Under this convention, the Stokes operator $S_0$ can be obtained by summing the signals from the two photodiodes, whereas $S_x$ can be obtained by taking their difference.

\begin{figure}[H]
    \centering
    \includegraphics[width=0.7\linewidth]{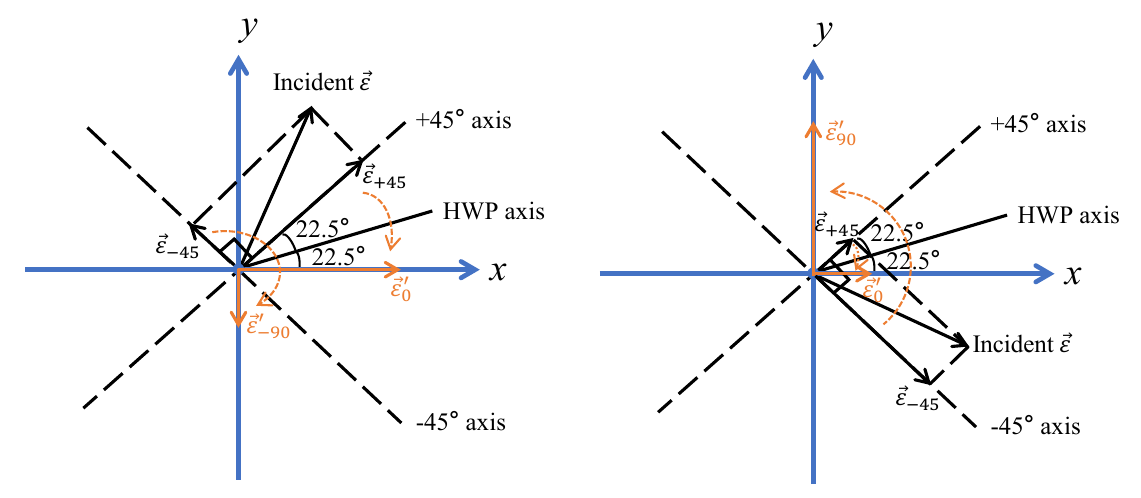}
    \caption{\textbf{Measurement principle for the Stokes operator $S_y$.} Note that both the left and right panels lead to the same conclusion.}
    \label{SyPrinciple}
\end{figure}

For the measurement of $S_y$, we need to add a half-wave plate with its optical axis set to $\pi / 8$ before the PBS. To illustrate the measurement principle more clearly, we provide a schematic diagram in Fig.~\ref{SyPrinciple}. Note that the left and right panels in Fig.~\ref{SyPrinciple} lead to exactly the same conclusion. Therefore, we describe the procedure using only the left panel. First, since $S_y$ measures the magnitude of the $+45^{\circ}$ component of the incident light field minus the $-45^{\circ}$ component, we can decompose the incident light field $\vec{\varepsilon}$ along $+45^{\circ}$ axis and $-45^{\circ}$ axis, denoting them as $\vec{\varepsilon}_{+45}$ and $\vec{\varepsilon}_{-45}$, respectively. After passing through the HWP, it is seen that the $\vec{\varepsilon}_{+45}$ component will be rotated to the $\vec{\varepsilon}_0^{\prime}$ component, which will align with the $x$ axis, while the $\vec{\varepsilon}_{-45}$ component will be rotated to the $\vec{\varepsilon}_{-90}^{\prime}$ component, which will align with the $y$ axis. In other words, the original $+45^{\circ}$ and $-45^{\circ}$ components of the incident light field, after passing through the HWP, are mapped to the $x$-axis and $y$-axis components. Therefore, by adding a PBS after the HWP, and subtracting the $y$-axis component from the $x$-axis component, we can obtain the $S_y$ operator of the incident light field.

\begin{figure}[H]
    \centering
    \includegraphics[width=0.7\linewidth]{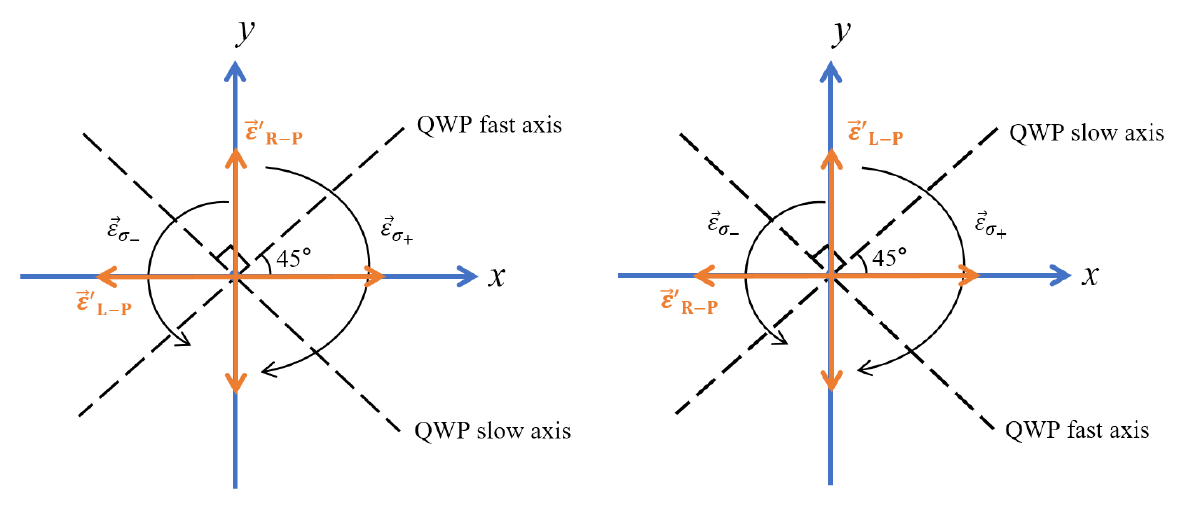}
    \caption{\textbf{Measurement principle for the Stokes operator $S_z$.} Note that the conclusions of the left and right panels differ only by a negative sign.}
    \label{SzPrinciple}
\end{figure}

For the measurement of $S_z$, we need to add a quarter-wave plate with its optical axis set to $\pi / 4$ before the PBS. To illustrate the measurement principle more clearly, we provide a schematic diagram in Fig.~\ref{SzPrinciple}. Note that the conclusions of the left and right panels in Fig.~\ref{SzPrinciple} differ only by a negative sign, so we just describe the procedure using only the left panel. First, since $S_z$ measures the magnitude of the $\sigma_{+}$ component of the incident light field minus the $\sigma_{-}$ component, we can decompose the incident light field $\vec{\varepsilon}$ into $\vec{\varepsilon}_{\sigma_{+}}$ and $\vec{\varepsilon}_{\sigma_{-}}$. We assume that the $+45^{\circ}$-axis is the fast axis of the QWP and the $-45^{\circ}$-axis is the slow axis. After passing through the QWP, $\vec{\varepsilon}_{\sigma_{+}}$ component will transform into $\vec{\varepsilon}_{\mathrm{R}-\mathrm{P}}^{\prime}$ component, and $\vec{\varepsilon}_{\sigma_{-}}$ component will transform into $\vec{\varepsilon}_{\mathrm{L}-\mathrm{P}}^{\prime}$. In other words, the original components of the light field, $\vec{\varepsilon}_{\sigma_{+}}$ and $\vec{\varepsilon}_{\sigma_{-}}$, after passing through the QWP, are mapped to the $y$-axis and $x$-axis components, respectively. Therefore, by adding a PBS after the QWP, and subtracting the $x$-axis component from the $y$-axis component, we can obtain the $S_z$ operator of the incident light field.

Based on the above discussion, we now introduce how to measure $S_{0,x, y, z}$ under the same experimental conditions. As shown in Fig.~\ref{StokesMeaSche}, by removing or replacing the waveplate in front of the PBS, one can measure different Stokes operators. For example, suppose we want to measure $S_{0,x, y, z}$ under a given experimental condition, we would proceed with the following steps: (1) Without placing any waveplate, record the two PD voltage signals — this allows us to obtain $S_0$ and $S_x$. (2) Insert a HWP with its optical axis set to $\pi / 8$, then record the two PD voltage signals — this allows us to obtain $S_y$. (3) Remove the HWP and insert a QWP with its optical axis set to $\pi / 4$, then record the two PD voltage signals — this allows us to obtain $S_z$.


\subsection{Measurement of the effective coupling strength $\Omega^{(i)}$}\label{supp:secOmegaMea}

The goal of our experiment is to estimate the effective coupling strength between light and nuclear spins $\Omega^{(i)}$, based on the measured PD voltages proportional to the light signal $X_S$ and other precise calibrations of the experimental setup.

To begin, let us call $V_1$ and $V_2$ the voltage signals from the two PDs shown in Fig.~\ref{StokesMeaSche}. These are related to the light signal as
\begin{equation}
\begin{aligned}
X_S & =\frac{\delta S_y}{\sqrt{\left\langle S_x\right\rangle_s}} \\
& =\frac{\frac{1}{2}\left[{n}_{\mathrm{ph}}\left(45^{\circ}\right)-{n}_{\mathrm{ph}}\left(-45^{\circ}\right)\right]}{\sqrt{\frac{1}{2}\left[{n}_{\mathrm{ph}}(x)-{n}_{\mathrm{ph}}(y)\right]}} \\
& =\frac{\sqrt{2}}{2} \cdot \frac{{n}_{\mathrm{ph}}\left(+45^{\circ}\right)-{n}_{\mathrm{ph}}\left(-45^{\circ}\right)}{\sqrt{{n}_{\mathrm{ph}}(x)}} \\
& =\frac{\sqrt{2}}{2} \cdot \frac{{n}_{\mathrm{ph}}\left(+45^{\circ}\right)-{n}_{\mathrm{ph}}\left(-45^{\circ}\right)}{\sqrt{{n}_{\mathrm{ph}}\left(45^{\circ}\right)+{n}_{\mathrm{ph}}\left(-45^{\circ}\right)}} \\
& =\frac{\sqrt{2}}{2} \cdot \frac{\frac{1}{\hbar \omega_p R_\text{pd} G}\left(V_1-V_2\right)}{\sqrt{\frac{1}{\hbar \omega_p R_\text{pd} G}\left(V_1+V_2\right)}} \\
& =\frac{\sqrt{2}}{2} \cdot \frac{1}{\sqrt{\hbar \omega_p R_\text{pd} G}} \cdot \frac{V_1-V_2}{\sqrt{V_1+V_2}},
\end{aligned}
\label{expXs}
\end{equation}
where $R_\text{pd}$ is the responsivity of the PD (A/W), $G$ is the transimpedance gain of the amplifier (V/A), $\omega_p$ is the angular frequency of the light field. 
As mentioned in the main text, assuming at $t=0$ we perform a slight tilt by an angle $\theta$ of the nuclear spin from the $x$ to the $z$ direction, which, in the presence of a small static magnetic field along $x$ results in a Larmor precession of the nuclear spin on the $yz$-plane at frequency $\omega_I$. This is described by
\begin{equation}
P_I(t)=P_I(0) \cos \left(\omega_I t\right),
\end{equation}
where $P_I(0)=\sin \theta \sqrt{M N_{\mathrm{cell}} / 2}$ is the initial nuclear spin state that depends on the tilt angle. 
Such a time evolution of the nuclear spin is mapped into the light field by the Faraday interaction, thus resulting in the signal
\begin{equation}
X_S(t)=\Omega^{(i)} P_I(0) \cos \left(\omega_I t\right).
\label{Xst}
\end{equation}
By combining Eqs.~\eqref{expXs} and \eqref{Xst}, and assuming that $V_1-V_2$ has the form of free-induction-decay signal $A e^{-\gamma t}\sin(\omega t + \phi)$, we can finally obtain the experimental calibration expression for the coupling strength as
\begin{equation}
\Omega^{(i)}=\frac{A_\text{FID}}{\sqrt{\hbar \omega_p R_\text{pd} G} \cdot \sqrt{V_1+V_2} \cdot \sqrt{M N_\text{cell}} \cdot \sin \theta},
\end{equation}
where $A_\text{FID}$ is the FID signal amplitude, $M$ is the nuclear polarization, $N_\text{cell}$ is the total number of ground state atoms, $\theta$ is the tilt angle of nuclear spin.


\subsection{Calibration of the responsivity $R_\text{pd}$ of the photodiodes}

From Eq.~\eqref{expOmega} in the main text, it can be seen that the responsivity $R_\text{pd}$ is an important parameter when we experimentally calibrate $\Omega^{(i)}$. Therefore, it is necessary to calibrate the responsivity $R_\text{pd}$ of the four PDs used in our experiments (two for the single-pass cell setup and two for the multi-pass cell setup). The calibration method we use is as follows: a laser beam is directed into the PD under test (note that each PD is equipped with a convex lens in front for focusing). Before entering the PD, the optical power of the beam is measured using a power meter. After entering, the resulting PD voltage produced by the PD and the transimpedance amplifier is recorded. By varying the incident optical power, we obtain a series of corresponding PD voltage values. A linear fit is then applied to these data to determine the responsivity $R_\text{pd}$ of the PDs.

\begin{figure}[H]
    \centering
    \includegraphics[width=0.95\linewidth]{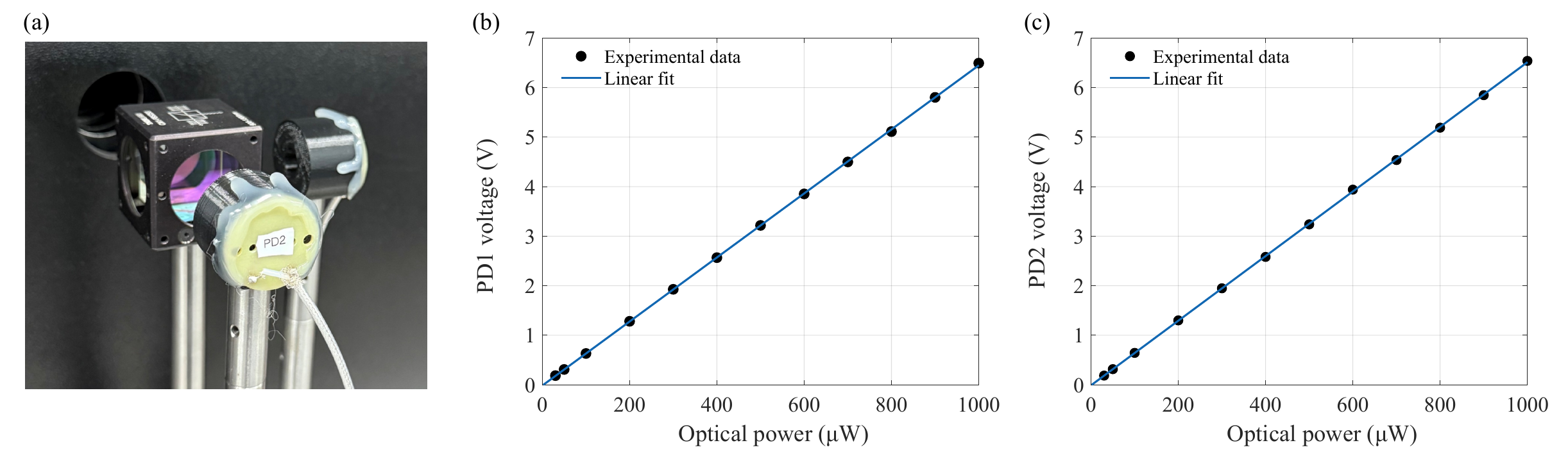}
    \caption{\textbf{Photodiodes calibration for the single-pass cell setup.} (a) Photograph of the two PDs used in the single-pass cell setup. (b) Calibration results for PD1. (c) Calibration results for PD2.}
    \label{SinglePD}
\end{figure}

For the two PDs used in the single-pass cell setup, the calibration results are shown in Fig.~\ref{SinglePD}, where the $R_\text{pd}$ of PD1 is 0.6466 A/W, and the $R_\text{pd}$ of PD2 is 0.6526 A/W.

\begin{figure}[H]
    \centering
    \includegraphics[width=0.95\linewidth]{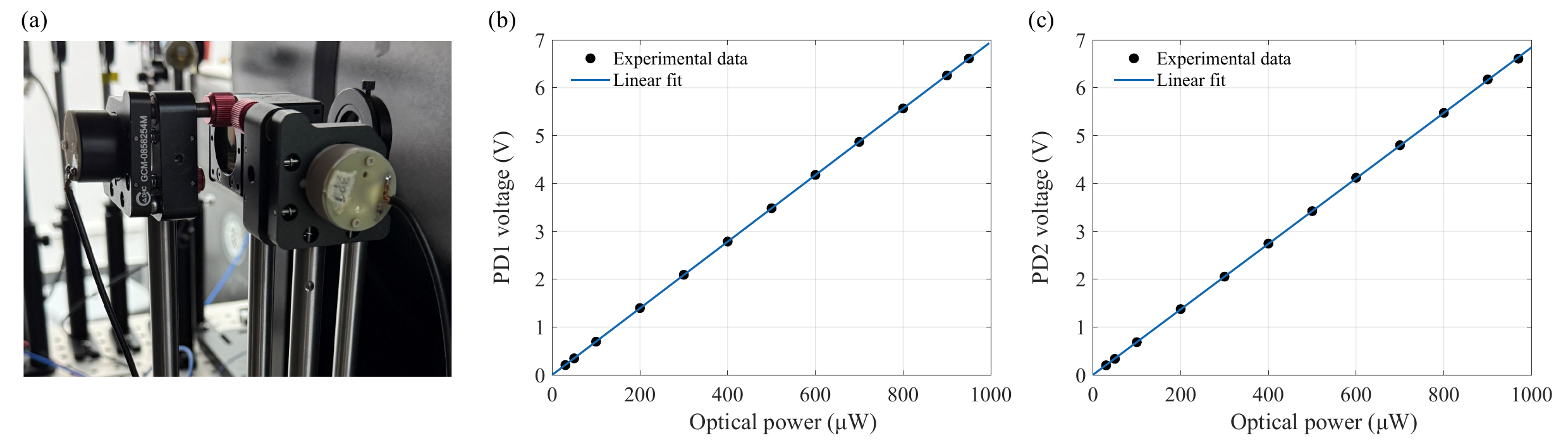}
    \caption{\textbf{Photodiodes calibration for the multi-pass cell setup.} (a) Photograph of the two PDs used in the multi-pass cell setup. (b) Calibration results for PD1. (c) Calibration results for PD2.}
    \label{MultiPD}
\end{figure}

For the two PDs used in the multi-pass cell setup, the calibration results are shown in Fig.~\ref{MultiPD}, where the $R_\text{pd}$ of PD1 is 0.6955 A/W, and the $R_\text{pd}$ of PD2 is 0.6839 A/W.

\clearpage
\newpage

\section{\label{calrffield}Calibration of the rf magnetic field pulse}

In this section, we describe the calibration procedure of the rf magnetic field pulse used in our experiment to perform Rabi rotations of the nuclear spin and discuss how the pulse setting are set under different holding magnetic fields.
The rf magnetic field pulse is generated by applying a current from a voltage-controlled current source with conversion coefficient of 50 mA/V, driven by a signal generator, to a coil with coil constant $\approx 70 \mathrm{~nT/mA}$. Therefore, amplitude $B_{rf}$, duration $t$ and frequency of the rf field can be adjusted simply by varying the sinusoidal voltage output $V_{cont}$ from the signal generator.

\begin{figure}[h]
    \centering
    \includegraphics[width=0.45\linewidth]{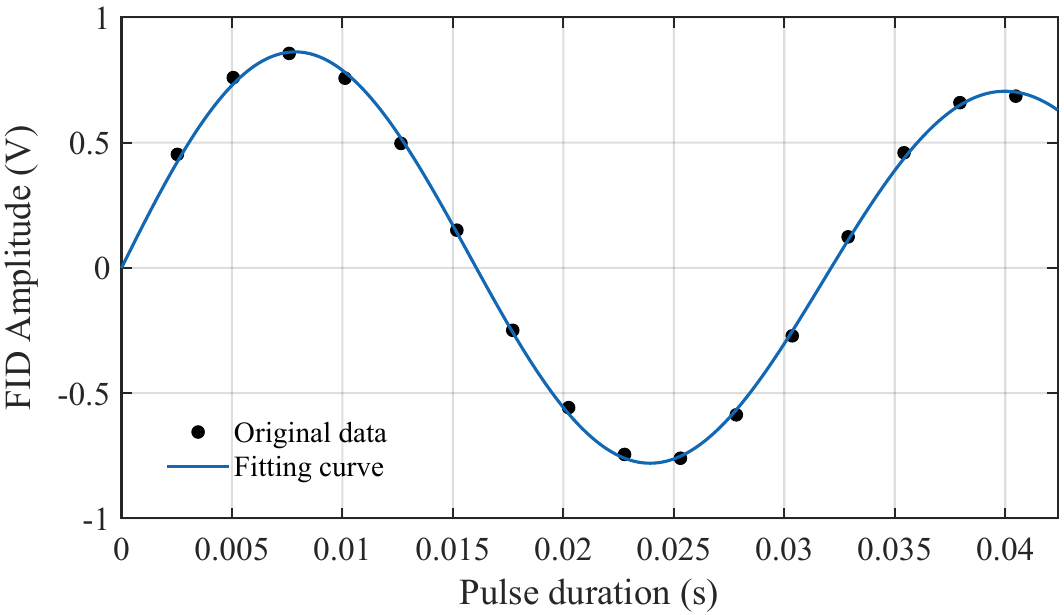}
    \caption{\textbf{Dependence of the FID signal amplitude after the pulse on the pulse duration.} Black points are the measured FID amplitudes, while the blue line is a fit to a decaying sinusoid.}
    \label{rffieldcal}
\end{figure}

In our calibration procedure, the pulse frequency $f_m$ is fixed at the Larmor frequency $f_L$. Taking the holding field to $B_0\approx 12000 \mathrm{~nT}$ as an example, we fix the pulse frequency at about $f_m=$395 Hz, and set the signal generator output voltage to $0.5~V_{pp}$. 
A measurement of how the FID signal amplitude after the pulse depends on the pulse duration $t$ is shown in Fig.~\ref{rffieldcal}. 
Fitting the original data with a damped sinusoid function, the calibration result indicates that the FID amplitude reaches its maximum at a pulse duration of 8.03 ms, which corresponds to the $\pi/2$ pulse duration. 
Then, we reduce the pulse duration to one period of the sine wave for convenience, i.e., $t=2.53\mathrm{~ms}$, and thus obtain a nuclear spin tilt angle of $\theta=28.4^{\circ}$.

\begin{figure}[h]
    \centering
    \includegraphics[width=0.78\linewidth]{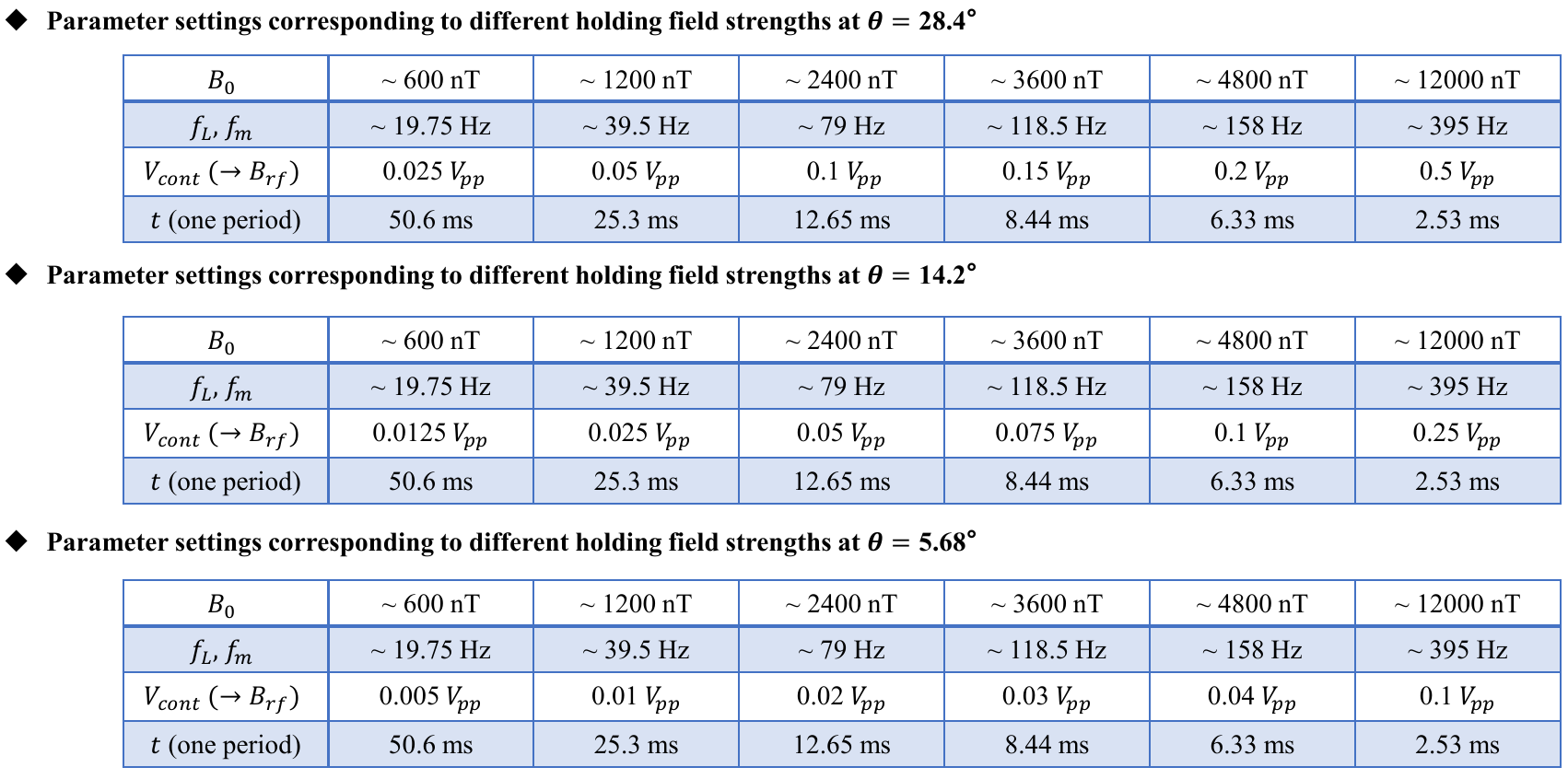}
    \caption{\textbf{Parameter settings summary.} Experimental parameter settings corresponding to different holding field strengths at different nuclear spin tilt angle.}
    \label{SingParaSet}
\end{figure}

Taking the above parameters as a reference, we discuss the following two cases. First, if we want to keep the holding magnetic field fixed and change the nuclear spin tilt angle, we just need to change $V_{cont}$ ($B_{rf}$) and keep all other parameters fixed, as shown in each column of Fig.~\ref{SingParaSet}. Second, if we want to keep the nuclear spin tilt angle fixed and change the holding field strength, as shown in each row of Fig.~\ref{SingParaSet}, we need to change the pulse frequency $f_m$ (for $f_m=f_L$) and the pulse duration $t$ (for one period of the sine wave), and then proportionally change the $V_{cont}$ ($B_{rf}$) to ensure $\theta=\gamma B_{rf} t$ is unchanged. In our experiments, we will choose three typical values of tilt angles $\theta\in\{5.68^\circ,14.2^\circ,28.4^\circ\}$, and change the holding fields in the range of $600 \mathrm{~nT} - 12000 \mathrm{~nT}$. Parameter settings corresponding to different holding field strengths at different nuclear spin tilt angle are shown in Fig.~\ref{SingParaSet}.

\clearpage
\newpage

\section{Metastable density measurement}\label{supp:secMetaNumMea}

\begin{figure}[h!]
    \centering
    \includegraphics[width=0.7\linewidth]{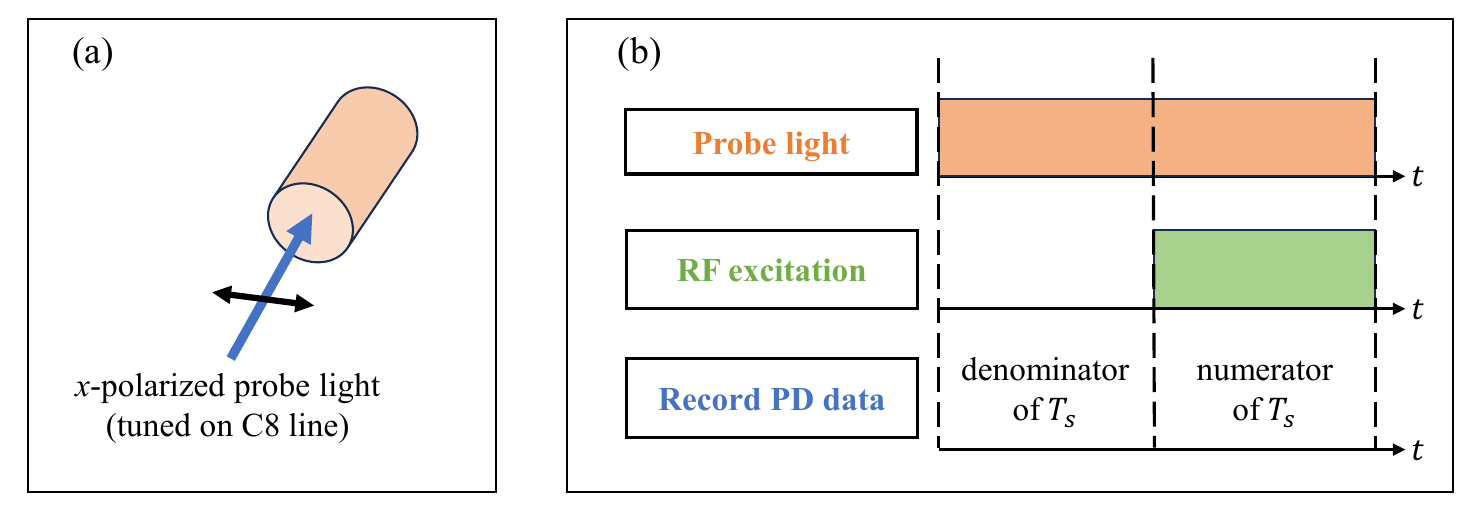}
    \caption{\textbf{Measurement of the metastable density.} (a) Illustration of the metastable density measurement scheme (Note that $B\approx 0$). A $x$-polarized probe beam on resonance with the \he $C_8$ line goes through the cell and its intensity at the output is measured by a photodiode to compute the transmission coefficient $T_s$. (b) Timing diagram for the measurement.}
    \label{nMeasingle}
\end{figure}

\subsection{Theory of metastable density measurement}

The measurement of metastable density follows the method described in Ref.~\cite{batzthesis} and will be briefly reviewed here. 
An illustration of the experiment is shown in Fig.~\ref{nMeasingle}.
We consider a probe beam propagating in the $z$ direction and assume the beam to be weak enough such that absorbed intensity is proportional to the local intensity at any point along the beam path (linear regime). 
Then, according to Lambert-Beer law, the absorbed light intensity per unit length is given by
\begin{equation}
\frac{d I(z)}{d z}=-k_a(z) I(z),
\label{lambbeer}
\end{equation}
where $k_a(z)$ is the local absorption rate and $z$ the corresponding linear coordinate along the beam path. Based on Eq.~\eqref{lambbeer}, the probe transmission coefficient $T_s$ is
\begin{equation}
T_s=\frac{I_{\mathrm{T}}}{I_0}=\exp \left(-\int_0^{L} k_a(z) d z\right),
\label{Ts}
\end{equation}
where $I_{\mathrm{T}}$ and $I_0$ are the transmitted and the incident probe intensities, respectively. For absorption measurements performed in the absence of pump light, the form of the local absorption rate is
\begin{equation}
k_a(z)=n_{\mathrm{m}}(z) \sum_{i, j} \hbar \omega_{i j} ~\overline{\Gamma}_{i j} ~a_i^{\mathrm{ST}}(M),
\label{localabsrate}
\end{equation}
where $n_{\mathrm{m}}(z)$ is the local metastable density, $\omega_{i j}$ is the angular frequency for the $\mathrm{A}_i \rightarrow \mathrm{B}_j$ line component of the $2^3 \mathrm{S}-2^3 \mathrm{P}$ transition, and $a_i^{\mathrm{ST}}(M)$ is the population of the state $\mathrm{A}_i$ in the spin-temperature distribution. 
The latter depends on the nuclear polarization $M$, and for $M=0$ we have $a_i(0)=1 / 6$. Finally, $\overline{\Gamma}_{i j}$ is the coefficient associated to the Maxwell-averaged optical transition rate $\overline{\gamma}_{i j}$ for the $\mathrm{A}_i \rightarrow \mathrm{B}_j$ line component, namely
\begin{equation}
\overline{\Gamma}_{i j}=\frac{\sqrt{\pi} \alpha_\text{FS} f}{m_e \omega D} ~T_{i j}(B) ~\mathrm{e}^{-\left(\delta_{\mathrm{L}}^{i j} / D\right)^2},
\label{normaltranrate}
\end{equation}
where we have used the definition $\overline{\Gamma}_{i j}=\overline{\gamma}_{i j}(z) / I(z)$. 
Here, $\alpha_\text{FS}$ is the fine structure constant, $f=0.5391$ is the oscillator strength of the $2^3 \mathrm{S}-2^3 \mathrm{P}$ transition, $m_e$ is the electron mass, $\omega$ is the light frequency, $D$ is the Doppler width for the $2^3 \mathrm{S} \rightarrow 2^3 \mathrm{P}$ transition, $\delta_{\mathrm{L}}^{i j}$ is the light detuning with respect to the $\mathrm{A}_i \rightarrow \mathrm{B}_j$ transition, $T_{ij}(B)$ is the transition matrix element that depends on magnetic field $B$~\cite{batzthesis,NacherJP50}.
We now rewrite the Eq.~\eqref{Ts} as
\begin{equation}
-\ln T_s=\int_0^{L} k_a(z) dz=L~n_{\mathrm{m}}^{\mathrm{S}} \sum_{i, j} \hbar \omega_{i j} ~\overline{\Gamma}_{i j} ~a_i^{\mathrm{ST}}(M),
\label{absorb}
\end{equation}
where
\begin{equation}
n_{\mathrm{m}}^{\mathrm{S}}=\frac{1}{L} \int_0^{L} n_{\mathrm{m}}(z) dz
\end{equation}
is the average metastable number density along the probe beam path. Thus, based on Eq.~\eqref{absorb}, we obtain the expression for $n_{\mathrm{m}}^{\mathrm{S}}$ as
\begin{equation}
n_{\mathrm{m}}^{\mathrm{S}}=\sigma^{-1} \frac{\left(-\ln T_s\right)}{L} \frac{1}{\sum_{i, j} a_i(M) ~T_{i j}~ \mathrm{e}^{-\left(\delta_{\mathrm{L}}^{i j} / D\right)^2}},
\label{nmMulti}
\end{equation}
where $\sigma=\hbar \omega \frac{\sqrt{\pi} \alpha_\text{FS} f}{m_e \omega D}=\unit{6.7979 \times 10^{-16} \cdot \left(\sqrt{T/\unit{300}{K}}\right)^{-1}}{m^{2}}$. Note that Eq.~\eqref{nmMulti} is the general expression for the case of multi-component transitions. If we consider the case of single-component transition, we can simplify Eq.~\eqref{nmMulti} into
\begin{equation}
n_{\mathrm{m}}^{\mathrm{S}}=\sigma^{-1} \frac{\left(-\ln T_s\right)}{L} \frac{1}{a_i(M) ~T_{i j} ~\mathrm{e}^{-\left(\delta_{\mathrm{L}}^{i j} / D\right)^2}}.
\end{equation}

\begin{figure}[h!]
    \centering
    \includegraphics[width=0.65\linewidth]{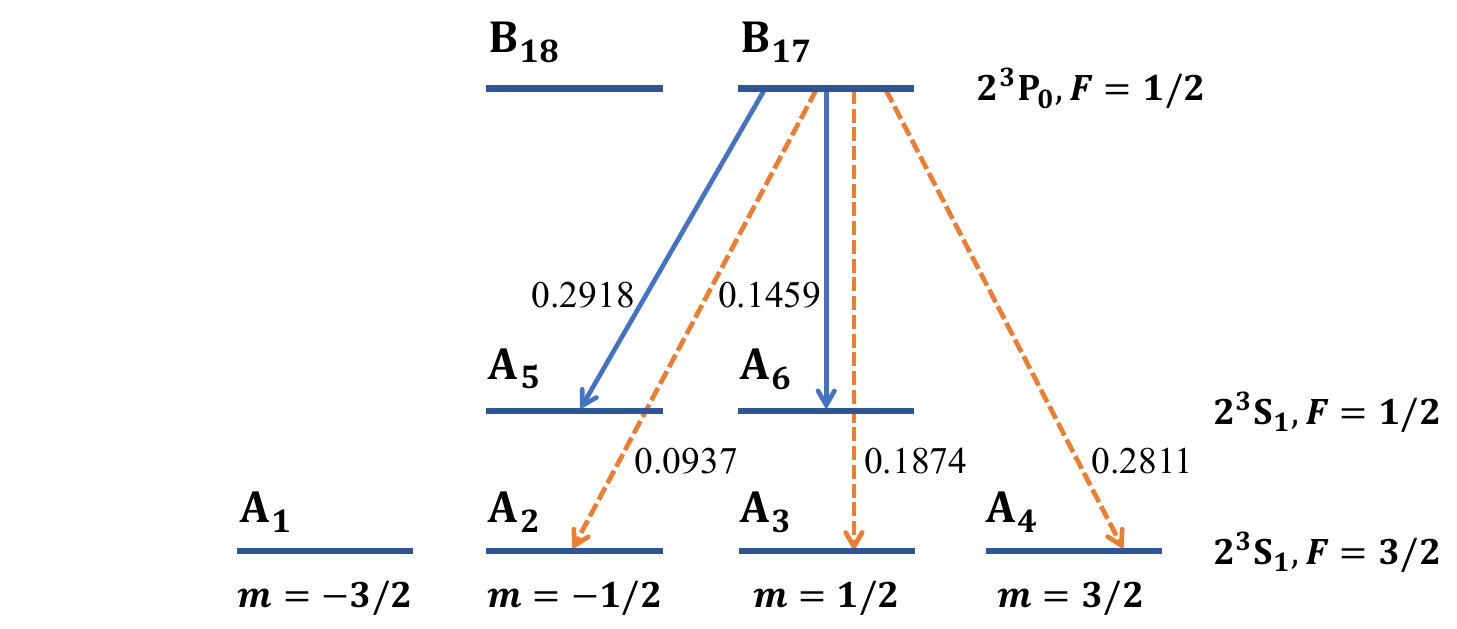}
    \caption{\textbf{$C_8$ and $C_9$ transition matrix elements.} Transition matrix elements for the \he $C_8$ (solid) and $C_9$ (dashed) transitions at $B=0$ field.}
    \label{nMeaEnergy}
\end{figure}

For a weak linearly-polarized probe on resonance with the $C_8$ transition and passing through an optically thin and unpolarized ensemble ($M=0$), we have
\begin{align}
    n_{\mathrm{m}}^{\mathrm{S}} &= \sigma^{-1} \dfrac{(-\ln(T_s))}{L} \dfrac{1}{ (\alpha_+ a_{5} T_{5,17} +\alpha_- a_{6} T_{6,18}) } \label{C8densitygen} \\
    & \approx 3.0 \dfrac{(-\ln(T_s))}{L} \cdot 10^{16} \unit{}{m^{-2}}
    \;,\label{C8density}
\end{align}
where we used $\alpha_+=\alpha_-=1/2$, $a_5=a_6=1/6$, and $T_{5,17}=T_{6,18}=0.291847$ are the transition matrix element for $C_8$ at $B=0$, see Fig.~\ref{nMeaEnergy} \cite{batzthesis}. 

\subsection{Metastable density measurement in the single-pass cell}

In this subsection, we present the results of the metastable density measurement in our single-pass cell. 
It is worth noting that during the measurement, the \he gas cell is placed inside a magnetic shielding cylinder, and no static magnetic field is applied, i.e., $B=0$. 
We use a linearly polarized light tuned on $C_8$ line to probe the optical absorption and calculate the metastable density through Eq.~\eqref{C8density}.
The probe transmission coefficient $T_s$ is estimated through the protocol shown in Fig.~\ref{nMeasingle}.
For each measurement, the first step is to turn off the rf discharge source and then record the reference PD voltage, which enters in the denominator of the probe transmission coefficient expression Eq.~\eqref{Ts}. 
The second step is to turn on the rf discharge source, set the rf power to the desired value, and then record the PD voltage again, which enters in the numerator of the probe transmission coefficient expression Eq.~\eqref{Ts}. 

In the single-pass cell measurements, we investigate the measured metastable density under two configurations: with and without an expanded probe beam, corresponding to beam waists of $\approx\unit{14}{mm}$ and $\approx\unit{1.4}{mm}$, respectively.
These two measurements serve as a check that saturation effects are not limiting or affecting our density measurements, due to a break-down of the assumptions behind the Lambert-Beer law.

\subsubsection{Case 1: Measurement of metastable density using an expanded probe beam}

\begin{figure}[H]
    \centering
    \includegraphics[width=0.85\linewidth]{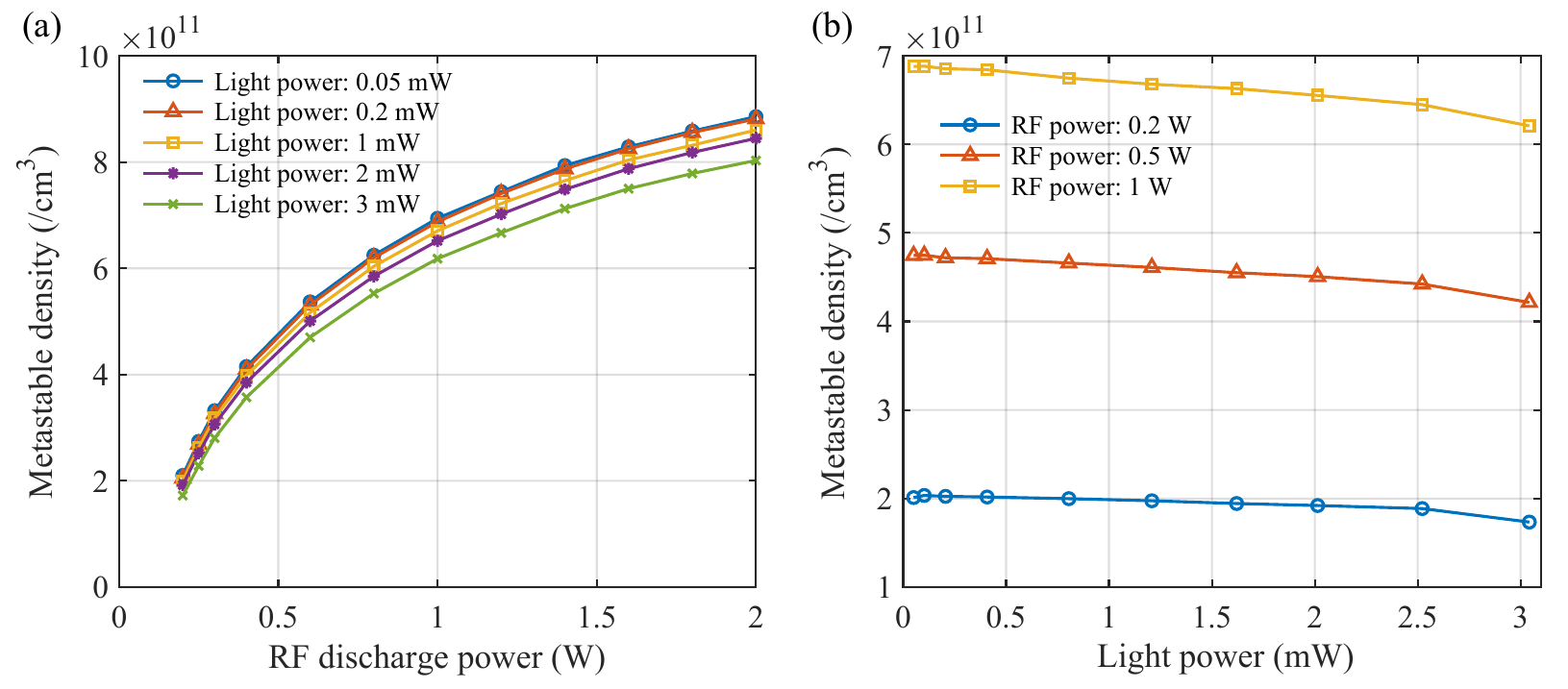}
    \caption{\textbf{Measurement results of metastable number density based on the $C_8$ probe using an expanded light beam.} (a) Metastable density results obtained by varying the rf excitation power under different incident light powers. (b) Metastable density results obtained by varying the incident light power under different rf excitation powers.}
    \label{nMeaSinC8}
\end{figure}

Measurement results taken with the expanded probe beam are shown in Fig.~\ref{nMeaSinC8}. 
In Fig.~\ref{nMeaSinC8}(a), we observe that the metastable density gradually increases and converges to a certain value with increasing rf discharge power, which aligns with our expectations. 
In Fig.~\ref{nMeaSinC8}(b), we observe that within a certain range of incident light power, the measured metastable density remains nearly constant. Only when the incident light power exceeds a specific threshold does the measured metastable density begin to change, possibly due to a break-down of the weak-light assumption behind the Lambert-Beer law.

\subsubsection{Case 2: Measurement of metastable density using an unexpanded probe beam}

\begin{figure}[H]
    \centering
    \includegraphics[width=0.85\linewidth]{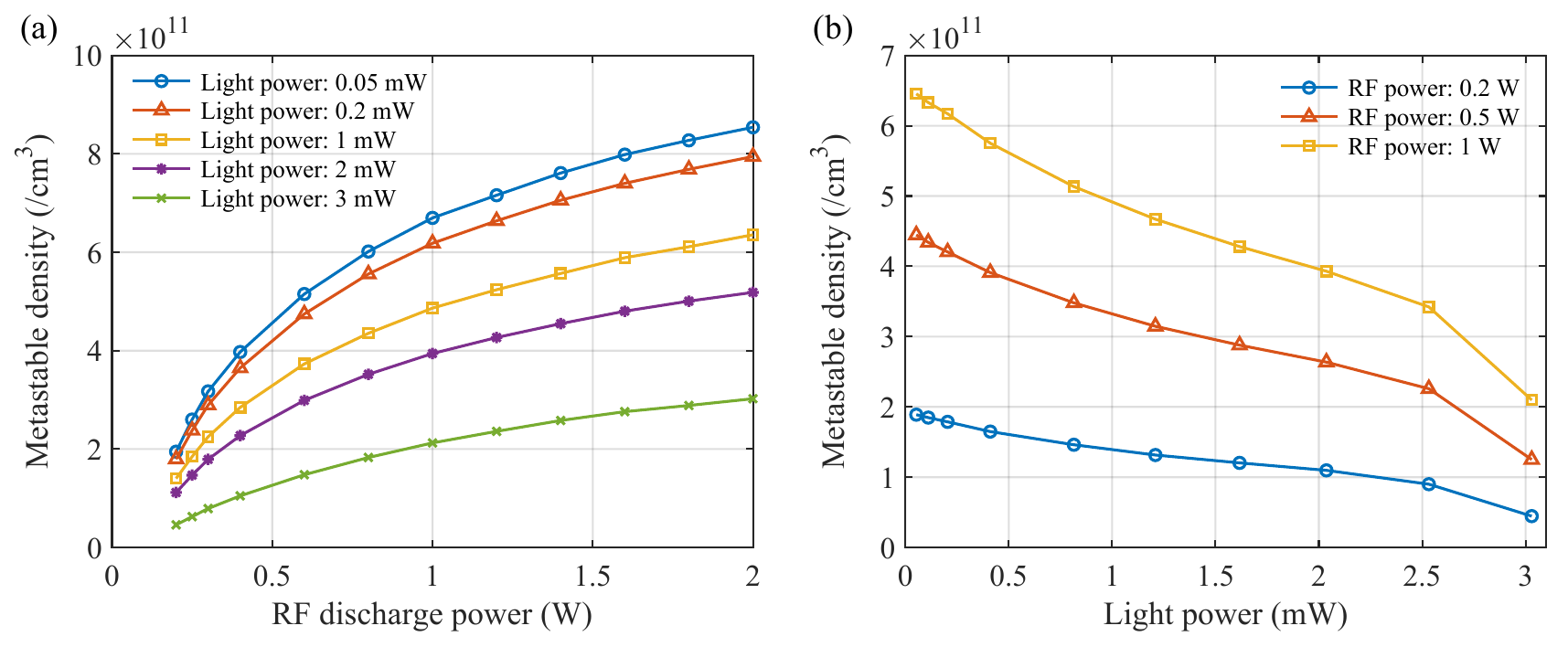}
    \caption{\textbf{Measurement results of metastable number density based on the $C_8$ probe using an unexpanded light beam.} (a) Metastable density results obtained by varying the rf excitation power under different incident light powers. (b) Metastable density results obtained by varying the incident light power under different rf excitation powers.}
    \label{nMeaSinC8-2}
\end{figure}

Measurement results taken with the unexpanded probe beam are shown in Fig.~\ref{nMeaSinC8-2}. 
From Fig.~\ref{nMeaSinC8-2}(a), it can be seen that the metastable density gradually increases and converges to a certain value with increasing rf discharge power, similarly to the case shown in Fig.~\ref{nMeaSinC8}(a). 
In Fig.~\ref{nMeaSinC8-2}(b), on the other hand, we observe that the measured metastable density exhibits a significant dependence on the incident light power, which stands in stark contrast to the results presented in Fig.~\ref{nMeaSinC8}(b). 
Let us emphasize here that the only difference between the two measurements lies in the beam spot size, meaning in the light intensity. 

This result further supports the requirement that, when using optical absorption method to measure metastable density, it is essential to ensure that the light intensity remains within the weak-light regime. Within this regime, the measured metastable density remains essentially constant. However, once the light intensity exceeds the weak-light threshold, the measured metastable density becomes strongly dependent on the incident light intensity, with higher intensities leading to lower measured densities.

Based on the above discussion, we conclude that when using the optical absorption method to measure the metastable density, the value obtained in the limit of zero incident optical power can be considered reliable. For example, by comparing Fig.~\ref{nMeaSinC8}(b) and Fig.~\ref{nMeaSinC8-2}(b), we observe that the blue curve in Fig.~\ref{nMeaSinC8}(b) remains consistently around $2 \times 10^{11} \mathrm{~cm}^{-3}$ throughout the entire weak-light regime, and the blue curve in Fig.~\ref{nMeaSinC8-2}(b) also yields a value of $2 \times 10^{11} \mathrm{~cm}^{-3}$ near zero incident power. Therefore, we may reasonably assume that the metastable density in the single-pass cell is approximately $2 \times 10^{11} \mathrm{~cm}^{-3}$ when the rf power is \unit{0.2}{W}. 

\subsection{Metastable density measurement in the multi-pass cell}

In this subsection, we present the results of the metastable density measurement in our multi-pass cell. The approach we follow is the same as for the single-pass cell.
Here, however, the probe beam passes through the hole in the front mirror to enter the multi-pass cell with an incident angle of $5^{\circ}$, traversing $\Npass=22$ times between the two mirrors, and then exits from the same hole. 
After that, a photodetector is employed to measure the transmitted light intensity.

\begin{figure}[h!]
    \centering
    \includegraphics[width=0.85\linewidth]{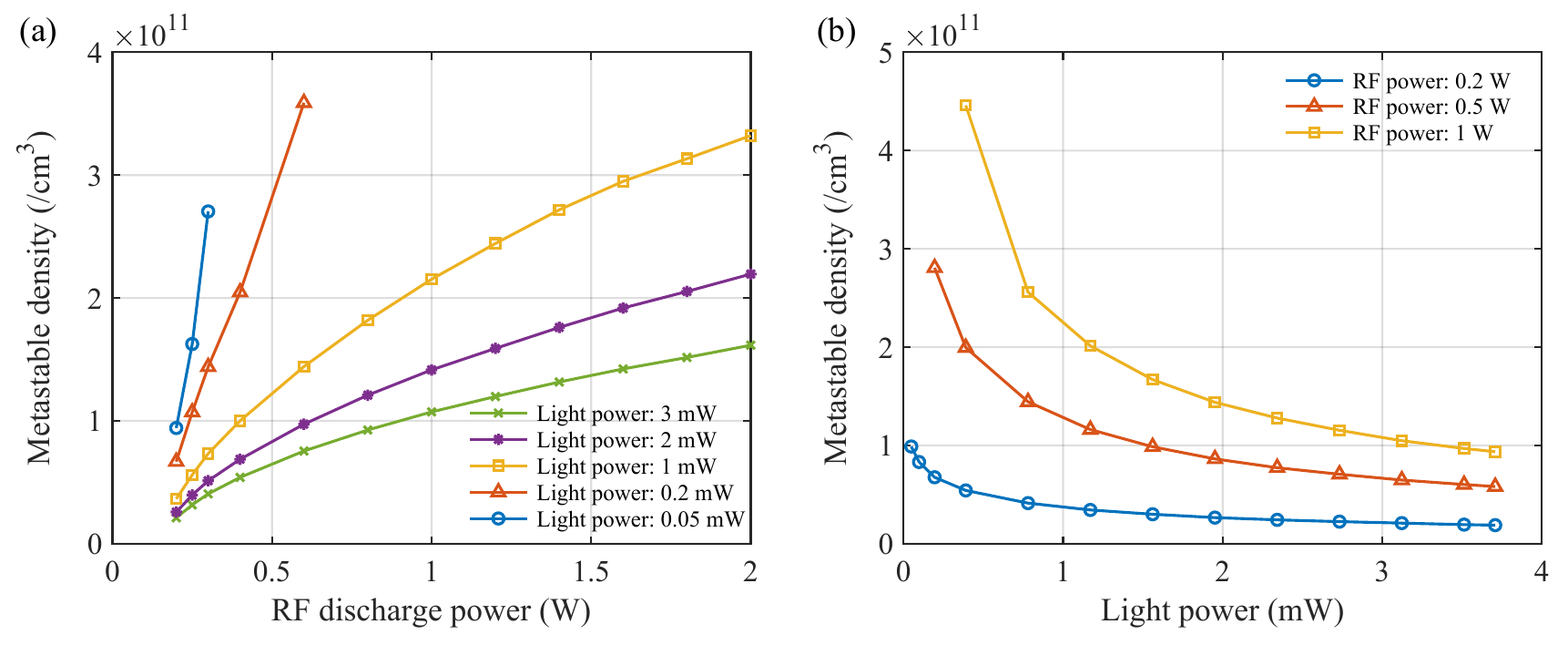}
    \caption{\textbf{Measurement results of metastable number density based on the $C_8$ probe.} (a) Metastable density results obtained by varying the rf excitation power under different incident light powers. (b) Metastable density results obtained by varying the incident light power under different rf excitation powers.}
    \label{nMeaMultiC8}
\end{figure}

Measurement results of the metastable density are shown in Fig.~\ref{nMeaMultiC8}. 
In Fig.~\ref{nMeaMultiC8}(b), we observe that the measured metastable number density shows a strong dependence on the incident light power. This phenomenon is very similar to the results shown in Fig.~\ref{nMeaSinC8-2}(b), because the light intensity has exceeded the weak-light regime. However, as discussed earlier, we believe that the metastable density value obtained in the limit of zero incident optical power can still provide a reasonable estimate. 
Thus, we conclude that the metastable density in the multi-pass cell is approximately $1 \times 10^{11} \mathrm{~cm}^{-3}$ when the rf power is \unit{0.2}{W}.

\clearpage
\newpage

\section{Nuclear polarization measurement}

\subsection{Theory of nuclear polarization measurement}

\begin{figure}[h!]
    \centering
    \includegraphics[width=0.6\linewidth]{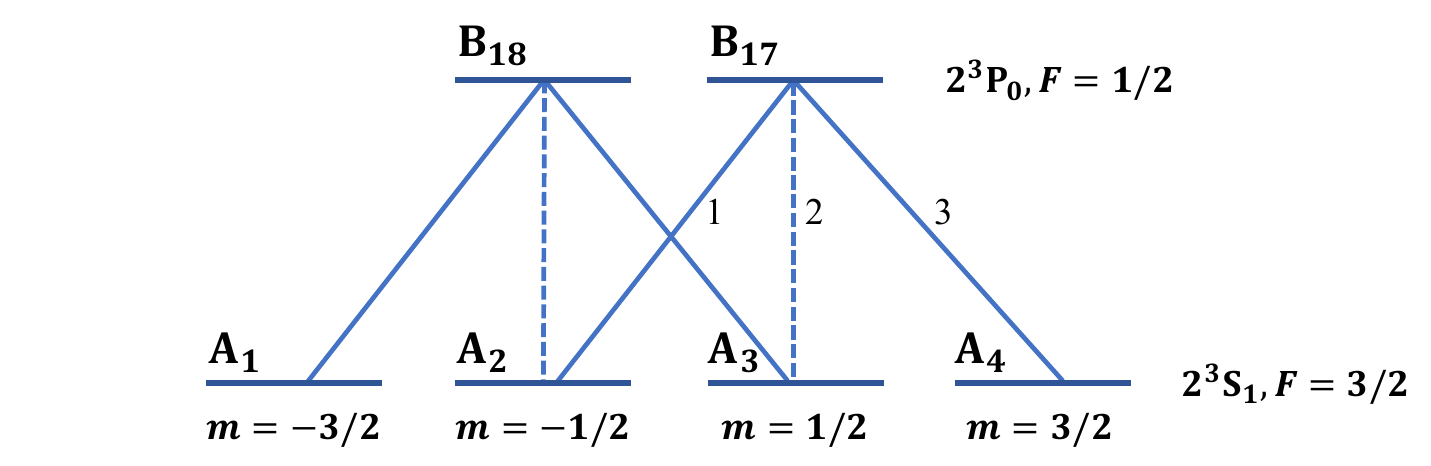}
    \caption{\textbf{Optical transitions relevant to the nuclear polarization measurements.} Magnetic sublevels involved in the optical detection using the $C_9$ component of the He-3 transition ($2{ }^3 \mathrm{S}_1, F={3}/{2} \rightarrow 2{ }^3 \mathrm{P}_0$). The solid lines are $\sigma$ transitions, the dotted lines are $\pi$ transitions. The numbers indicated along the transitions are the relative oscillator strengths.}
    \label{C9level}
\end{figure}

In order to measure the \he nuclear polarization we follow the method described in Ref.~\cite{He3polmea}, which we review here for completeness. 
Such method often called ``Transverse probe scheme", referring to the fact that the probe beam propagation is perpendicular to the magnetic field \cite{batzthesis}. 
In our experiment, we use a linearly polarized probe light tuned on resonance with the $C_9$ transition, see Fig.~\ref{C9level}, to perform a light-absorption measurements. 
The polarization axis of the light is at an angle $\theta = 45^{\circ}$ relative to the quantization axis, and thus can be decomposed into two equal components, $\varepsilon_{\pi}$ (polarization along $B_0$) and $\varepsilon_{\sigma}$ (polarization perpendicular to $B_0$). 
In this configuration, the formula for the light absorption signal is \cite{batzthesis}
\begin{equation}\label{lightabseq}
A_k(M) \propto n_{\mathrm{m}}^{\mathrm{S}}(M) \sum_{i, j} \hbar \omega_{i j} T_{i j}(B) e^{-\left(\delta_{\mathrm{L}}^{i j} / D\right)^2} a_i^{\mathrm{ST}}(M), \quad k=\pi, \sigma
\end{equation}
where $M$ is the nuclear polarization, $n_{\mathrm{m}}^{\mathrm{S}}$ is the average density of metastable atoms along the probe beam path, $\omega_{i j}$ is the angular frequency for the $\mathrm{A}_i \rightarrow \mathrm{B}_j$ line component of the $2^3 \mathrm{S}-2^3 \mathrm{P}$ transition, $T_{ij}(B)$ is the transition matrix element that depends on magnetic field $B$, $\delta_{\mathrm{L}}^{i j}$ is the light detuning with respect to the $\mathrm{A}_i \rightarrow \mathrm{B}_j$ transition, $D$ is the Doppler width for the $2^3 \mathrm{S} \rightarrow 2^3 \mathrm{P}$ transition, and $a_i^{\mathrm{ST}}(M)$ is the population of the state $\mathrm{A}_i$ in the spin-temperature distribution.

\begin{figure}[h]
    \centering
    \includegraphics[width=0.8\linewidth]{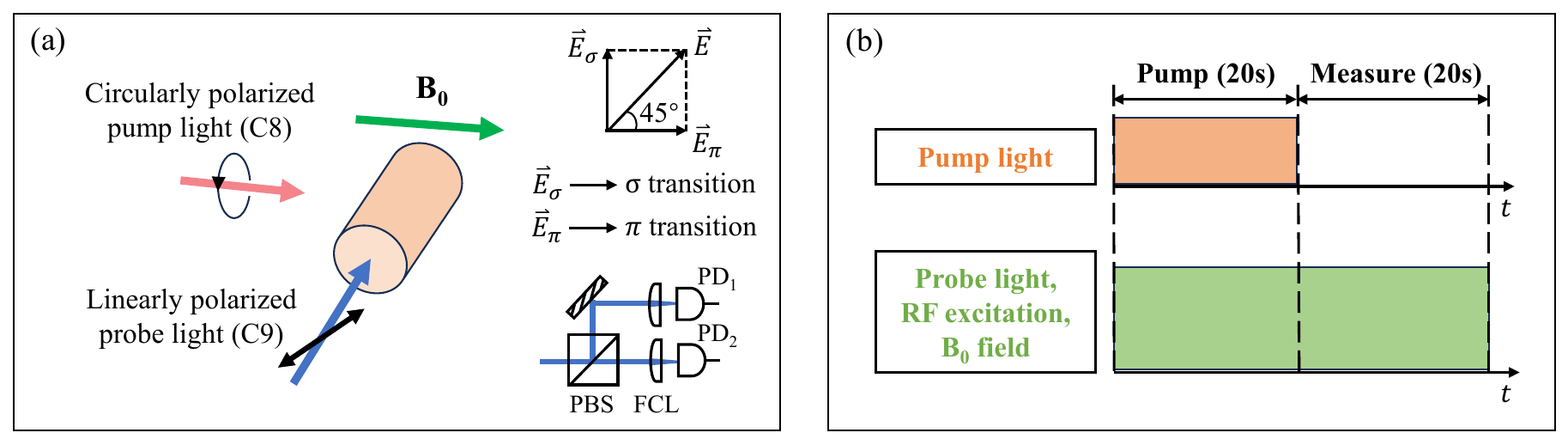}
    \caption{\textbf{Measurement of the nuclear polarization.} (a) Illustration of the nuclear polarization measurement scheme. A linearly-polarized probe beam on resonance with the \he $C_9$ line goes through the polarized cell and the intensity at the output of different polarization components is measured by two photodiodes to compute the absorption signal $A_k(M)$. (b) Timing diagram for the measurement.}
    \label{PMSsingle}
\end{figure}

After passing through the cell, the probe light is split by a PBS into two beams with perpendicular polarizations, whose intensities are then measured by two photodiodes, see Fig.~\ref{PMSsingle}.
This give the two readout voltages $V_k(M)$, one for each polarization $k=\pi, \sigma$.
By recording these voltages for an ensemble with polarization $M$ and normalizing them by the voltages $V_{k,\text{off}}$ recorded in the absence of rf discharge (see later Fig.~\ref{TypPolSignal}), we can write
\begin{equation}
\begin{aligned}
& A_\pi(M)=-\ln \left(V_\pi(M) / V_{\pi,\text{off}}\right) \\
& A_\sigma(M)=-\ln \left(V_\sigma(M) / V_{\sigma, \text{off}}\right) \\
& A_\pi(M=0)=-\ln \left(V_\pi(M=0) / V_{\pi, \text{off}}\right) \\
& A_\sigma(M=0)=-\ln \left(V_\sigma(M=0) / V_{\sigma, \text{off}}\right).
\label{absdef}
\end{aligned}
\end{equation}

Based on Eq.~\eqref{lightabseq}, we can define the light absorption ratio
\begin{equation}
R_L=\frac{A_\pi(M)}{A_\sigma(M)}=\frac{2\left(T_{2,18} a_2+T_{3,17} a_3\right)}{T_{1,18} a_1+T_{2,17} a_2+T_{3,18} a_3+T_{4,17} a_4} \;.
\label{lightabsratio}
\end{equation}
Since the population distribution in $2^3 \mathrm{S}$ is spin-temperature distribution, we have
\begin{equation}
\frac{a_1}{a_2}=\frac{a_2}{a_3}=\frac{a_3}{a_4}=\frac{1-M}{1+M}=e^\beta.
\label{STdis}
\end{equation}
Substituting Eq.~\eqref{STdis} into Eq.~\eqref{lightabsratio}, we get
\begin{equation}
R_L=\frac{2\left(2 e^{2 \beta}+2 e^\beta\right)}{3 e^{3 \beta}+e^{2 \beta}+e^\beta+3}=\frac{1-M^2}{1+2 M^2}.
\label{R}
\end{equation}
Based on Eq.~\eqref{R}, we find that
\begin{equation}
M=\sqrt{(1-R_L) /(1+2 R_L)}.
\label{Meq}
\end{equation}
This means that as long as we measure $R_L$, we can extract the magnitude of the ground-state polarization $M$. In our experiment, we will measure the normalized absorption ratio
\begin{equation}
\left[{A}_\pi(M) /{A}_\pi(M=0)\right] /\left[{A}_\sigma(M) / {A}_\sigma(M=0)\right]
\label{normR}
\end{equation}
as mentioned in Ref.~\cite{He3polmea}, to eliminate the dependence of small variations in probe laser intensity as well as variations in the density of metastable atoms.

\subsection{Nuclear polarization measurement in the single-pass and multi-pass cells}

In this subsection, we will introduce the nuclear polarization measurement results in our single-pass and multi-pass cells. The measurement scheme and the timing diagram for both cases are shown in Fig.~\ref{PMSsingle} (note that the cylindrical cell we have depicted in Fig.~\ref{PMSsingle}(a) is merely a representation, and can be directly replaced with the multi-pass cell).

First, we use a circularly polarized pump light resonant with the $C_8$ line to generate atomic polarization, with a pumping duration of 20 seconds to ensure that the ground-state polarization reaches a steady-state value. Then, the pump beam is turned off, and a linearly polarized probe beam propagating transversely is used to detect the optical absorption signal. The polarization axis of the probe beam is set at $45^{\circ}$, and the beam is resonant with the $C_9$ line. After traversing the cell, we use a PBS to separate the $\sigma$ and $\pi$ components of the light and detect the signals by two PDs. The measurement duration is also set to 20 seconds.

\begin{figure}[H]
    \centering
    \includegraphics[width=0.5\linewidth]{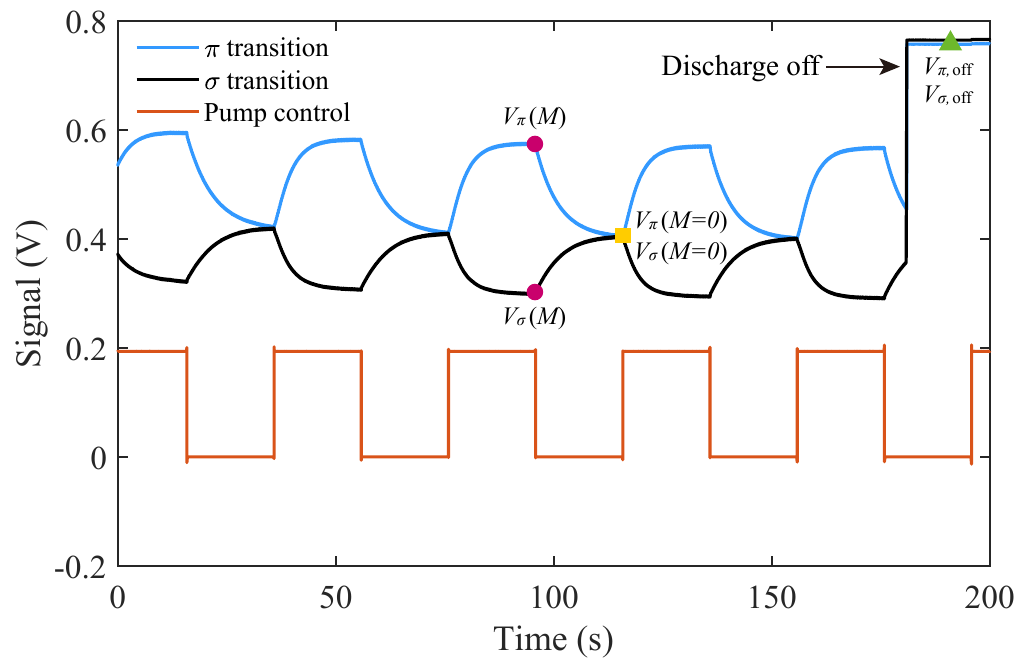}
    \caption{\textbf{Measurement signal for a typical nuclear polarization measurement.} The rf discharge is always on until the last $\approx\unit{20}{s}$. The blue line represents the PD voltage signal corresponding to the $\pi$ component of the probe light, the black line represents the PD voltage signal corresponding to the $\sigma$ component of the probe light, and the red line represents the pump light control signal (high level indicates on, low level indicates off). The data points marked by purple circles on the blue (black) curve correspond to $V_\pi(M)$ ($V_\sigma(M)$), those marked by yellow squares correspond to $V_\pi(M=0)$ ($V_\sigma(M=0)$), and those marked by green triangles correspond to $V_{\pi, \text { off }}$ ($V_{\sigma, \text { off }}$), respectively.}
    \label{TypPolSignal}
\end{figure}

A typical example of the experimental signal is shown in Fig.~\ref{TypPolSignal}.
Using the values of $V_\pi(M)$, $V_\sigma(M)$, $V_\pi(M=0)$, $V_\sigma(M=0)$, $V_{\pi, \text { off }}$, $V_{\sigma, \text { off }}$, we can calculate the nuclear polarization $M$ based on Eq.~\eqref{absdef}, Eq.~\eqref{Meq}, and Eq.~\eqref{normR}. 

\begin{figure}[H]
    \centering
    \includegraphics[width=0.5\linewidth]{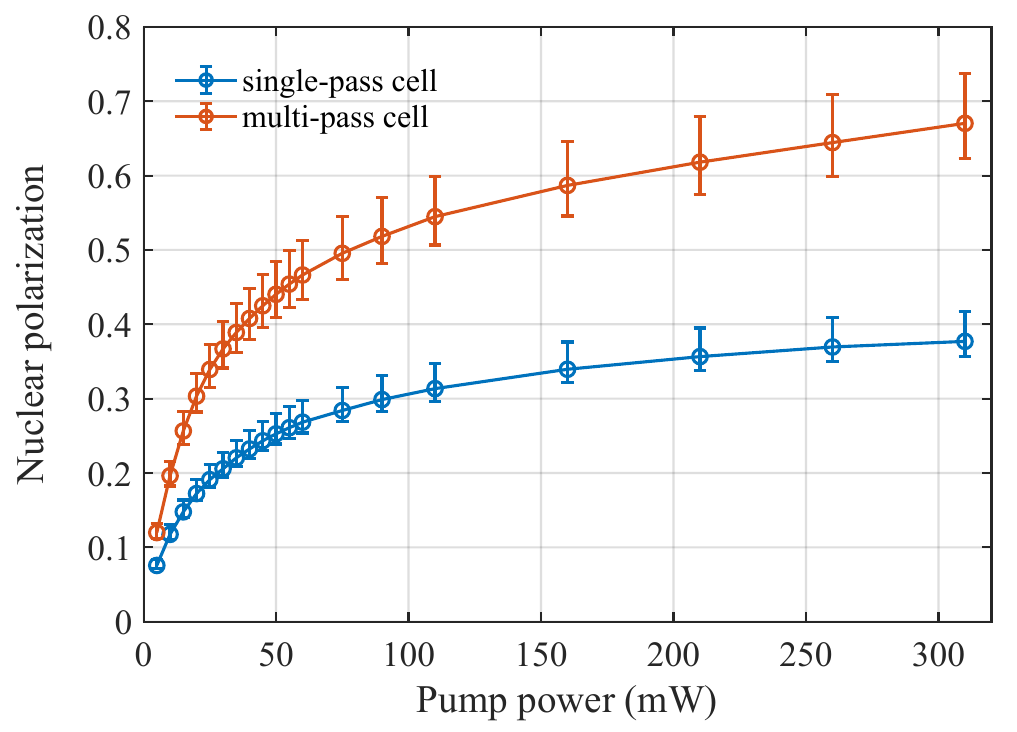}
    \caption{\textbf{Nuclear polarization as a function of pump power.} The nuclear polarization measurement results of the single-pass (blue line) and multi-pass (red line) cells. Error bars represent the variation of the measured results arising from slight changes in the probe power and detuning.}
    \label{Polcompare}
\end{figure}

The experimental parameters that we employed in the measurement are as follows: (1) the pump light power is the variable, and the pump wavelength is resonant with the $C_8$ line; (2) the probe light power is 0.3 mW, and the probe wavelength is resonant with the $C_9$ line; (3) The bias magnetic field $B_0$ is 12000 nT; (4) The rf discharge power is 0.2 W; (5) The pump duration is 20 seconds, and the probe duration is 20 seconds. Under these parameters, the nuclear polarization measurement results with different pump light powers of the single-pass and multi-pass cells are shown in Fig.~\ref{Polcompare}. 

\clearpage
\newpage

\section{Magnetic-field dependence of the light signal and effective coupling strength}\label{supp:secMagdepend}

\begin{figure}[H]
    \centering
    \includegraphics[width=0.7\linewidth]{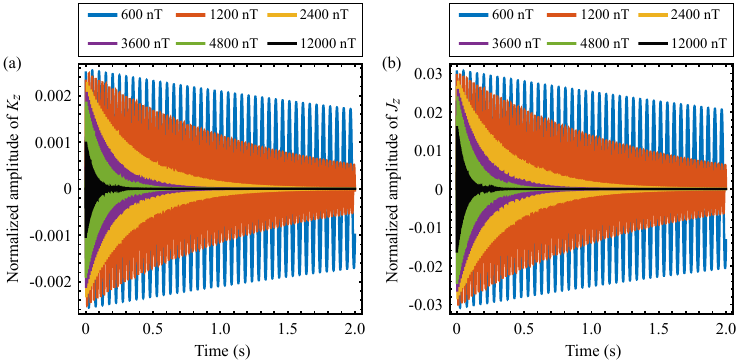}
    \caption{\textbf{Time evolution of the metastable transverse spin components $K_z$ (a) and $J_z$ (b), obtained from simulations based on the full model, for different magnetic-field strengths.} The values of both $K_z$ and $J_z$ are normalized to $n_{\text{cell}}$. The simulation parameters are: $\theta=2.9^\circ$, $M=0.38$, $T=300~\mathrm{K}$, $N_{\mathrm{cell}}=8.6\times10^{16}$, $n_{\mathrm{cell}}=7.2\times10^{11}$, and $V=3.6~\mathrm{cm}^3$.}
    \label{MagDependence}
\end{figure}

In this section, we discuss the dependence of the light signal $S_y$ and the effective coupling strength $\Omega^{(i)}$ on the magnetic-field strength. As shown in Fig.~\ref{fig:2}(c) of the main text, the measured FID amplitude $A_{\text{FID}}$ gradually decreases as the holding magnetic field increases, leading to the reduction of the calibrated effective coupling strength $\Omega^{(i)}$ shown in Fig.~\ref{fig:2}(d). To understand the origin of this behavior, we use the full model presented in Sec.~\ref{supp:secEqsatomic} to simulate the time evolution of the transverse spin components $K_z$ in the metastable $F=1/2$ manifold and $J_z$ in the metastable $F=3/2$ manifold at different magnetic-field strengths. The results are shown in Fig.~\ref{MagDependence}. It can be seen that the amplitudes of both $K_z$ and $J_z$ decrease as the magnetic field increases. According to Eq.~\eqref{eq:LMSyEOM}, this reduction directly leads to a smaller measured $S_y$ signal amplitude, thereby explaining the behavior observed in Fig.~\ref{fig:2}(c) of the main text.

The physical mechanism underlying the reduction of the transverse metastable collective spins with increasing magnetic field can be understood as follows. Owing to the large mismatch between the gyromagnetic ratios of the metastable and ground-state manifolds, the effect of this mismatch becomes increasingly pronounced as the magnetic field increases. As a result, although frequent MECs lock the metastable and ground-state spins together so that they precess at a common Larmor frequency, the mismatch simultaneously leads to a reduction of the transverse metastable spins and a shortening of the ground-state transverse relaxation time. It is worth noting that the latter effect is precisely why, as discussed in the section ``Towards nuclear spin squeezing" of the main text, an additional MEC-induced relaxation must be taken into account under nonzero magnetic-fields.
\begin{figure}[H]
    \centering
    \includegraphics[width=0.45\linewidth]{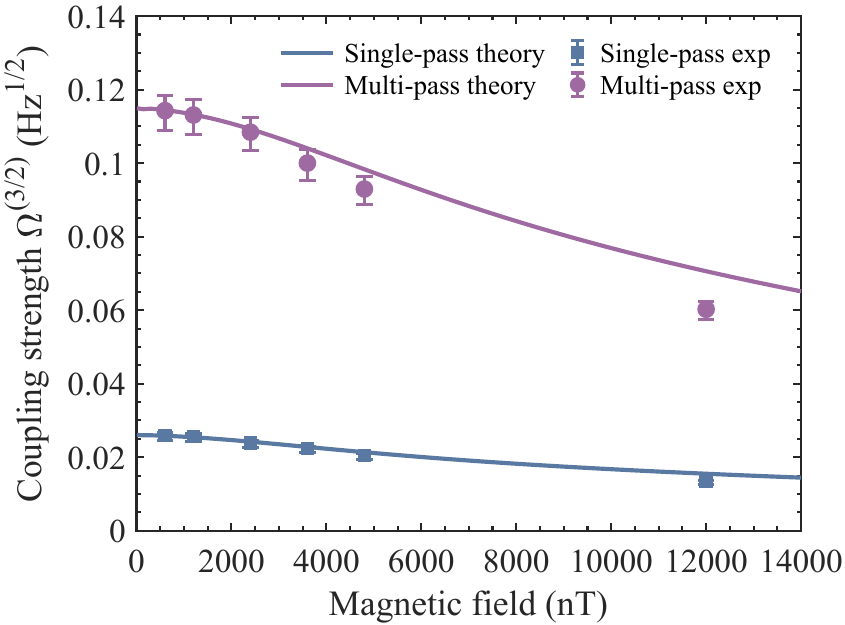}
    \caption{\textbf{Dependence of the effective coupling strength $\Omega^{(3/2)}$ on the bias magnetic field}. Square (circle) markers represent measurements from the single-pass (multi-pass) cell with $M=0.38$ ($M=0.67$) at $\theta=5.68^\circ$, while the curves show the simulation results based on the full model. The simulation parameters are the same as the experimental parameters. Error bars mostly come from uncertainty in the value of $M$.}
    \label{Coup_MagDepend}
\end{figure}

Having qualitatively explained the reduction of the $S_y$ signal amplitude with increasing magnetic field, we now quantitatively evaluate the magnetic-field dependence of the effective coupling strength $\Omega^{(i)}$ and compare the theoretical predictions with the experimental results shown in Fig.~\ref{fig:2}(d) of the main text. Taking Config.2 as an example, the results obtained for the single-pass and multi-pass cells are shown in Fig.~\ref{Coup_MagDepend}. The theoretical values are extracted by first solving the full model to obtain the amplitude of the light signal $X_S(t)$ and then dividing it by $P_I(0)$ (see Eq.~\eqref{eq:lightsig} of the main text). It should be noted that, as mentioned in the main text, the theoretical and experimental values are in good agreement for the single-pass cell, whereas the theoretical value for the multi-pass cell is approximately twice as large as the experimental one. Therefore, when plotting Fig.~\ref{Coup_MagDepend}, we multiply the theoretical values by correction factors of approximately 1.1 for the single-pass cell and 0.5 for the multi-pass cell, respectively, so that the comparison focuses only on the magnetic-field dependence of the effective coupling strength, i.e., its relative variation with magnetic field.

As can be seen from Fig.~\ref{Coup_MagDepend}, the magnetic-field dependence of the experimentally measured effective coupling strength exhibits excellent agreement with the theoretical prediction in the low-field regime. At higher magnetic fields, however, the experimental values decrease more rapidly than predicted by the theoretical model. A possible explanation is that magnetic-field gradients become increasingly important at large fields, leading to an additional reduction of the transverse metastable spin components $K_z$ and $J_z$, and consequently to a further decrease of the measured light signal amplitude.

\clearpage
\newpage

\section{Demonstration of the switchability of the Faraday interaction}\label{supp:secSwitchability}

The effective Faraday interaction between light and the nuclear spins of helium-3 mediated by MECs has a particularly attractive feature, namely, that it can be conveniently switched on and off by controlling the rf discharge. The physical mechanism is straightforward. When the rf discharge is turned on, a finite population of metastable helium atoms, $n_\text{cell}$, is maintained inside the cell. As a result, according to Eq.~\eqref{eq:OmegaThCorrection} in the main text, a finite effective coupling strength can be established between the light field and the ground-state helium-3 nuclear spins. In contrast, when the rf discharge is turned off, the metastable population rapidly decays to zero on the timescale of the metastable lifetime. Consequently, the MEC-mediated effective Faraday interaction between light and the ground-state nuclear spins is also switched off. To demonstrate this switchability in a more detailed and explicit way, in this supplementary section we present corresponding experimental results.

We first briefly describe the method used in the experiment to switch the rf discharge on and off. Taking the single-pass cell as an example, the minimum rf power required to sustain a stable discharge is about 0.2 W. Therefore, when the rf power is reduced to 0.1 W, the discharge is extinguished. On the other hand, after the discharge has been turned off for a certain period of time, the plasma can be re-established by increasing the rf power from 0.1 W to 1 W. Such a relatively large rf power is sufficient to re-establish the discharge without any additional ignition procedure. Based on this, in order to realize an on-off-on sequence of the rf discharge in the experiment, we manually adjust the output power of the rf excitation source to switch according to the sequence: a desired operating power greater than or equal to 0.2 W, then 0.1 W, and finally 1 W. Meanwhile, the measured $S_y$ signal is continuously recorded during this process for subsequent analysis.

\begin{figure}[H]
    \centering
    \includegraphics[width=0.9\linewidth]{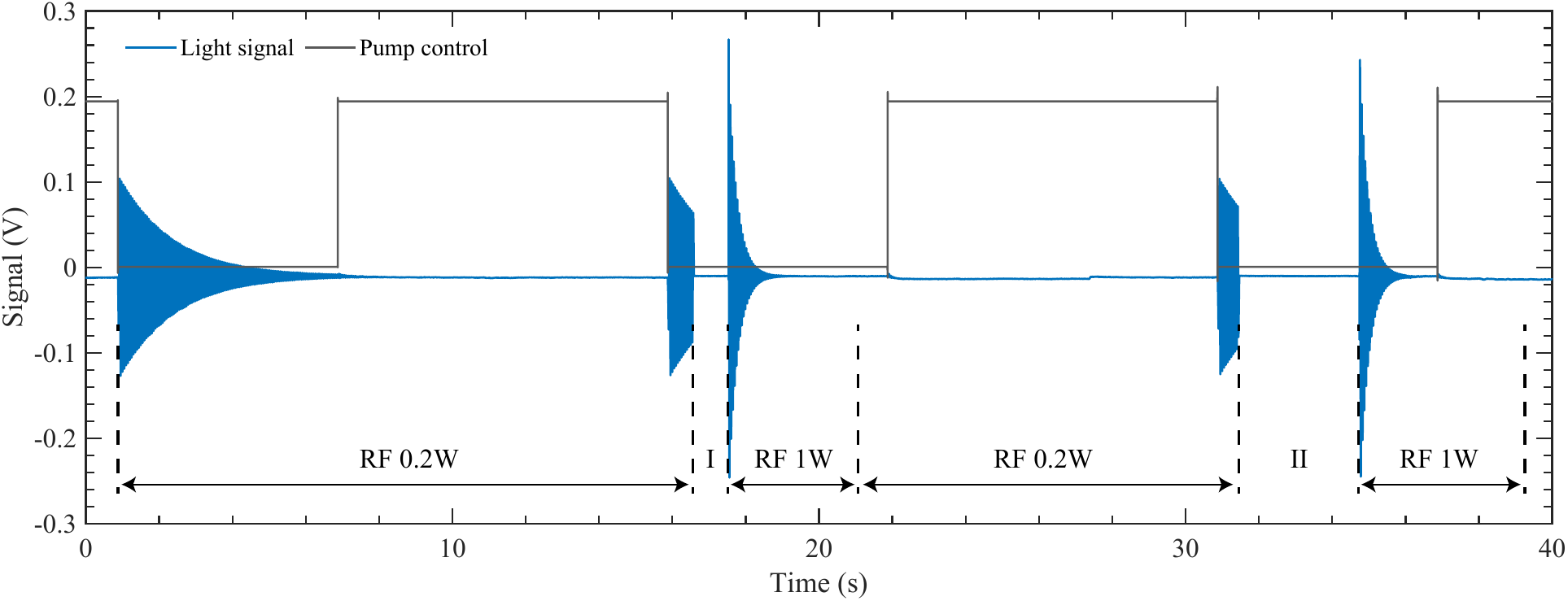}
    \caption{\textbf{Measurement signal obtained during a sequence where the rf discharge power is changed.} The blue trace represents the measured $S_y$ signal, while the gray trace shows the control signal of the pump light (high level indicates on, low level indicates off). The durations of the pumping and probing stages are set to 9 s and 6 s, respectively. For clarity, the control signal of the rf magnetic-field pulse is not shown. In the two time windows labeled “I” and “II”, the applied rf power is set to 0.1 W, for which the rf discharge is turned off.}
    \label{signal-0.2-0.1-1}
\end{figure}

Figure~\ref{signal-0.2-0.1-1} shows the raw signal obtained during the rf power switching sequence. In the first probing window, the rf power is kept at 0.2 W, for which the effective Faraday interaction is on and a complete FID signal is observed. In the second probing window, the rf power is first kept at 0.2 W for about 0.66 s and is then switched to 0.1 W for about 0.94 s. In this interval, the rf discharge is extinguished and the metastable population rapidly decays to zero. As a consequence, the MEC-mediated effective Faraday interaction between light and the ground-state nuclear spins is turned off, and the measured $S_y$ signal vanishes. After that, the rf power is increased to 1 W, which re-establishes the rf discharge and also the metastable population. These initially unpolarized metastable atoms rapidly acquire the polarization information from the ground state through MEC, thereby re-establishing the effective Faraday interaction between light and the ground-state nuclear spins. The metastable spin is dynamically locked to the ground-state nuclear spin, so that both precess together at the Larmor frequency and generate the reappearing FID signal.

It is worth noting that the amplitude of the reappearing FID signal is larger than that observed before the discharge was turned off, when the rf power was 0.2 W. This is because an rf power of 1 W leads to a significantly larger metastable atomic density than 0.2 W. According to Eq.~\eqref{eq:OmegaThCorrection} in the main text, this results in a significantly stronger effective coupling, and hence in a larger FID amplitude. At the same time, it should also be noted that increasing the rf power enhances various collisional processes in the plasma, which shortens the relaxation time of the ground-state nuclear spins.

Overall, the second probing window provides a complete demonstration of the effective Faraday interaction being switched from on to off and then back on again, thereby directly verifying its switchability. Finally, by comparing the FID amplitudes in the third and second probing windows after the discharge is re-established, one can see that the signal amplitude in the third window is smaller than that in the second one. This is because the signal amplitude depends not only on the metastable atomic density, but also on the ground-state nuclear polarization, as indicated by Eqs.~\eqref{eq:OmegaThCorrection} and \eqref{eq:lightsig} in the main text. The ground-state nuclear polarization is jointly determined by the duration of nuclear-spin decay before the discharge is turned off and by the duration of nuclear-spin decay after the discharge is turned off (where the relaxation rates in these two time intervals are significantly different), which ultimately leads to the difference in the reappearing FID amplitude between the two probing windows.

\end{document}